\documentclass[11pt,a4paper]{article}
\pdfoutput=1
\usepackage{ifpdf}
\usepackage{jheppub}
\usepackage{rotating}
\usepackage[bb=boondox]{mathalfa}
\usepackage{comment}

\usepackage{tikz-cd}

\graphicspath{ {./figures/} }
\usepackage{mathrsfs}
\usepackage{amsmath,amsfonts,amssymb}
\usepackage{mathtools}
\usepackage{dcolumn}% Align table columns on the decimal point
\usepackage{bm}% bold math
\usepackage[mathlines]{lineno}% Enable numbering of text and display math
\usepackage{float}
\usepackage{blindtext}
\usepackage{titlesec}
\title{Sections and Chapters}
\usepackage[toc,page]{appendix}
\usepackage{cancel}
\usepackage[thinc]{esdiff}
\usepackage[euler]{textgreek}
\usepackage{braket}
\usepackage{physics}
\usepackage{xfrac}
\usepackage{soul}
\usepackage{stmaryrd}
\usepackage{pifont}
\usepackage{gensymb}
\usepackage{makecell}
\usepackage{orcidlink}
\usepackage{slashed}
\usepackage{multirow}
\usepackage[normalem]{ulem}
\usepackage{booktabs}

\newcommand{\n}{\nonumber}

\newcommand{\lhp}{\lambda_{HS_3}}
\newcommand{\lsp}{\lambda_{S_2S_3}}

\allowdisplaybreaks

\usetikzlibrary{decorations.markings,decorations.pathmorphing}

\tikzset{
fermion/.style={
postaction={decorate},
decoration={markings,mark=at position 0.55 with {\arrow{>}}}
},
antifermion/.style={
postaction={decorate},
decoration={markings,mark=at position 0.55 with {\arrow{<}}}
}
}

\title{\textcolor{black}{Connecting Vector-like Muons, pNGB Dark Matter and Electroweak Phase Transition through Collider and Gravitational Wave}}

\author[a]{Jaydeb Das\,\orcidlink{0000-0001-6335-9377}\,,}
\author[b]{\!\!\,\,Saurabh Niyogi\,\orcidlink{0009-0008-2355-8847},\,}
\author[c,d]{Amir Subba\,\orcidlink{https://orcid.org/0000-0002-0126-1419\,}}
\affiliation[a]{Department of Physics, Indian Institute of Technology Guwahati,
North Guwahati, Assam-781039, India.}
\affiliation[b]{Department of Physics, Gokhale Memorial Girls' College, Kolkata, West Bengal-700020, India.}
\affiliation[c]{Wilczek Quantum Center, Shanghai Institute for Advanced Studies, Shanghai 201315, China}
\affiliation[d]{ University of Science and Technology of China, Hefei 230026, China }
\emailAdd{jaydebphys@rnd.iitg.ac.in}
\emailAdd{saurabhphys@gmail.com}
\emailAdd{amirsubba@ustc.edu.cn}
\abstract{We study an extension of the Standard Model with two complex scalar singlets, $S_2$ and $S_3$, charged under $\mathbb{Z}_2$ and $\mathbb{Z}_3$, respectively, and a muon-philic vector-like lepton $\psi$ carrying the same $\mathbb{Z}_3$ charge as $S_3$. The pseudoscalar associated with the $\mathbb{Z}_2$ sector remains stable due to the CP symmetry of the potential and serves as the dark matter (DM) candidate, with its mass generated through the corresponding soft breaking of the global $U(1)$ symmetry. The vector-like muon couples to the $\mathbb{Z}_3$ scalar and renders the second pseudoscalar unstable, thereby realizing an effectively single-component pNGB DM scenario. The pNGB nature strongly suppresses the tree-level spin-independent direct-detection cross section, while viable parameter regions reproduce the observed relic abundance and satisfy LHC monojet constraints. We further compute the one-loop contribution to DM-nucleon scattering and find that the resulting cross section remains below current experimental limits while being potentially accessible to future direct-detection experiments.
 At a multi-TeV muon collider, scalar mediated $ t$-channel processes can significantly enhance vector-like-muon pair production. We perform a detailed multivariate analysis for a future muon collider at 3 TeV center-of-mass energy. We also identify viable benchmark points exhibiting strong first-order electroweak phase transitions (SFOEWPT) with successful bubble nucleation, which can generate potentially observable stochastic gravitational wave (GW) signals. These results highlight the complementarity of dark matter searches, muon-collider probes, and SFOEWPT, with the resulting GW signals providing an additional probe of the extended scalar sector.}
\keywords{Muon collider, LHC Monojet searches, pNGB dark matter, Electroweak phase transition, Gravitational wave.}

\begin{document}
\maketitle
\flushbottom

%\preprint{}

%-----------------------------------
%.  NEW SECTION
%----------------------------------
\section{Introduction}
\label{sec:intro}

Cosmological observations require a non-baryonic matter component accounting for approximately $26\%$ of the present energy density of the Universe~\cite{Planck:2018vyg}, while no viable particle dark matter (DM) candidate exists within the Standard Model (SM). Among the simplest and most extensively explored possibilities is the weakly interacting massive particle (WIMP)~\cite{Steigman:1984ac,Lee:1977ua,Goodman:1984dc,Griest:1990kh,Jungman:1995df,Bertone:2004pz,Steigman:2012nb,Arcadi:2017kky,Bandyopadhyay:2022tsf} paradigm, in which the relic abundance is generated through thermal freeze-out. However, increasingly stringent limits from direct-detection experiments, including XENONnT~\cite{XENON:2024hup}, PandaX-4T~\cite{PandaX:2018wtu} and LUX-ZEPLIN~\cite{LZ:2022lsv,LZ:2024zvo}, strongly constrain conventional WIMP scenarios in which the same interaction controls both DM annihilation and elastic scattering off nuclei. This motivates frameworks in which the direct-detection amplitude is naturally suppressed without compromising efficient thermal annihilation in the early Universe.

Pseudo-Nambu--Goldstone boson (pNGB) dark matter provides an elegant realization of this idea~\cite{Gross:2017dan,Azevedo:2018exj,Ishiwata:2018sdi,Karamitros:2019ewv,Kannike:2019wsn,Alanne:2020jwx,Abe:2024vxz,Kannike:2019mzk,Biekotter:2025vxl,Bernal:2025qkj,Sheikh:2025qym,Liu:2022evb,Ghosh:2024ing,Okada:2020zxo,Abe:2020iph}. In the simplest construction, a complex scalar possesses an approximate global U(1) symmetry which is spontaneously broken, while a soft explicit breaking generates a mass for its pseudoscalar component. The pNGB nature of the DM enforces a cancellation between the CP-even scalar mediated contributions to DM--nucleon scattering, causing the tree-level amplitude to vanish in the zero-momentum-transfer limit~\cite{Gross:2017dan,Azevedo:2018exj,Ishiwata:2018sdi,Karamitros:2019ewv,Kannike:2019wsn}. Importantly, this cancellation is specific to the $t$-channel kinematics relevant for direct detection and does not suppress $s$-channel DM annihilation at finite center-of-mass energy. The pNGB framework can therefore reconcile a standard thermal relic with stringent direct-detection constraints without requiring unnaturally small portal interactions.

A seemingly independent cosmological puzzle is the origin of the baryon asymmetry. Electroweak baryogenesis requires the Sakharov conditions~\cite{Sakharov:1967dj}, including a sufficiently strong departure from thermal equilibrium. This can be realized if electroweak symmetry breaking proceeds through a strong first-order electroweak phase transition (SFOEWPT). For the measured Higgs-boson mass, however, nonperturbative studies establish that the electroweak phase transition in the SM is a smooth crossover~\cite{Kajantie:1996mn,Kajantie:1996qd}. New dynamics in the scalar sector is therefore required to modify the finite-temperature Higgs potential and generate a barrier between the symmetric and broken phases.

Scalar singlet extensions are particularly attractive in this respect because they provide one of the minimal modifications of the SM Higgs sector while simultaneously offering a natural connection to dark-sector physics~\cite{Choi:1993cv,Profumo:2007wc,Espinosa:2011ax,Profumo:2014opa,Curtin:2014jma,Kurup:2017dzf,Chiang:2017nmu}. Singlet vacuum expectation values (VEVs) and Higgs-portal interactions can substantially alter the vacuum structure and may lead to one-step or multi-step thermal histories, including transitions through an intermediate singlet-breaking phase. In conventional renormalizable singlet scenarios, however, obtaining a sufficiently strong transition is often correlated with sizable portal couplings, appreciable scalar mixing, or relatively light additional scalars. These features are increasingly constrained by Higgs signal-strength measurements, searches for additional scalar resonances and multi-Higgs production at the LHC. The simultaneous requirements of vacuum stability, perturbative unitarity, collider compatibility, viable DM and a completed SFOEWPT therefore make the interplay between the dark and electroweak sectors highly non-trivial.

Motivated by this complementarity, we consider an extension of the SM containing two $SU(2)_L$ complex singlet scalars, $S_2~ \rm{and}~ S_3$, transforming non-trivially under $\mathbb Z_2$ and $\mathbb Z_3$ symmetries, respectively. Under the $\mathbb{Z}_2$ symmetry, $S_2 \to -S_2$, whereas $S_3$ and the SM fields remain even; under the $\mathbb{Z}_3$ symmetry, $S_3 \to e^{2\pi i/3} S_3$, while all SM fields and $S_2$ transform trivially.
%Under $\mathbb Z_2$, $S_2\to -S_2$, whereas $S_3$ and the SM fields remain even; under $\mathbb Z_3$, $S_3\to e^{2\pi i/3}S_3$.
After spontaneous symmetry breaking, the pseudoscalar components of the two complex fields emerge as pNGB states, with their masses generated by explicit breaking of the corresponding global symmetries. The scalar sector therefore provides both the ingredients required for pNGB dark matter and additional field directions capable of producing a non-trivial thermal history.

An important distinction arises between the $\mathbb Z_2$ and $\mathbb Z_3$ sectors. The zero-momentum cancellation of the pNGB direct-detection amplitude is exact only for particular forms of explicit symmetry breaking. If the global $U(1)$ symmetry is broken by operators of dimension higher than two, as can occur in the $\mathbb Z_3$ sector, a non-vanishing tree-level spin-independent scattering amplitude can reappear~\cite{Gross:2017dan,Abe:2021byq,Abe:2020iph,Kannike:2019mzk}. Consequently, a genuine two-component DM realization would be subject to restrictive direct-detection (DD) constraints, especially in regions with appreciable scalar mixing.

 Therefore, in order to avoid a two-component pNGB DM scenario, we instead introduce a vector-like lepton with muon-type quantum numbers and suitable discrete charges, allowing a renormalizable Yukawa interaction with $S_3$. This interaction renders the pseudoscalar coming from $S_3$ unstable by opening its decay into muons, while the pseudoscalar originating predominantly from $S_2$ remains the stable DM candidate. The vector-like muon (VLM) is therefore not merely an auxiliary ingredient introduced to remove a second DM component: it provides an additional phenomenological connection between the extended scalar sector and the visible sector. Vector-like leptons are independently well-motivated extensions of the SM and are actively constrained through collider searches \cite{ATLAS:2024mrr,CMS:2024bni}, while their preferential connection to the muon sector makes the present construction particularly relevant for both LHC phenomenology and future muon-collider studies.

The resulting framework thus brings together several otherwise distinct probes of new physics. The pNGB structure suppresses conventional direct-detection signatures, the extended scalar potential can support a SFOEWPT, and the VLM sector provides an additional collider probe. Although the DM–nucleon spin-independent (SI) scattering cross section is suppressed $(\mathcal{O}(10^{-80}\text{cm}^2))$ at tree level due to the pNGB nature of the DM candidate, a non-vanishing contribution arises at one loop~\cite{Alanne:2020jwx,Abe:2024vxz,Azevedo:2018exj,Ishiwata:2018sdi,Glaus:2020ihj,Abe:2022mlc}. In this work, we calculate the loop-induced SI DM–nucleon scattering cross section and study the resulting direct-detection signal. Despite the tree-level suppression, the loop-level contribution can yield a potentially observable signal in direct-detection experiments across the phenomenologically viable parameter space. Moreover, a first-order cosmological phase transition can generate a stochastic gravitational-wave~(GW) background through sound waves in the thermal plasma, magneto-hydrodynamic turbulence and depending on the transition dynamics, expanding and colliding bubbles~\cite{Hindmarsh:2013xza,Caprini:2015zlo,Caprini:2019egz,Ghorbani:2018yfr,Kang:2017mkl,Choudhury:2026jus,Choudhury:2026laq,Das:2026zuo,Srivastava:2025oer,Chaudhuri:2022sis,Lu:2026byr,Borah:2024emz,Chiang:2019oms,Borah:2023zsb,Behera:2026tmo,Bhattacharyya:2026esv,Gazi:2024boc}. The GW spectrum carries information about the transition temperature, released vacuum energy and characteristic time scale, thereby offering a direct probe of the thermal history that is complementary to laboratory measurements.

In this work, we systematically investigate this interplay. We impose the relevant theoretical consistency conditions together with collider and dark-matter constraints and study the zero- and finite-temperature effective potential, including thermal corrections and daisy resummation. We identify regions supporting a SFOEWPT and examine the associated thermal trajectories, paying particular attention to multi-step transitions involving intermediate singlet-breaking minima. Since the existence of a thermodynamically favorable first-order transition does not by itself guarantee its cosmological completion, we further calculate the bubble-nucleation dynamics. For viable benchmark scenarios, we determine the resulting stochastic GW spectra and assess their prospects at future interferometers, including LISA~\cite{LISA:2017pwj}, BBO~\cite{Yagi:2011wg} and DECIGO~\cite{Nakayama:2009ce}, together with complementary sensitivities of future gravitational-wave facilities. Our analysis therefore illustrates how dark-matter phenomenology, collider searches and gravitational-wave observations can jointly probe the structure of an extended scalar sector and its role in the thermal evolution of the early Universe.

The paper is organized as follows. The theoretical framework and underlying symmetry are discussed in Sec.\,\ref{model}, along with the relevant theoretical and LHC constraints. In Sec.\,\ref{muon-coll}, we present a detailed collider study of the VLM. The DM phenomenology, including the DM--nucleon scattering cross section at one-loop level, is discussed in Sec.\,\ref{dm-pheno}. In Sec.\,\ref{ewpt}, we study the SFOEWPT and the resulting stochastic GW background. Finally, we summarize our results and conclude in Sec.\,\ref{summary}.

%%%%%%%%%%%%%%%%%
\section{Theoretical Framework}
\label{model}
We extend the SM scalar and fermion sectors by introducing two complex singlet scalars, $S_2$ and $S_3$, together with a vector-like muon $\psi$. The new fields are distinguished by two discrete symmetries, $\mathbb{Z}_2$ and $\mathbb{Z}_3$: $S_2$ carries a non-trivial $\mathbb{Z}_2$ charge, whereas $S_3$ and $\psi$ transform non-trivially under $\mathbb{Z}_3$. In contrast, all SM fields are taken to be neutral under both discrete symmetries. The resulting transformation properties of the BSM fields are summarized as follows: 
\begin{eqnarray}
 && \mathbb{Z}_2:\qquad  S_2 \longrightarrow -S_2,\qquad S_3 \longrightarrow S_3,\quad \psi\longrightarrow \psi,\nonumber\\
&&\mathbb{Z}_3:\qquad S_2 \longrightarrow S_2 ,\qquad  S_3 \longrightarrow e^{\frac{2\pi i}{3}}S_3,\quad  \psi \longrightarrow e^{\frac{2\pi i}{3}}\psi.
\end{eqnarray}
The renormalizable Lagrangian comprises the singlet scalar fields and a vector-like muon
\begin{eqnarray}
    \mathcal{L}\supset\, |\partial_\mu S_2|^2 + |\partial_\mu S_3|^2 + \overline{\psi}(i\slashed{D}-m_\psi)\psi +\mathcal{L}_{\rm int}-V(H,S_2,S_3),
\end{eqnarray}
where $m_\psi$ is the bare mass parameter of $\psi$. The renormalizable potential invariant under the SM gauge group and discrete symmetry group $\mathbb{Z}_2 \otimes \mathbb{Z}_3$ can be written as
\begin{eqnarray}\label{eq:LagsoftU1}
   V(H,S_2, S_3)  &=&   \mu_H^{2}|H|^{2} + \lambda_H |H|^{4}+ \mu_{S_2}^2|S_2|^{2} + \lambda_{S_2}|S_2|^4 +  \lambda_{HS_2 }|H|^2 |S_2|^2  + \mu_{S_3}^2 |S_3|^2 \nonumber \\
  &  +& \lambda_{S_3}|S_3|^4 + \lambda_{HS_3} |H|^2|S_3|^2
   +\lambda_{ S_2S_3} |S_2|^2 |S_3|^2 \nonumber\\
   &+&\frac{\mu_2}{2}(S_2^2+S_2^{\ast 2}) +\frac{\sqrt{2}\mu_3}{3}(S_3^3+ S_3^{\ast 3}).
\end{eqnarray}
Here, the SM Higgs doublet $H$ and complex singlet scalar fields $S_2$ and $S_3$ are parametrized as
\begin{equation}
   H= \frac{1}{\sqrt{2}}\begin{pmatrix}
        \chi_1 - i \chi_2\\
          h + i \chi_3
    \end{pmatrix}, \quad S_2 = \frac{1}{\sqrt{2}}\left(s_2+i \eta_2 \right),\quad S_3 = \frac{1}{\sqrt{2}}\left(s_3+i \eta_3 \right),
\end{equation}
The corresponding zero-temperature VEVs are defined as $v\equiv\sqrt{2}\langle H \rangle \approx 246$ GeV, $v_2\equiv\sqrt{2}\,\langle S_2 \rangle$, and $v_3\equiv\sqrt{2}\,\langle S_3 \rangle$.
The fields $\chi_i$ correspond to the Nambu-Goldstone bosons associated with the SM Higgs doublet, and remain massless at zero temperature. Here, $\eta_1$ and $\eta_2$ are pseudo-Goldstone bosons that acquire masses through soft breaking of the global $U(1)$ symmetry, characterized by the parameters $\mu_2$ and $\mu_3$, respectively. 

 After spontaneous symmetry breaking (SSB) induced by the VEVs of the scalar fields, the scalar potential continues to respect the discrete transformations $\eta_2 \to -\eta_2$ and $\eta_3 \to -\eta_3$. This invariance is a consequence of the CP symmetry of the potential and the hermiticity properties of the potential. As a result, $\eta_2$ is stable and can be considered a viable dark matter candidate.

In contrast, $\eta_3$ cannot serve as a dark matter candidate, since it can decay into a pair of muons through Yukawa interactions involving $S_3$ and $\psi$, as discussed below.

We introduce a muon-philic vector-like lepton \(\psi\) with the same hypercharge and electric charge as the right-handed muon, transforming as a \textit{singlet} under \(SU(2)_L\). Since \(\psi\) carries the same \(\mathbb{Z}_3\) charge as the singlet scalar \(S_3\), gauge invariance and the discrete symmetry allow only a single renormalizable interaction,
\begin{equation}
\mathcal{L}_{\rm int} =-\, y_\psi\, \overline{\psi}_L\, \mu^\prime_R\, S_3 \;+\; \text{h.c.},
\end{equation}
where \(y_\psi\) is a dimensionless Yukawa coupling. Note that the global $U(1)$ charge of $\psi$ is the same as that of the singlet scalar field $S_3$. In general, $\mu_2$ and $\mu_3$ may be complex, but their phases can be absorbed into $S_2$ and $S_3$, respectively, allowing them to be taken as real without loss of generality.

\subsection*{Scalar mass matrix}
Expanding the potential in Eq.~\eqref{eq:LagsoftU1} around the background fields yields the tree-level potential:
\begin{eqnarray}\label{eq:treesoftU1}
V_0(h,s_2,s_3) &=& \frac{ \mu_H^2}{2}h^2+\frac{\lambda_H}{4}h^4 + \frac{\mu_{S_2}^2}{2} s_2^2 + \frac{\lambda_{S_2}}{4}s_2^4 + \frac{\mu_{S_3}^2}{2} s_3^2 + \frac{\lambda_{S_3}}{4}s_3^4 + \frac{\lambda_{HS_2}}{4}h^2 s_2^2 \nonumber\\ 
&+& \frac{\lhp}{4}h^2s_3^2 + \frac{\lsp}{4}s_2^2s_3^2 +\frac{\mu_2}{2}s_2^2+\frac{\mu_3}{3}s_3^3 ,
\end{eqnarray}
From the stationary conditions of the tree-level potential, we get the following relations at the electroweak vacuum $x_0=(v,v_2,v_3)$:
\begin{eqnarray}
   &&\mu_H^2 = \frac{1}{2} \left(-2 \lambda_{H} v^2 -\lambda_{ HS_2 } v_2^2 - \lambda_{HS_3} v_3^2\right),\nonumber\\
&& \mu_{S_2}^2 =\frac{1}{2} \left(-2 \mu_2 - \lambda_{HS_2} v^2-2 \lambda_{S_2} v_2^2-\lambda_{S_2S_3}v_3^2\right),\\
  && \mu_{S_3}^2 =\frac{1}{2} \left(-2 \mu_3v_3 - \lambda_{HS_3} v^2-2 \lambda_{S_3} v_3^2-\lambda_{S_2S_3}v_2^2\right).\nonumber
\end{eqnarray}
 At zero temperature, the CP-even scalar fields mix, and the corresponding mass-squared mixing matrix evaluated at $x_0$ is given by
\begin{equation}\label{eq:massmatrix}
   M^2(x_0) = \begin{pmatrix}
        M^2_{hh}& M^2_{hs_2}& M^2_{hs_3}\\
      M^2_{hs_2}& M^2_{s_2s_2}& M^2_{s_2s_3}\\
       M^2_{hs_3}& M^2_{s_2s_3}& M^2_{s_3s_3}
    \end{pmatrix},
\end{equation}
where the matrix elements of the above mass-squared matrix are
\begin{eqnarray}
 &&   M_{hh}^2 = 2 v^2\lambda_H, \quad     M_{hs_2}^2 = vv_2\lambda_{HS_2},\quad  M_{hs_3}^2 =  v v_3\lambda_{HS_3},  \nonumber\\
  &&  M_{s_2s_2}^2 = 2 v_2^2\lambda_{S_2}, \quad     M_{s_2s_3}^2 = v_2v_3\lambda_{S_2S_3},\quad M_{s_3s_3}^2 = 2 v_3^2\lambda_{S_3}+v_3\mu_3 .
\end{eqnarray}
To obtain the physical scalar masses, the mass matrix is diagonalized, and the mass eigenstates $(h_1, h_2, h_3)$ are obtained as
\begin{equation}
    \begin{pmatrix}
    h_1\\
    h_2\\
    h_3\\
    \end{pmatrix}
    =
    \mathcal{O}(\alpha_1,\alpha_2,\alpha_3)
    \begin{pmatrix}
    h\\
    s_2\\
    s_3\\
    \end{pmatrix} \, ,
\end{equation}
where $3\times 3$ rotational matrix $\mathcal{O}(\alpha_1,\alpha_2,\alpha_3)$ can be written as a product of three orthogonal matrices:
\begin{eqnarray}
 \mathcal{O}(\alpha_1,\alpha_2,\alpha_3) =\mathcal{R}(\alpha_3)\mathcal{R}(\alpha_2)\mathcal{R}(\alpha_1) ,  
\end{eqnarray}
with
\begin{equation}\hspace*{-0.3cm} \label{eq:Euler}
{\mathcal R}({\alpha_1}) = \begin{pmatrix} 
c_{\alpha_{1}} & s_{\alpha_{1}} & 0 \\ -s_{\alpha_{1}} & c_{\alpha_{1}} & 0 \\ 0 & 0 & 1  \end{pmatrix},~~
{\mathcal R}({\alpha_2}) = \begin{pmatrix} c_{\alpha_{2}} & 0 & s_{\alpha_{2}} \\ 0 & 1 & 0 \\ -s_{\alpha_{2}} & 0 & c_{\alpha_{2}}  \end{pmatrix},~~ {\mathcal R}({\alpha_3}) = \begin{pmatrix} 1 & 0 & 0 \\  0 & c_{\alpha_{3}} & s_{\alpha_{3}} \\ 0 &   -s_{\alpha_{3}} & c_{\alpha_{3}}\end{pmatrix}.
\end{equation}
Here, we have used shorthand notations $c_{\alpha_i} = \cos(\alpha_i)$ and $s_{\alpha_i} = \sin(\alpha_i)$, where $i=1,2,3$. The relation between the physical mass basis ($h_1,h_2,h_3$) with the unphysical gauge basis ($h,s_2,s_3$) is given by,
\begin{eqnarray}
   && h = (c_{\alpha_1} c_{\alpha_2}) h_1 -(c_{\alpha_1}s_{\alpha_2}s_{\alpha_3}+s_{\alpha_1}c_{\alpha_3}) h_2 + (s_{\alpha_1}s_{\alpha_3}-c_{\alpha_1}s_{\alpha_2}c_{\alpha_3}) h_3, \\
   && s_2 = (s_{\alpha_1}c_{\alpha_2}) h_1 + (c_{\alpha_1}c_{\alpha_3}-s_{\alpha_1}s_{\alpha_2}s_{\alpha_3}) h_2 -(c_{\alpha_1}s_{\alpha_3}+s_{\alpha_1}s_{\alpha_2}c_{\alpha_3}) h_3, \\
    && s_3 =  (s_{\alpha_2}) h_1 + (c_{\alpha_2} s_{\alpha_3}) h_2 + (c_{\alpha_2}c_{\alpha_3}) h_3.
\end{eqnarray}
The mass matrices are related through an orthogonal transformation, 
\begin{eqnarray}\label{eq:diag}
    M^2_{\rm diag} = \mathcal{O}(\alpha_1,\alpha_2,\alpha_3) M^2(x_0) \mathcal{O}^T(\alpha_1,\alpha_2,\alpha_3),
\end{eqnarray}
where $M^2_{\rm diag}=\mathrm{diag}(m^2_{h_1}, m^2_{h_2},m^2_{h_3})$. Throughout this analysis, we assume the following mass hierarchy among the CP-even scalar states: 
\begin{eqnarray}
    m^2_{h_3} > m^2_{h_2} > m^2_{h_1}.
\end{eqnarray}
The lightest CP-even scalar $(h_1)$ is the SM Higgs boson, \textit{i.e.,} $m_{h_1}=125$ GeV. 
Using Eqs.\,\eqref{eq:massmatrix} and \eqref{eq:diag}, the dimensionless model parameters in terms of inputs:
\begin{eqnarray}
 &&  \lambda_H = \frac{c_{\alpha_1}^2 c_{\alpha_2}^2 m^2_{h_1}}{2 v^2}+\frac{m^2_{h_2} (c_{\alpha_1} s_{\alpha_2} s_{\alpha_3}+c_{\alpha_3} s_{\alpha_1})^2}{2 v^2}+\frac{m^2_{h_3} (c_{\alpha_1} c_{\alpha_3} s_{\alpha_2}-s_{\alpha_1} s_{\alpha_3})^2}{2 v^2},\\
&&   \lambda_{S_2} =\label{eq:ls2} \frac{s_{\alpha_1}^2 c_{\alpha_2}^2 m^2_{h_1}}{2 v_2^2}+\frac{m^2_{h_2} (c_{\alpha_1} c_{\alpha_3} -s_{\alpha_1} s_{\alpha_2}s_{\alpha_3})^2}{2 v_2^2}+\frac{m^2_{h_3} (c_{\alpha_3} s_{\alpha_1} s_{\alpha_2}+c_{\alpha_1} s_{\alpha_3})^2}{2 v_2^2} ,\\
&&    \lambda_{S_3} = \frac{c_{\alpha _2}^2 \left(m^2_{h_3} c_{\alpha _3}^2+m^2_{h_2} s_{\alpha _3}^2\right)+ m^2_{h_1}  s_{\alpha _2}^2-  v_3\mu_3}{2 v_3^2} ,\\
&&   \lambda_{HS_2} = \label{eq:lh2}\frac{c_{\alpha _1} s_{\alpha _1} \left(m^2_{h_1} c_{\alpha _2}^2+c_{\alpha _3}^2 \left(m^2_{h_3} s_{\alpha _2}^2-m^2_{h_2}\right)\right)+c_{\alpha _1} s_{\alpha _1} s_{\alpha _3}^2 \left(m^2_{h_2} s_{\alpha _2}^2-m^2_{h_3}\right)}{v v_2}\nonumber\\
&&  \hspace{12mm} -\frac{c_{\alpha _3} s_{\alpha _2} s_{\alpha _3}(m^2_{h_2}-m^2_{h_3})  \left(c_{\alpha _1}^2-s_{\alpha _1}^2\right)}{v v_2}, \\
&&   \lambda_{HS_3} = \frac{c_{\alpha _2} \left(c_{\alpha _1} s_{\alpha _2} \left(m^2_{h_1}-m^2_{h_3} c_{\alpha _3}^2-m^2_{h_2} s_{\alpha _3}^2\right)+c_{\alpha _3}s_{\alpha _1} s_{\alpha _3} (m^2_{h_3}-m^2_{h_2}) \right)}{v v_3},\\
&&  \lambda_{S_2S_3} = \frac{c_{\alpha _2} \left(s_{\alpha _1} s_{\alpha _2} \left(m^2_{h_1}-m^2_{h_3} c_{\alpha _3}^2-m^2_{h_2} s_{\alpha _3}^2\right)+c_{\alpha _1} c_{\alpha _3}s_{\alpha _3} (m^2_{h_2}-m^2_{h_3}) \right)}{v_2 v_3}.
\end{eqnarray}
The masses of the CP-odd particles at $T=0$ are given as
\begin{eqnarray}
   m_{\chi_i}^2 = 0,\quad  m_{\eta_2}^2 = -2 \mu_2,\quad  m_{\eta_3}^2 = -3 \mu_3 v_3.
\end{eqnarray}
From the above equation, it is evident that the pseudo Goldstone bosons $\eta_2$ and $\eta_3$ acquire their masses from the softly $U(1)$-breaking terms introduced in Eq.\,\eqref{eq:LagsoftU1}. By inverting the above relations, the model parameters $\mu_2$ and $\mu_3$ can be expressed in terms of $m_{\eta_2}$, $m_{\eta_3}$, and $v_3$.
%Expressed in terms of the physical masses $m_{\eta_2}$ and $m_{\eta_3}$ of the DM candidates, the corresponding model parameters $\mu_2$ and $\mu_3$ can be written as
%\begin{eqnarray}
%    \mu_2 = -\frac{m_{\eta_2}^2}{2},\quad \mu_3 = -\frac{m_{\eta_3}^2}{3v_3}.
%\end{eqnarray}
\subsection*{Fermionic mixing matrix}
Since complex scalar field $S_3$ acquires a VEV at zero temperature, mixing arises between $\mu^\prime$ and $\psi$. The relevant mixing term is
\begin{eqnarray}\label{eq:massmat}
    \mathcal{L} \supset -( \overline{\mu^\prime}_{L}\,\, \overline{\psi}_{L})M 
    \begin{pmatrix}
    \mu^\prime_{R}\\
    \psi_{R}
   \end{pmatrix} + \text{h.c},\quad \text{with}\,\,\, 
   M = 
\begin{pmatrix}
\frac{ y_{\mu^\prime} v}{\sqrt{2}} & 0 \\
\frac{y_\psi v_3}{\sqrt{2}} & m_\psi
\end{pmatrix},
\end{eqnarray}
where $y_{\mu^\prime}$ is the SM muon Yukawa coupling.
The mass matrix is diagonalized by a bi-unitary transformation,
\begin{eqnarray}\label{eq:mixF}
    \mathcal{L} &\supset& -( \overline{\mu^\prime}_{L}\,\, \overline{\psi}_{L})M 
    \begin{pmatrix}
    \mu^\prime_{R}\\
    \psi_{R}
    \end{pmatrix} + \text{h.c} 
    = -( \overline{\mu^\prime}_{L}\,\, \overline{\psi}_{L})U_L^\dagger U_L M U_R^\dagger U_R
    \begin{pmatrix}
    \mu^\prime_{R}\\
    \psi_{R}
    \end{pmatrix} + \text{h.c}\nonumber\\
    &=& -( \overline{\mu}_{L}\,\, \overline{vl}_{L})M^\mu_{\rm diag} 
    \begin{pmatrix}
    \mu_{R}\\
    vl_{R}
    \end{pmatrix} + \text{h.c},
\end{eqnarray}
where we have used the identities $U_{L/R}^\dagger U_{L/R}= I$.
The relation between gauge and mass eigenstates is
\begin{equation}
    \begin{pmatrix}
    \mu_{L/R}\\
    vl_{L/R}
    \end{pmatrix}
    = U_{L/R} 
    \begin{pmatrix}
    \mu^\prime_{L/R}\\
    \psi_{L/R}
    \end{pmatrix}, \qquad
    U_{L/R} = 
    \begin{pmatrix}
    \cos \theta_{L/R} & -\sin \theta_{L/R} \\
    \sin \theta_{L/R} & \cos \theta_{L/R}
    \end{pmatrix}.
\end{equation}
In general, the diagonalizing matrix contains complex phases. Since CP violation is not considered in this analysis, these phases are neglected. The mass eigenstates are therefore
\begin{eqnarray}
    \mu_{L/R} &=& \mu^\prime_{L/R}\cos\theta_{L/R} + \psi_{L/R}\sin\theta_{L/R},\\
    vl_{L/R} &=& \psi_{L/R}\cos\theta_{L/R} - \mu^\prime_{L/R}\sin\theta_{L/R}.
\end{eqnarray}
From the second and third equality in Eq.\,\eqref{eq:mixF}, one obtains
\begin{eqnarray}\label{eq:phymass}
    M^\mu_{\rm diag} =
    \begin{pmatrix}
    m_\mu & 0 \\
    0 & m_{vl}
    \end{pmatrix}
    = U_L M U_R^\dagger 
    \Longrightarrow 
    M = U_L^\dagger 
    \begin{pmatrix}
    m_\mu & 0 \\
    0 & m_{vl}
    \end{pmatrix}
    U_R,
\end{eqnarray}
where $m_\mu$ and $m_{vl}$ denote the physical masses of mass eigenstates $\mu$ and $vl$, respectively.

Using Eqs.~\eqref{eq:phymass} and \eqref{eq:massmat}, the bare parameters can be expressed in terms of physical ones as
\begin{eqnarray}\label{eq:yuwVL}
&&y_{\mu^\prime} = \dfrac{\sqrt{2}m_\mu}{v}
\left( \cos \theta_L \cos \theta_R + \dfrac{m_{vl}}{m_\mu} \sin \theta_L \sin \theta_R\right), \nonumber\\
&&y_\psi = \dfrac{\sqrt{2}m_{vl}}{v_3}
\left( \cos \theta_L \sin \theta_R - \dfrac{m_\mu}{m_{vl}} \sin \theta_L \cos \theta_R\right), \nonumber\\
&&m_\psi  = m_{vl}\left( \cos \theta_L \cos \theta_R + \frac{m_\mu}{m_{vl}} \sin \theta_L \sin \theta_R\right).
\end{eqnarray}
Only one mixing angle is independent, with
\begin{eqnarray}
    \tan\theta_L = \frac{m_\mu}{m_{vl}}\tan\theta_R.
\end{eqnarray}
In the limit \( m_{vl} \gg m_\mu \), the mixing angle \( \sin\theta_L \) approaches zero. Consequently, observables associated with left-handed charged current transitions remain essentially unchanged in this regime.

Taking $\theta_R$ as the independent parameter, the inputs in this sector are the muon mass
$m_\mu = 0.105~\text{GeV}$, the mass of $vl$ (from a few GeV to the TeV range), and the mixing angle $\sin\theta_R$. For this scenario, the complete set of input parameters is summarized in Tab.\,\ref{tab:input_params}.

\begin{table}[h]
\centering
\renewcommand{\arraystretch}{1.2}
\begin{tabular}{|c|c|}
%\rowcolor{violet!10}
\hline
Model Parameters & Inputs \\
\hline
%\rowcolor{green!10}
\makecell{
$\mu_{H}^2,\, \mu_{S_2}^2,\, \mu_{S_3}^2, \,\lambda_H,\, \lambda_{S_2},\, \lambda_{S_3},$\\
$\lambda_{HS_2},\, \lambda_{HS_3},\, \lambda_{S_2S_3},$\,
$\mu_2,\, \mu_3$\\
$m_\psi,\, y_\psi,\, y_{\mu^\prime}$
}
&
\makecell{
$m_{h_1},\, m_{h_2},\, m_{h_3},\, v,\, v_2, \,v_3,$\\
$s_{\alpha_1},\, s_{\alpha_2},\,s_{\alpha_3},\,$
$m_{\eta_2}, \,m_{\eta_3}$\\
$m_\mu,\, m_{vl},\, \sin\theta_R$
}
\\
\hline
\end{tabular}
\caption{List of model and input parameters associated with the scalar and muon sector.}
\label{tab:input_params}
\end{table}
\subsection{Theoretical constraints}
\subsubsection*{Perturbative unitarity constraints}

The perturbativity condition sets constraints on the quartic couplings in the tree-level potential as~\cite{Kannike:2012pe}
\begin{equation*}
\left(|\lambda_H|,\hspace{1mm} |\lambda_{S_2}|,\hspace{1mm} |\lambda_{S_3}|,\hspace{1mm} |\lambda_{HS_2}|,\hspace{1mm} |\lambda_{HS_3}|,\hspace{1mm} |\lambda_{S_2 S_3}|\right) \le 4\pi.
\end{equation*}

\subsubsection*{Vacuum stability conditions}
The constraints on the parameter space from the vacuum stability condition~\cite{Kannike:2012pe}:
\begin{eqnarray}
&&\lambda_H>0,\quad \lambda_{S_2}>0,\quad \lambda_{S_3}>0,\quad  a_1>0,\quad a_2>0,\quad a_3>0,\\
\nonumber\\
&&2\left(\lambda_{HS_2}\sqrt{\lambda_{S_3}}+\lambda_{S_2S_3}\sqrt{\lambda_H}+\lambda_{HS_3}\sqrt{\lambda_{S_2}}
+2\sqrt{\lambda_H \lambda_{S_2} \lambda_{S_3}}+ \sqrt{a_1a_2a_3}\right)>0,
\end{eqnarray}
with 
\begin{eqnarray}
    a_1= \lambda_{HS_2}+2\sqrt{\lambda_H\lambda_{S_2}},\quad  a_2= \lambda_{HS_3}+2\sqrt{\lambda_H\lambda_{S_3}},\quad  a_3= \lambda_{S_2S_3}+2\sqrt{\lambda_{S_2}\lambda_{S_3}}.
\end{eqnarray}

\subsection{Constraints from Mono-jet searches}
\label{sec:collider}
At the Large Hadron Collider (LHC), dark matter (DM) production is probed through final states characterized by large missing transverse momentum 
($\slashed{E}_T$) recoiling against visible Standard Model (SM) radiation.  Among these, mono-jet  signature provide the most promising channels over a wide range of mediator and DM mass hypotheses. Searches for new phenomena in events with jets and $\slashed{E}_T$ at $\sqrt{s}=13$ TeV have been previously performed by the CMS~\cite{CMS:2021far,CMS:2017zts} and ATLAS~\cite{ATLAS:2018nda,ATLAS:2021kxv} collaborations.

To assess the collider constraints on the scalar sector considered in this work, we perform a dedicated mono-jet analysis at the $\sqrt{s}=13$ TeV LHC. The model is implemented in \textsc{FeynRules}~\cite{Alloul:2013bka} using the Lagrangian defined in Eq.\,\eqref{eq:LagsoftU1}, and exported in \textsc{Universal FeynRules Output} (UFO)~\cite{Degrande:2011ua,Darme:2023jdn} format. QCD corrections at one-loop order are incorporated, with the required ultraviolet counterterms and rational $R_2$ terms generated using \textsc{NLOCT}~\cite{Degrande:2014vpa}.

\begin{figure}[!t]
    \centering
    \includegraphics[width=0.40\linewidth, height=5cm]{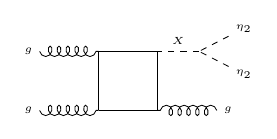}
    \includegraphics[width=0.40\linewidth, height=5cm]{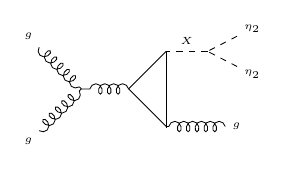}
    \caption{Leading order Feynman diagram representing the production of a pair of DM associated with QCD jet. In both panels, ``X'' represents the SM Higgs boson along with two new scalars. }
    \label{fig:feyndm}
\end{figure}

Parton-level events are generated with \textsc{MadGraph5\_aMC@NLO} (MG5 henceforth)~\cite{Alwall:2014hca} within the $\alpha$-scheme~\cite{Biekotter:2023xle}. The events are subsequently interfaced with \textsc{Pythia}~\cite{Sjostrand:2000wi,Bierlich:2022pfr} for parton showering and hadronization. Jets are reconstructed using the anti-$k_T$~\cite{Cacciari:2008gp} clustering algorithm with a radius parameter $R=0.5$. Finally, detector effects are modeled using a fast detector simulation framework with \textsc{Delphes}~\cite{deFavereau:2013fsa} to account for experimental resolutions and efficiencies. The input parameters are
\begin{align}
    m_b = 4.7~\mathrm{GeV},~m_t &= 173.0~\mathrm{GeV},~m_Z=91.19~\mathrm{GeV},~m_{h_1}=125.0~\mathrm{GeV}, \nonumber \\\alpha_{\mathrm{EW}} &= 127.9,~G_F=1.16637\times 10^{-5},~\alpha_S=0.1184.\nonumber
\end{align}
The signal process corresponds to loop pair production of dark matter particles in association with a hard QCD jet, leading to a characteristic mono-jet plus missing transverse energy signature. 
Representative Feynman diagrams contributing to the signal and dominant 
SM backgrounds are shown in Fig.\,\ref{fig:feyndm}. 
The primary irreducible background originates from 
$Z(\nu\bar{\nu})$+jets production, while $W(\ell\nu)$+jets and top-quark 
processes contribute when charged leptons escape detection.

We implement event selection criteria following ATLAS mono-jet analysis~\cite{ATLAS:2021kxv}. In particular, we require zero isolated leptons (electrons or muons), missing transverse energy $\slashed{E}_T > 200~\text{GeV}$, and at least one jet with leading transverse momentum $p_T > 100~\text{GeV}$ and pseudorapidity $|\eta_j| \le 2.5$. A cut on the difference in azimuthal angle between the $\slashed{E}_T$ and the first four leading jets of 
$\Delta\phi(\mathrm{jet}$, $\slashed{E}_T) > 0.5$ is also applied. This is done to suppress the QCD background from miss-measurements of jet momentum or detector noise, which could introduce missing transverse momentum in the event, in the same direction as the miss-measured jet. 

Further, the selected events are binned into $13$ different disjoint regions of $\slashed{E}_T$ defined as
\begin{align}
    \slashed{E}_T \in \left[200,250,300,350,400,500,600,700,800,900,1000,1100,1200,\infty\right]~\mathrm{GeV}\nonumber.
\end{align}
The predicted background contributions along with the uncertainties and observed values are taken from Ref.\,\cite{ATLAS:2021kxv}. 

\begin{figure}[!htb]
    \centering
    \includegraphics[width=0.49\linewidth]{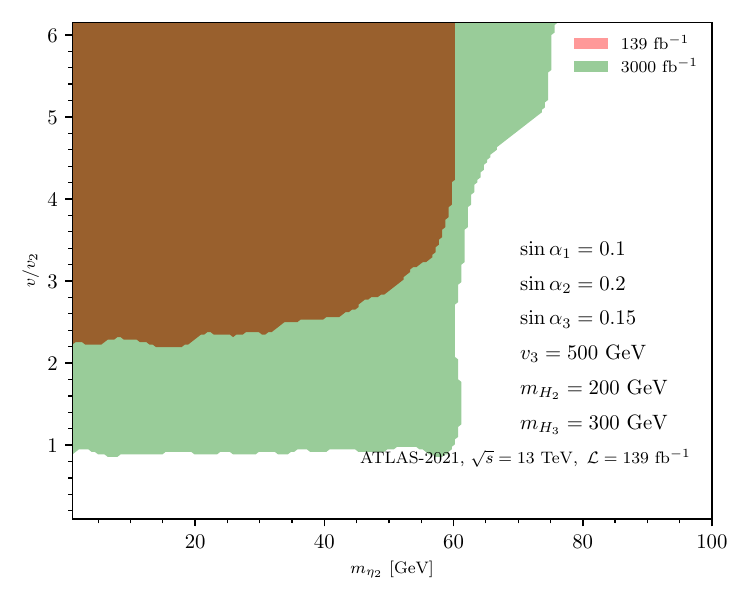}
    \includegraphics[width=0.49\linewidth]{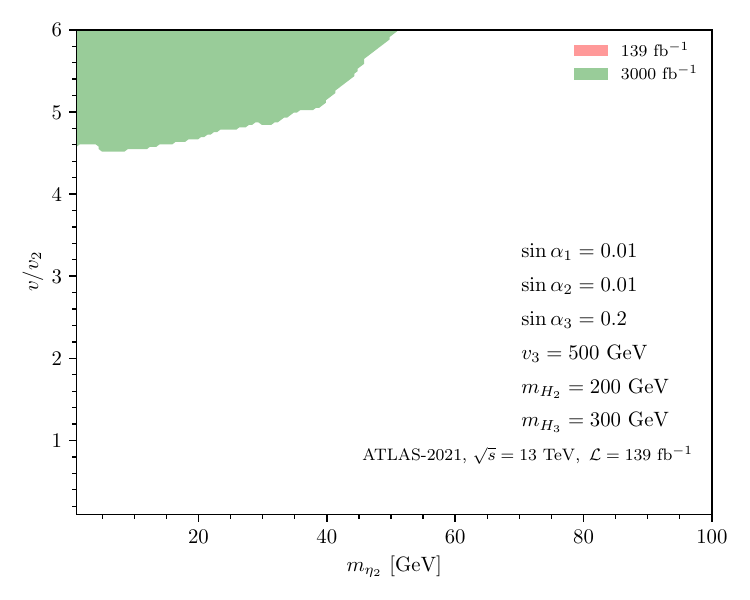}
    \caption{$95\%$ confidence level exclusion region of dark matter mass ($m_{\eta 2}$) and ratio of two vacuum expectation value ($v/v_2$) for two set of benchmark points. The exclusion is shown at two values of luminosity, viz. $150$ fb$^{-1}$ (light red) and $3000$ fb$^{-1}$ (shown in light red). }
    \label{fig:monojet}
\end{figure}

To derive the compatibility of the signal hypothesis with the observed data, we use the $CL_S$ method with test static
\begin{equation}
    q_\mu = -2\,\mathrm{ln}\left(\frac{\mathcal{L}\Big(\mathrm{data}|\mu,\hat{\theta}_\mu\Big)}{\mathcal{L}\Big(\mathrm{data}|\hat{\mu},\hat{\theta}\Big)}\right),
\end{equation}
where $\mu$ is the signal strength, $\theta_\mu$ are the nuisance parameters for a given $\mu$, $\hat{\mu}$ is the maximum likelihood value of $\mu$, and $\hat{\theta}_\mu$ are the corresponding nuisance parameters. In the counting experiment, the likelihood function is usually taken to be a Poisson distribution as
\begin{equation}
    \mathcal{L}(\mu,\theta) = \prod_{i \in \mathrm{bins}} \mathrm{Pois}\Big(n^i|\mu n_s^i + n_b^i + \theta^i\sigma_b^i\Big)\cdot \prod_{j \in \mathrm{nui.}}\mathcal{N}\Big(\theta^j|0,1\Big).
\end{equation}
The $CL_S$ is then defined as the ratio 
\begin{equation}
    CL_S = \frac{p_{s+b}}{1-p_b},
\end{equation}
where $p_{s+b}$ denotes the probability of finding a data set with a value of the test statistic equal to or larger under the signal plus background hypothesis, and $p_b$ is the probability of finding a data set equally or more incompatible with the background only hypothesis. We employ the \textsc{Spey}~\cite{Araz:2023bwx} statistical package for performing exclusion of parameter space. The resultant excluded parameter space on $m_{\eta_2}-v/v_2$ planes for two benchmark input parameters (the values are given within the figure) is shown in Fig.\,\ref{fig:monojet}, where the space in light red and green represents exclusion at $139$ fb$^{-1}$ and $3000$ fb$^{-1}$ of luminosity, respectively.  In the left panel of  Fig.\,\ref{fig:monojet}, we observed a sharp horizontal boundary at $v/v_2 \sim 1$ and a vertical exclusion at $m_{\eta_2}\sim 60$ GeV up to $v/v_2 \sim 4.5$. When we reduce the mixing angles $\alpha_i,i\in \{1,2\}$ by $\mathcal{O}(1)$ (shown in the right panel), the excluded region shrinks and lies at lower dark matter mass and higher $v/v_2$ values. In this case, the current LHC luminosity of $150$ fb$^{-1}$ does not exclude any region of parameter space.

\section{Muon Collider Study}
\label{muon-coll}
A high-energy muon collider provides a particularly suitable environment for probing the vector-like muon (VLM) sector of the present model. Since the new vector-like lepton \(vl\) couples preferentially to the muon sector through the singlet $S_3$, both associated and pair-production modes can in principle be considered. We therefore begin by identifying the production channel that offers the most promising discovery sensitivity. Two relevant possibilities are the pair production of VLMs and associated production of a VLM with a muon. The latter is particularly sensitive to the mixing between the SM muon and the vector-like state. As discussed in Sec.\,\ref{model}, the two chiral mixing angles satisfy
$$ \tan \theta_L = \frac{m_{\mu}}{m_{vl}}~\tan \theta_R.$$
Consequently, for a VLM in the TeV mass range, $m_\mu/m_{vl} \ll 1$, and the left-handed mixing angle $\theta_L$ is naturally very small. This strongly suppresses the off-diagonal interactions entering associated VLM production. In particular, the scalar-mediated amplitudes involve either the small left-handed mixing or the muon Yukawa coupling, while the off-diagonal \(Z\mu vl\) coupling is also proportional to the muon–VLM mixing. The resulting associated-production cross section is therefore considerably smaller than that of VLM pair production in the parameter region of interest. This motivates us to focus on $\mu^+ \mu^- \rightarrow vl^+ vl^-$ as the primary discovery channel. 

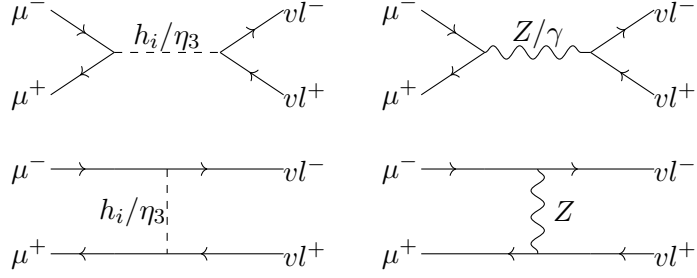
\begin{figure}[!htb]
\centering
\begin{tikzpicture}[scale=0.7]

\begin{scope}
\node at (-0.6,0.8) {$\mu^-$};
\draw[fermion] (-0.2,0.8) -- (1,0);
\node at (-0.6,-0.8) {$\mu^+$};
\draw[antifermion] (-0.2,-0.8) -- (1,0);
\draw[dashed] (1,0) -- (3,0);
\node at (2,0.3) {$h_i/\eta_3$};
\draw[fermion] (3,0) -- (4.2,0.8);
\node at (4.6,0.8) {$vl^-$};
\draw[antifermion] (3,0) -- (4.2,-0.8);
\node at (4.6,-0.8) {$vl^+$};
\end{scope}

\begin{scope}[xshift=7cm]
\node at (-0.6,0.8) {$\mu^-$};
\draw[fermion] (-0.2,0.8) -- (1,0);
\node at (-0.6,-0.8) {$\mu^+$};
\draw[antifermion] (-0.2,-0.8) -- (1,0);
\draw[decorate,decoration={snake}] (1,0) -- (3,0);
(1,0)--(3,0);
\node at (2,0.35) {$Z/\gamma$};
\draw[fermion] (3,0) -- (4.2,0.8);
\node at (4.6,0.8) {$vl^-$};
\draw[antifermion] (3,0) -- (4.2,-0.8);
\node at (4.6,-0.8) {$vl^+$};
\end{scope}

\begin{scope}[yshift=-3cm]
\node at (-0.6,0.8) {$\mu^-$};
\draw[fermion] (-0.2,0.8) -- (1,0.8);
\node at (-0.6,-0.8) {$\mu^+$};
\draw[antifermion] (-0.2,-0.8) -- (1,-0.8);
\draw[dashed] (2,0.8) -- (2,-0.8);
\node at (1.35,0) {$h_i/\eta_3$};
\draw[fermion] (1,0.8) -- (4.2,0.8);
\node at (4.6,0.8) {$vl^-$};
\draw[antifermion] (1,-0.8) -- (4.2,-0.8);
\node at (4.6,-0.8) {$vl^+$};
\end{scope}

\begin{scope}[xshift=7cm,yshift=-3cm]
\node at (-0.6,0.8) {$\mu^-$};
\draw[fermion] (-0.2,0.8) -- (1,0.8);
\node at (-0.6,-0.8) {$\mu^+$};
\draw[antifermion] (-0.2,-0.8) -- (3,-0.8);
\draw[decorate,decoration={snake}]
%\draw[decorate,decoration={snake,amplitude=1.5pt,segment length=5pt}]
(2,0.8)--(2,-0.8);
\node at (2.5,0) {$Z$};
\draw[fermion] (1,0.8) -- (4.2,0.8);
\node at (4.6,0.8) {$vl^-$};
\draw[antifermion] (3,-0.8) -- (4.2,-0.8);
\node at (4.6,-0.8) {$vl^+$};
\end{scope}

\end{tikzpicture}
\caption{Tree-level $s$- and $t$-channel contributions to $\mu^+ \mu^- \to vl^+ vl^-$ mediated by scalars and $Z/\gamma$ .}
\label{fig:tree22}
\end{figure}

The pair-production process receives both gauge- and scalar-mediated contributions, as shown in Fig.\,\ref{fig:tree22}. The usual $s$-channel $\gamma/Z$ diagrams provide an essentially model-independent gauge contribution. In addition, the extended scalar sector generates scalar-mediated diagrams involving $h_i$ and $\eta_3$. Although scalar $s$-channel contributions involving the SM muon are mixing suppressed, the scalar-mediated $t$-channel amplitude is qualitatively different: the relevant $\mu-vl-h_i(\eta_3)$ interaction does not necessarily carry the small $\sin\theta_L$ suppression. Consequently, for sufficiently large $y_\psi$, the scalar-mediated $t$-channel can dominate the total pair-production rate. In particular, the contributions involving the predominantly $S_3$-sector states $h_3$ and $\eta_3$ are generally more important than those mediated by $h_1$ and $h_2$. This enhancement constitutes one of the characteristic collider features of the present construction. The current manuscript already finds this behaviour explicitly in the decomposition of the pair-production cross section. 

\begin{table}[h!]
\centering 
\begin{tabular}{cccccccccccc}
\hline \hline
$\sqrt{s}$ & BP1 & BP2 & BP3 & BP4 & BP5 & BP6 & BP7 & BP8 & BP9 & BP10 & BP11\\
\hline
3 & 8.04 &  5.2 & 5.2 & 5.2 & 6.3 & 5.19 & 8.0 & 4.3 & 17.3 & 4.8 & 4.4\\
\hline
10 & 0.87 & 0.724 & 0.723 & 0.722 & 0.782 & 0.724 & 0.87 & 0.704 & 2.2 & 0.695 & 0.706\\
\hline \hline
\end{tabular}
\caption{LO cross sections (in fb) for vector-like muon pair production at $\sqrt{s} = 3~\rm{and}~10$ TeV muon collider for various choices of our benchmark points. In all points, the mass of the VLM is set to be $1.4$ TeV.}
\label{tab:vlpair-xsec}
\end{table}

We select eleven representative benchmark points whose complete parameter choices and associated phenomenological properties are summarized later in Table \ref{tab:BpTab}, to extensively study the properties of the pNGB DM, the collider phenomenology of the VLM, and possible observable GW signatures in future interferometer experiments. First, we illustrate the variation of the VLM pair-production rate across the phenomenologically viable parameter space. We calculate the leading-order cross section using MG5, employing the model discussed in Eq.~\eqref{model}. The pair-production cross sections for $m_{vl}=1~\mathrm{TeV}$ at $\sqrt s=3$ and $10~\mathrm{TeV}$ are summarized in Tab.\ref{tab:vlpair-xsec}. For the benchmark configurations considered, the rates at the $3~\mathrm{TeV}$ collider are larger, with BP9 yielding the largest cross section, $$\sigma(\mu^+ \mu^- \rightarrow vl^+ vl^-) = 17.3~\rm{fb}$$ for $m_{vl}=1.4$~TeV. We, therefore, adopt BP9 as the representative benchmark for the detailed collider analysis at $\sqrt{s}=3$ TeV.

\subsection{Signal topology}

Once produced, the vector-like muons predominantly decay through the scalar sector. For the benchmark considered here, the relevant decay is
$$ vl^\pm\rightarrow \mu^\pm h_3. $$
The heavy CP-even scalar $h_3$ subsequently decays predominantly into a pair of the stable pNGB dark-matter particles $\eta_2$,
$$ h_3 \rightarrow \eta_2\eta_2. $$
The complete pair-production decay chain is, therefore,
$$ 
\begin{aligned} 
\mu^+\mu^- &\rightarrow vl^+vl^-\\ 
&\rightarrow (\mu^+h_3)(\mu^-h_3)\\ 
&\rightarrow \mu^+\mu^-+ (\eta_2\eta_2)+(\eta_2\eta_2). 
\end{aligned} 
$$

\begin{figure}[!htb]
    \centering
    \includegraphics[width=0.49\linewidth]{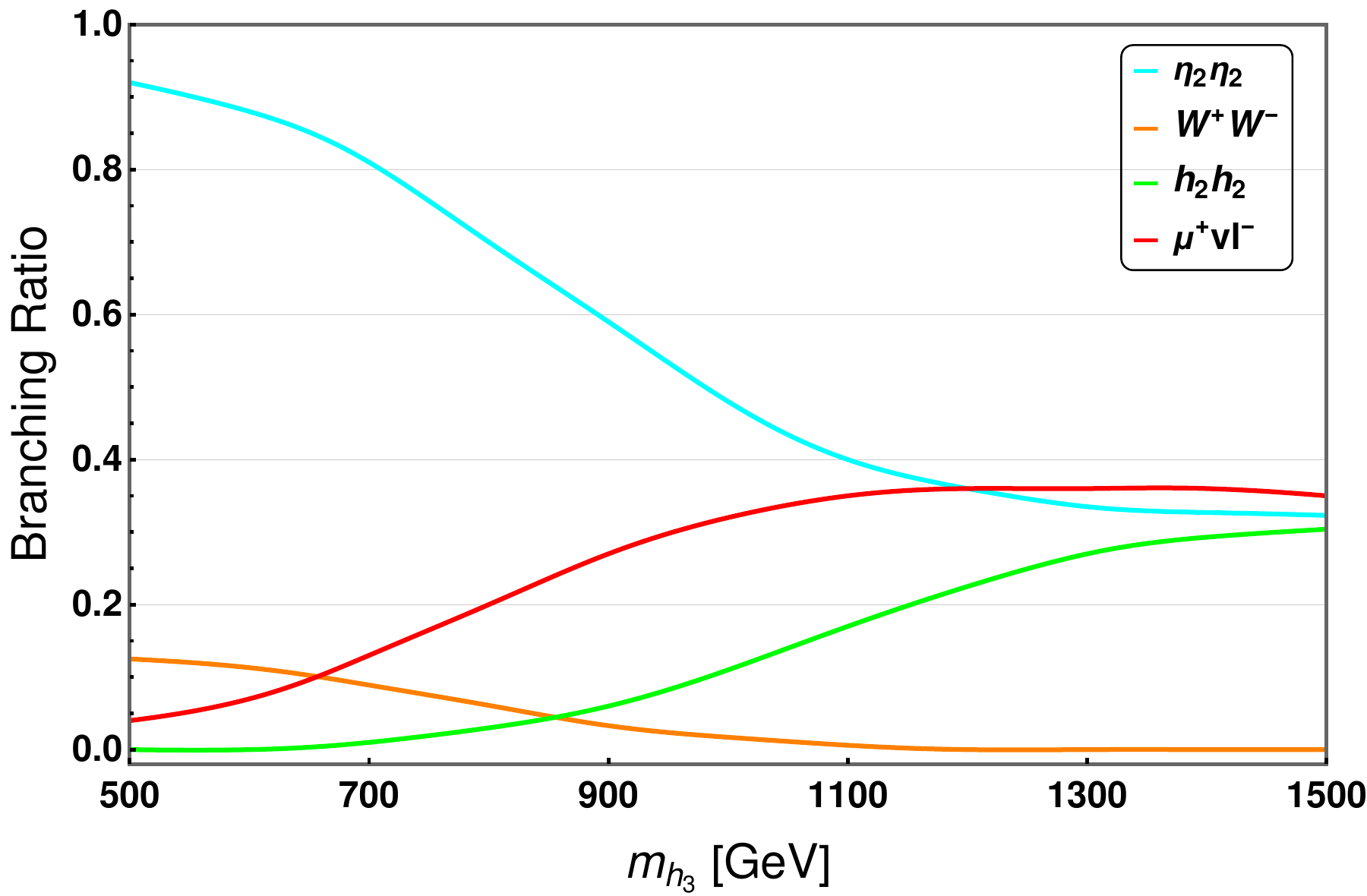}
    \includegraphics[width=0.49\linewidth]{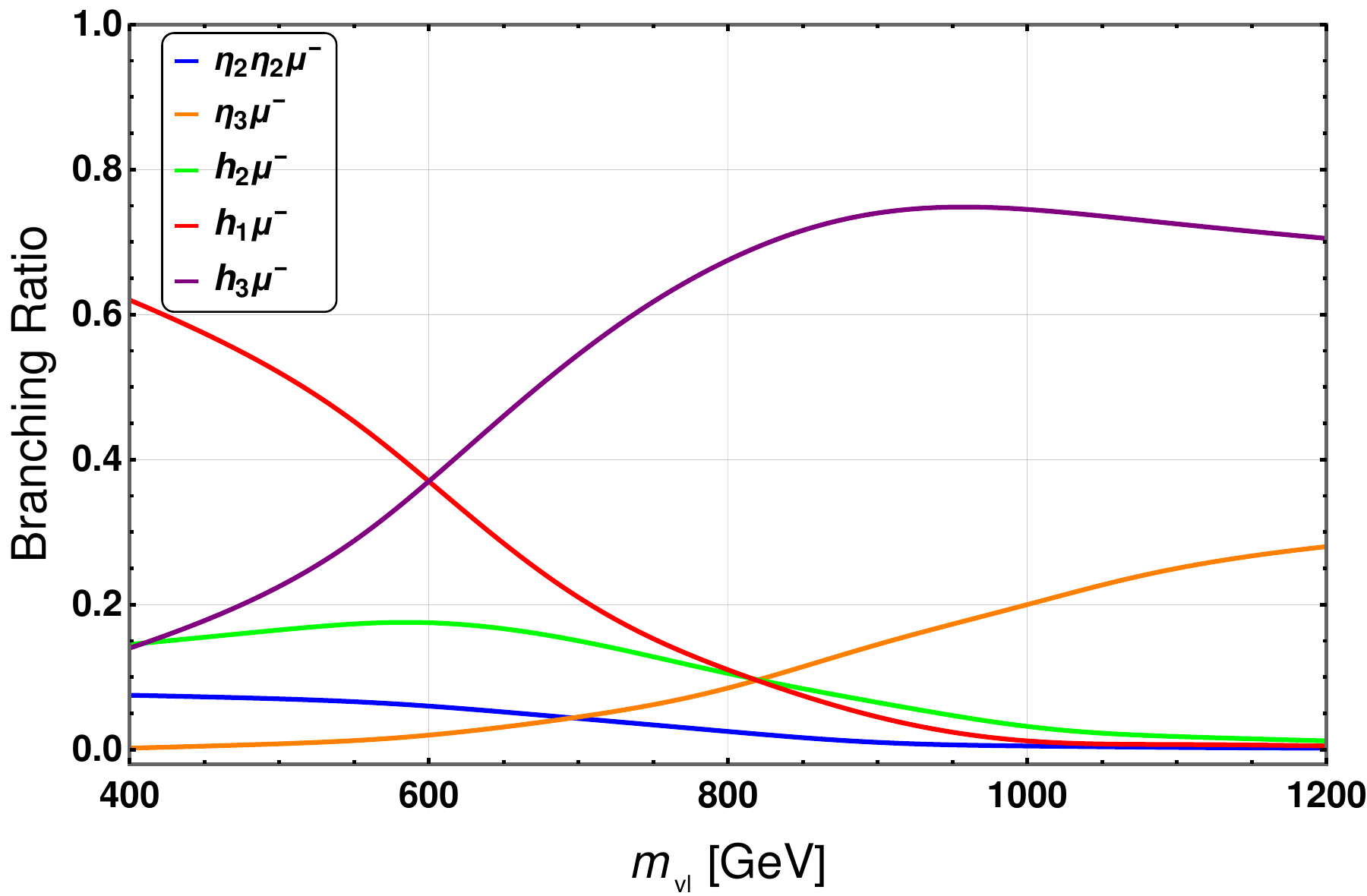}
    \caption{Variation of branching ratios with masses for the various decay channels of the heavy scalar $h_3$ (left panel) and the vector-like muon $vl^-$ (right panel).}
    \label{fig:br}
\end{figure}
\begin{figure}[!htb]
    \centering
    \includegraphics[width=0.49\linewidth]{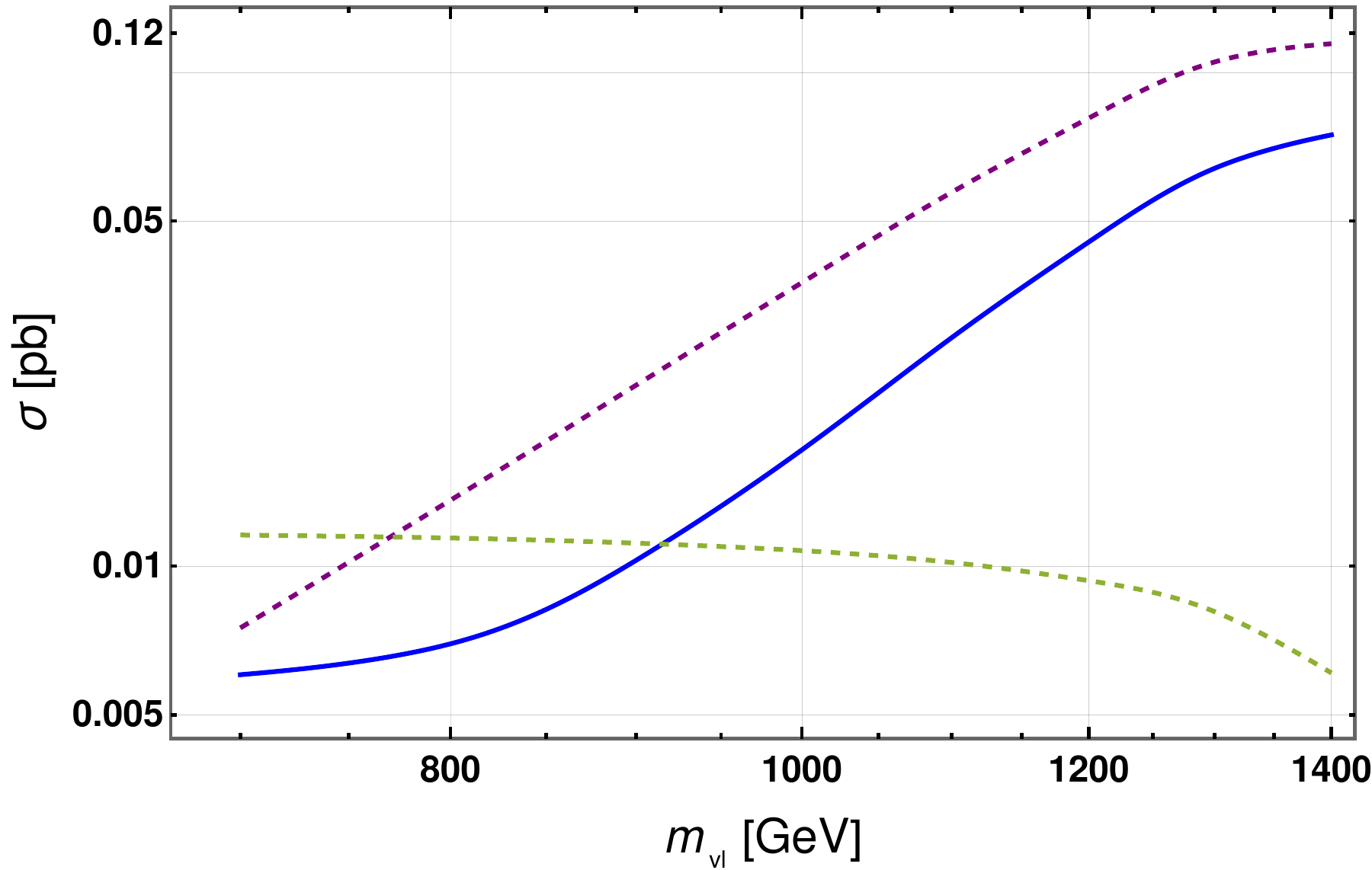}
    \includegraphics[width=0.49\linewidth]{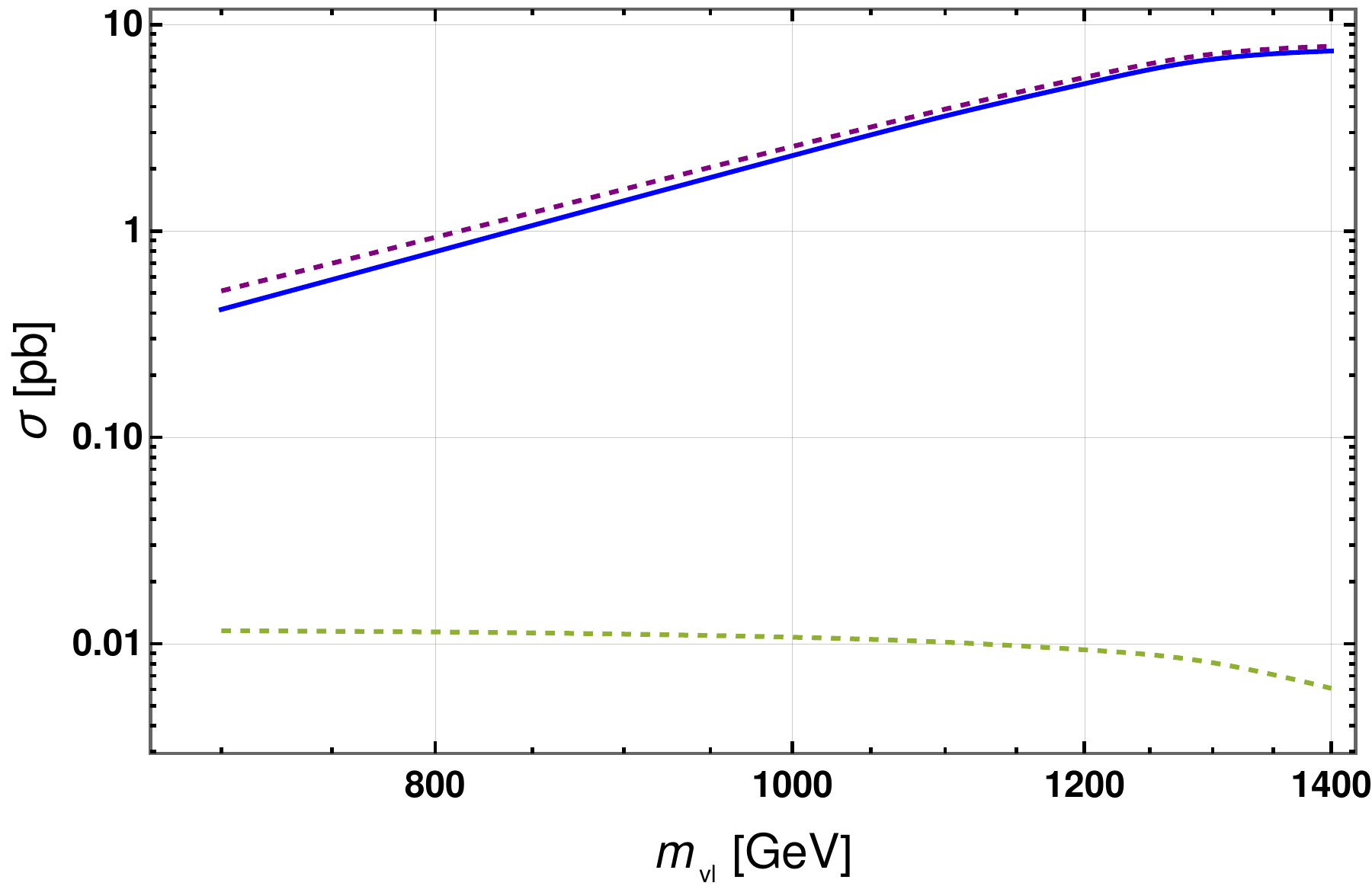}
    \caption{Pair-production cross section of the VLM as a function of $m_{vl}$, at a 3~TeV COM muon collider for  
    BP9. Solid blue, purple dashed, and gray dashed curves denote the total, scalar mediated $t$-channel, and gauge boson mediated $s$-channel contributions, respectively. The left (right) panel corresponds to $\sin\theta_R=0.1$ ($0.3$).}
    \label{fig:xsec-vlmpair}
\end{figure}

Since the $\eta_2$ states are invisible to the detector, the experimentally relevant signature is $\mu^+ \mu^- + \cancel{E_T}$ with two isolated, opposite-sign muons accompanied by substantial missing transverse momentum. Thus, although four DM particles occur in the full decay chain, their detector-level effect is captured by the missing-momentum distribution. The branching pattern of the VLM and $h_3$ is illustrated in Fig.\,\ref{fig:br}. Fig.\,\ref{fig:xsec-vlmpair} illustrates the dependence of the VLM pair-production cross section on $m_{vl}$ for BP9 at a $3~\mathrm{TeV}$ muon collider. A pronounced enhancement of the production rate is observed as the right-handed mixing angle is increased from $\sin\theta_R=0.1$ to $0.3$. This behaviour originates from the corresponding enhancement of the Yukawa coupling $y_\psi$ (see right panel of Fig.\,\ref{fig:g_2}), which controls the scalar mediated $t$-channel contribution. For $\sin\theta_R=0.3$, the $t$-channel contribution closely follows the total cross section over essentially the entire mass range and exceeds the gauge-mediated $s$-channel contribution by several orders of magnitude, demonstrating that VLM pair production is dominated by scalar exchange in this region. For the smaller mixing, $\sin\theta_R=0.1$, the gauge and scalar contributions are comparable at lower $m_{vl}$, whereas the scalar-mediated contribution becomes increasingly important as the VLM mass increases. The fact that the total rate lies below the isolated $t$-channel contribution in this case also indicates a destructive interference with the gauge-mediated amplitude. Interestingly, the production cross section increases with $m_{vl}$ over most of the displayed range despite the decreasing phase space.

The present manuscript already identifies $vl^\pm\to\mu^\pm h_3$ followed by invisible $h_3$ decay as the origin of the $2$-OS-muon $+\cancel{E_T}$ signature. We also apply a veto on events containing a jet to suppress jets faking as leptons and leptonic decays of events containing top and bottom quarks.

The dominant irreducible SM background to this topology is

$$ \mu^+\mu^-\rightarrow \mu^+\mu^-\nu_i\bar\nu_i, \qquad i=e,\mu,\tau, $$
where the neutrinos generate genuine missing transverse momentum. Other backgrounds, when included, are treated independently at the event-generation level and combined only after retaining their individual production cross sections and normalizations.

\subsection{Event generation and detector simulation}

Signal and background events are generated at the parton level and subsequently interfaced with \textsc{Pythia8} for showering, hadronization, initial- and final-state radiation, and unstable-particle decays. Detector effects are simulated with \textsc{Delphes} using the muon-collider detector configuration. Reconstructed jets are clustered using the anti-$k_T$ algorithm with radius parameter $R=0.4$, while electrons and muons are subjected to the corresponding transverse-momentum, pseudorapidity, and isolation requirements. The missing transverse momentum is reconstructed from the negative vector sum of the transverse momenta of the visible reconstructed objects. These detector-level objects are then used to construct the kinematic observables entering the multivariate analysis.

At the analysis level, events are required to contain two oppositely charged muons. The same final-state requirements are imposed on the signal and every background sample before writing the reduced analysis ROOT trees. This ensures that the subsequent multivariate classifier compares signal and background within the same experimentally defined final state.

\subsection{Multivariate analysis}

Although both the signal and the dominant SM background lead to the same visible topology,
$$ \mu^+\mu^-+\slashed{E}_T, $$
their kinematic distributions can be substantially different. In the signal, the observed muons originate from the two-body decays of heavy vector-like leptons,
$$ vl^{\pm}  \rightarrow \mu^{\pm} h_3,$$
whereas in the background the muons and missing momentum originate entirely from SM electroweak processes. The large mass scale associated with VLM production therefore leads to characteristic differences in the muon momenta, total event activity, missing transverse momentum, invariant masses, and angular correlations. Rather than imposing a sequence of independent rectangular cuts, we exploit these correlated differences using a Boosted Decision Tree.

For each selected event, we construct seven kinematic observables:
\begin{itemize}
    \item Effective mass : $M_{\rm eff}$ 
    \item The scalar sum of the $P_T$ of the four final state lepton : $L_T$
    \item Rapidity between two selected muons: $\Delta R_{\mu \mu}$
    \item The $p_T$ of the leading lepton: $p_T^{l_1}$
    \item Missing energy : $\slashed{E}_T$
    \item The invariant mass of the two leptons : $M_{\mu \mu}$
    \item Transverse mass of the two leptons: $M^{\mu\mu}_{T}$
\end{itemize}

\begin{figure}
    \centering
    \includegraphics[width=0.49\linewidth]{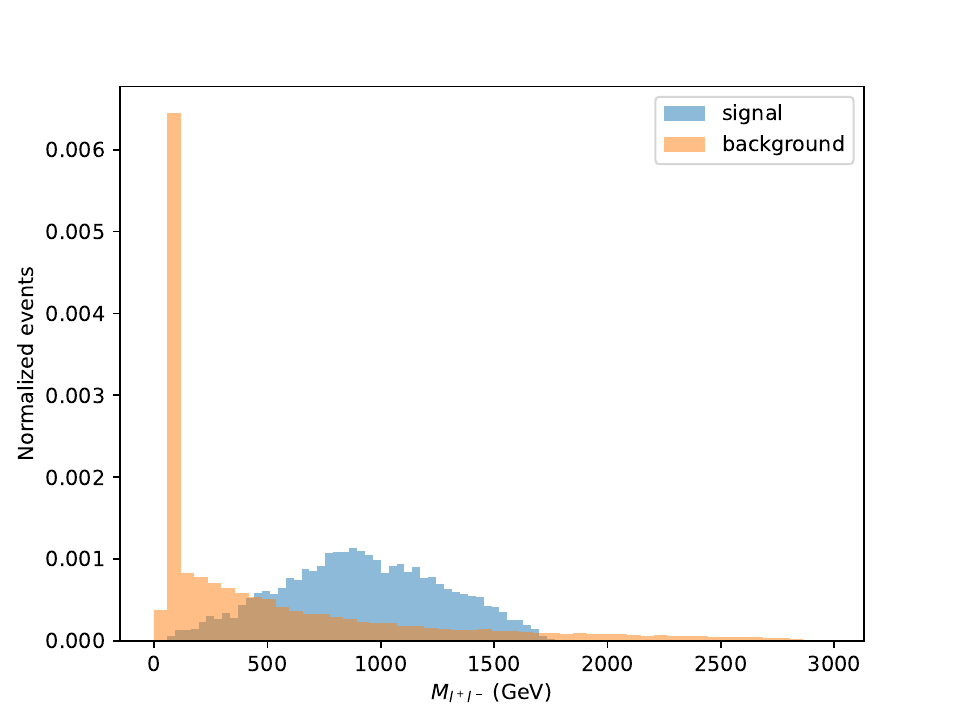}
    \includegraphics[width=0.49\linewidth]{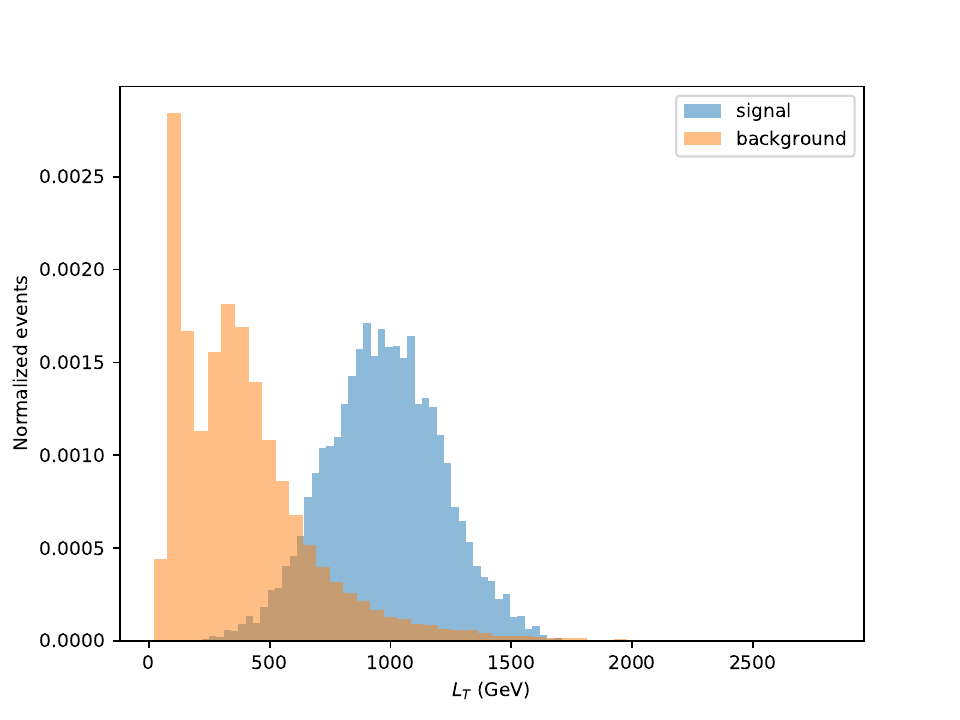}
    \includegraphics[width=0.49\linewidth]{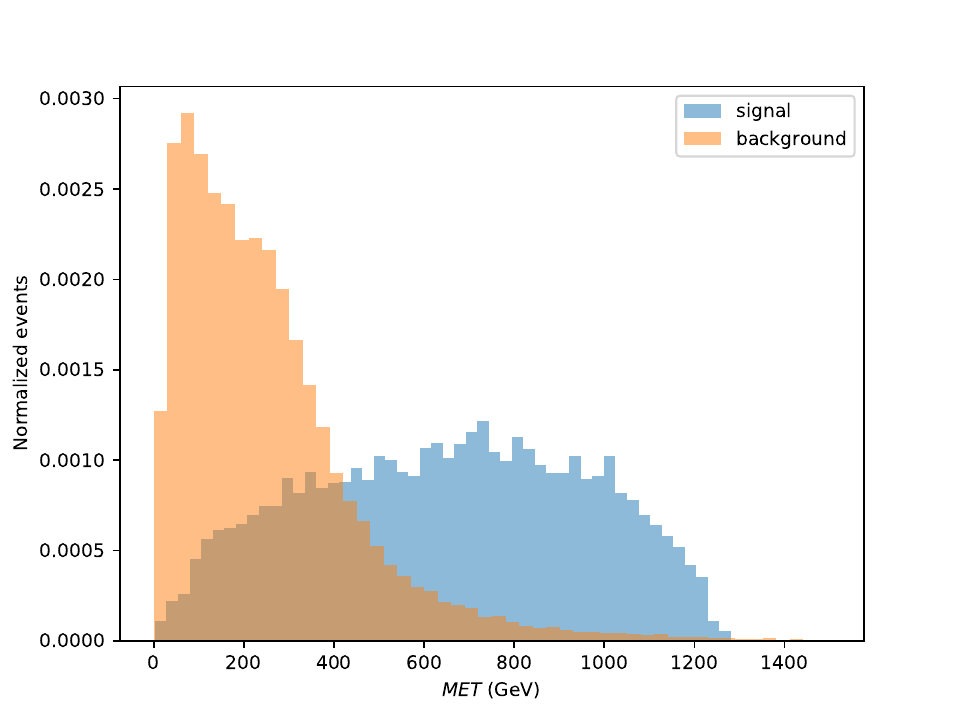}
    \includegraphics[width=0.49\linewidth]{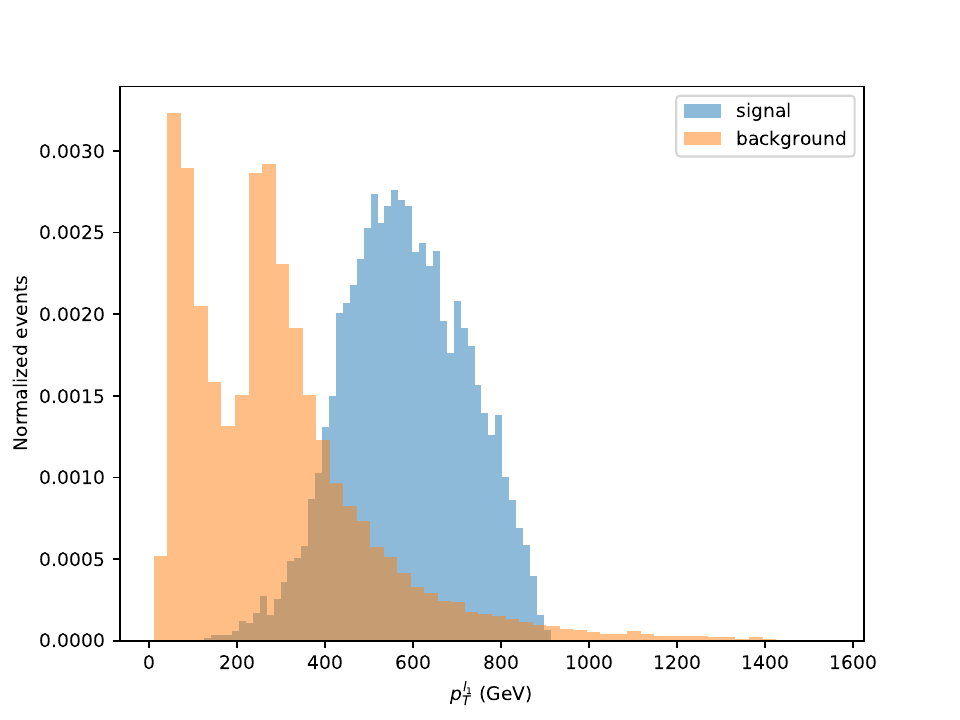}
    \caption{Normalized distributions of some of the important kinematic observables used for signal vs. background classification.}% 
\label{fig:distributions}
\end{figure}

Here $L_T$ characterizes the total transverse momentum carried by the selected leptons, and $M_{\rm eff}$ combines the visible transverse activity with the missing momentum. The dilepton invariant mass, transverse-mass observable, and angular variables provide complementary information about the topology and boost of the underlying event. Representative signal and background distributions for some of the kinematic observables are shown in Fig.\,\ref{fig:distributions}. 

For each process, the training samples are weighted to the luminosity as $$ w_i = \frac{1}{N_{MC}} \times \sigma_i \times \mathcal{L}, $$
where $N_{MC}$ are the number of events satisfying the kinematic and muon multiplicity cuts. 
Importantly, different background processes are retained separately when assigning these weights, so that their relative physical normalizations are preserved when they are subsequently combined.

The selected events are then divided into statistically independent training and testing samples, with $80\%$ of the events used for training and the remaining $20\%$ reserved for performance evaluation. %Signal events are assigned the binary label $y=1$, while all background processes correspond to $y=0$.
We employ boosted decision trees (BDT) based classifier implemented with XGBoost \cite{Chen:2016btl} using the above kinematic variables to construct a multivariate discriminant,
$$ 0 \le \mathcal{D_{\rm{BDT}}} \le 1$$
with larger values corresponding to increasingly signal-like events.

\begin{figure}[h]
    \centering
    \includegraphics[width=6cm, height=5.5cm]{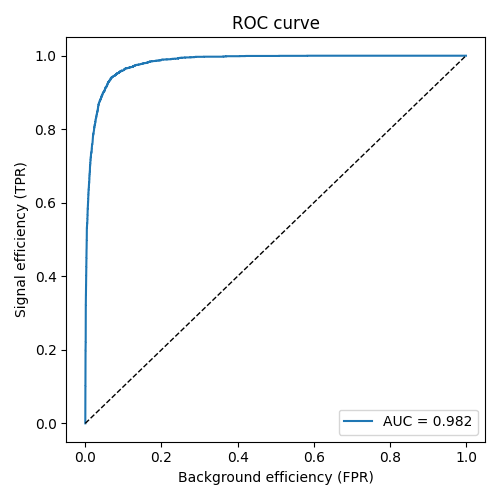}
    \caption{The receiver-operating-characteristic (ROC) curve constructed by plotting the true positive rate (TPR) against the false positive rate (FPR).}
\label{fig:roc}
\end{figure}

The discriminating power of the multivariate classifier is quantified through the receiver operating characteristic (ROC) curve, shown in Fig.\ref{fig:roc}. The ROC curve displays the signal efficiency, $\epsilon_S$ (true-positive rate), against the background efficiency, $\epsilon_B$ (false-positive rate), as the selection threshold on the BDT response is varied. A classifier with no discriminating power would follow the diagonal line, corresponding to an area under the curve (AUC) of 0.5, whereas an ideal classifier approaches the upper-left corner with $\mathrm{AUC}=1$. For the benchmark considered here, the XGBoost classifier yields a test-sample AUC of $0.982$, demonstrating a very strong separation between the VLM signal and the SM backgrounds.

Finally, the BDT output is converted into a collider sensitivity by applying a classifier threshold as
$$ \mathcal{D_{\rm{BDT}}} > \mathcal{D_{\rm{cut}}},$$
and determining the surviving signal and background yields, 
$$ S(\mathcal{D_{\rm{cut}}}) = \sigma_S~\mathcal{L}~\epsilon_S~(\mathcal{D_{\rm{cut}}}),$$
$$ B(\mathcal{D_{\rm{cut}}}) = \mathcal{L} \sum \sigma_B~\epsilon_B~(\mathcal{D_{\rm{cut}}}),$$
where $\epsilon_S$ and $\epsilon_{B}$ denote the BDT selection efficiencies of signal and background respectively. The surviving signal and background yields are evaluated using their corresponding event weights, and the statistical significance is estimated as
$$ Z = \frac{S}{\sqrt{S+B}}.$$
This procedure allows the correlated multidimensional kinematic information to be exploited more efficiently than a conventional cut-and-count analysis and provides the basis for assessing the VLM discovery reach of a future \(3~\mathrm{TeV}\) muon collider. The significance reaches its maximum for an optimized BDT requirement of
$$ \mathcal{D_{\rm BDT}} > 0.924. $$
For the assumed integrated luminosity of $500~{\rm fb}^{-1}$, this selection retains the bulk of the signal while reducing the SM background to negligible resulting in statistical significance $Z \sim 12$, demonstrating a strong discovery potential for the considered benchmark at a $3~{\rm TeV}$ muon collider. However, this significance corresponds to the purely statistical estimate and does not include systematic uncertainties on background prediction.

\subsection*{Muon anomalous magnetic moment}
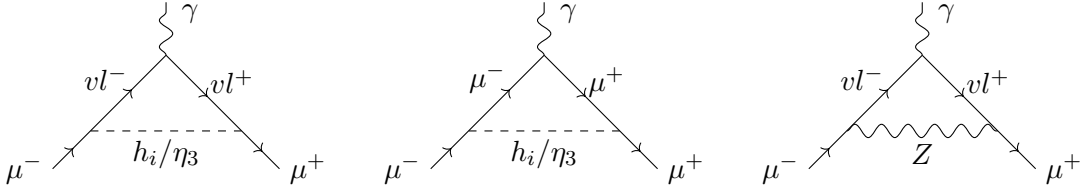
\begin{figure}[!htb]
\centering
\begin{tikzpicture}[scale=0.5]
\begin{scope}
\coordinate (L) at (0,0);
\coordinate (T) at (2,2);
\coordinate (R) at (4,0);
\draw[fermion] (-1,-1)--(L);
\node[left] at (-1,-1){$\mu^-$};
\draw[fermion] (L)--(T);
\node[left] at (1.25,1.4){$vl^-$};
\draw[fermion] (T)--(R);
\node[right] at (2.9,1.3){$vl^+$};
\draw[fermion] (R)--(5,-1);
\node[right] at (5,-1){$\mu^+$};
\draw[dashed] (L)--(R);
\node at (2,-0.6){$h_i/\eta_3$};
\draw[decorate,decoration={snake}] (T)--(2,3.4);
\node[right] at (2.15,3.1){$\gamma$};
\end{scope}
\begin{scope}[xshift=10cm]
\coordinate (L) at (0,0);
\coordinate (T) at (2,2);
\coordinate (R) at (4,0);
\draw[fermion] (-1,-1)--(L);
\node[left] at (-1,-1){$\mu^-$};
\draw[fermion] (L)--(T);
\node[left] at (1.25,1.4){$\mu^-$};
\draw[fermion] (T)--(R);
\node[right] at (2.9,1.3){$\mu^+$};
\draw[fermion] (R)--(5,-1);
\node[right] at (5,-1){$\mu^+$};
\draw[dashed] (L)--(R);
\node at (2,-0.6){$h_i/\eta_3$};
\draw[decorate,decoration={snake}] (T)--(2,3.4);
\node[right] at (2.15,3.1){$\gamma$};
\end{scope}
\begin{scope}[xshift=20cm]
\coordinate (L) at (0,0);
\coordinate (T) at (2,2);
\coordinate (R) at (4,0);
\draw[fermion] (-1,-1)--(L);
\node[left] at (-1,-1){$\mu^-$};
\draw[fermion] (L)--(T);
\node[left] at (1.25,1.4){$vl^-$};
\draw[fermion] (T)--(R);
\node[right] at (2.9,1.3){$vl^+$};
\draw[fermion] (R)--(5,-1);
\node[right] at (5,-1){$\mu^+$};
\draw[decorate,decoration={snake}] (L)--(R);
\node at (2,-0.6){$Z$};
\draw[decorate,decoration={snake}] (T)--(2,3.4);
\node[right] at (2.15,3.1){$\gamma$};
\end{scope}
\end{tikzpicture}
\caption{One-loop contributions to the muon $g$-2 mediated by new BSM fields.}
\label{fig:feyng-2}
\end{figure}
The muon anomalous magnetic moment, $a_\mu$, is in good agreement with the SM prediction, with recent high-precision lattice calculations~\cite{Boccaletti:2024guq,Borsanyi:2020mff,Aliberti:2025beg} showing agreement with the experimental measurement within $0.5\,\sigma$~\cite{Muong-2:2025xyk}. In our framework, the new degrees of freedom contribute to $a_\mu$ through the diagrams shown in Fig.\,\ref{fig:feyng-2}, giving an additional contribution $\Delta a_\mu$. The left diagram gives the dominant contribution, arising from the coupling of the VLM to the scalar states $h_3$ and $\eta_3$ propagating in the loop. The middle diagram is suppressed by either the muon Yukawa coupling or $\sin\theta_L$, while the right diagram is further suppressed by $\sin^2\theta_L$. 

\begin{figure}
    \centering
    \includegraphics[width=0.48\linewidth]{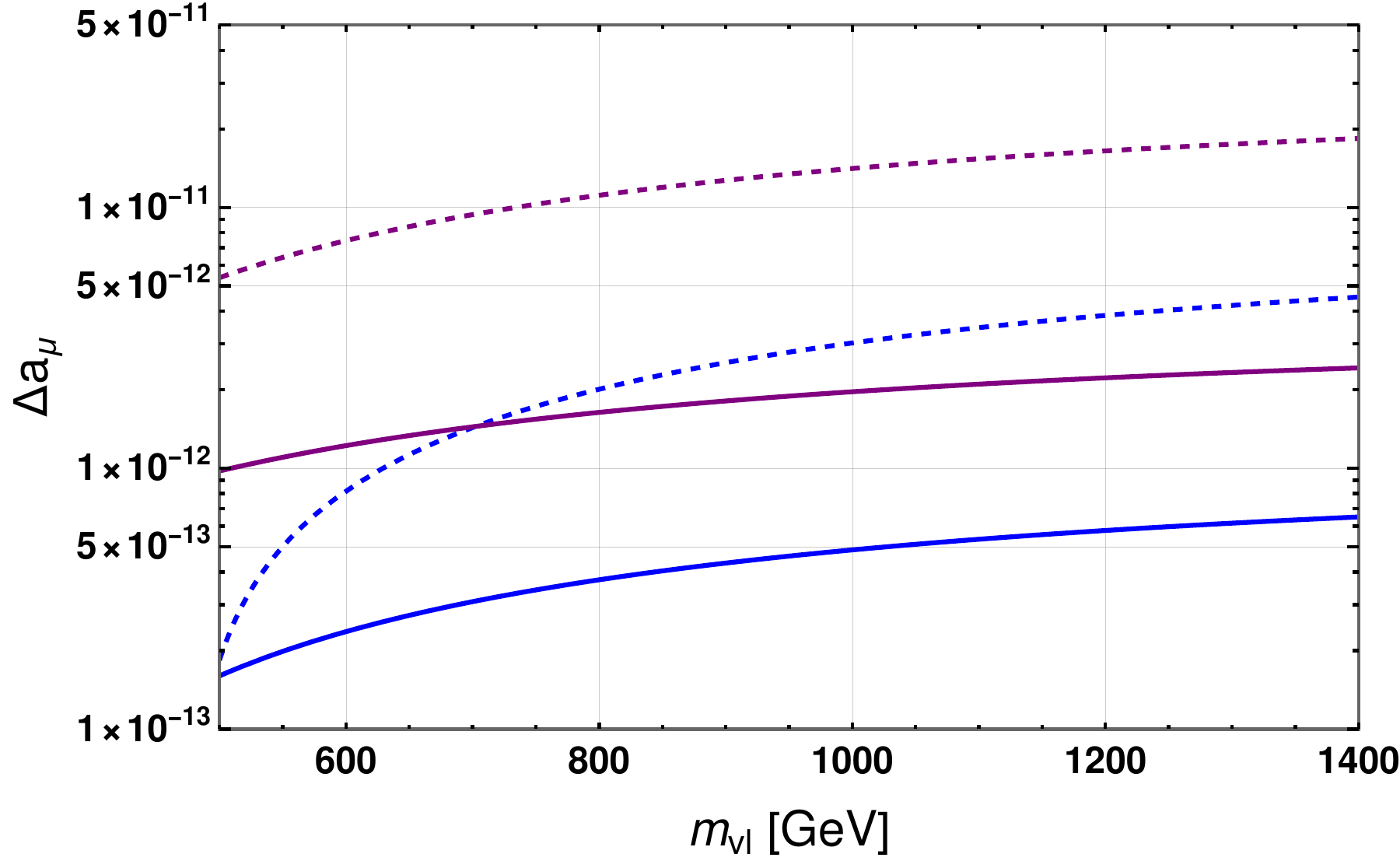}
    \includegraphics[width=0.48\linewidth]{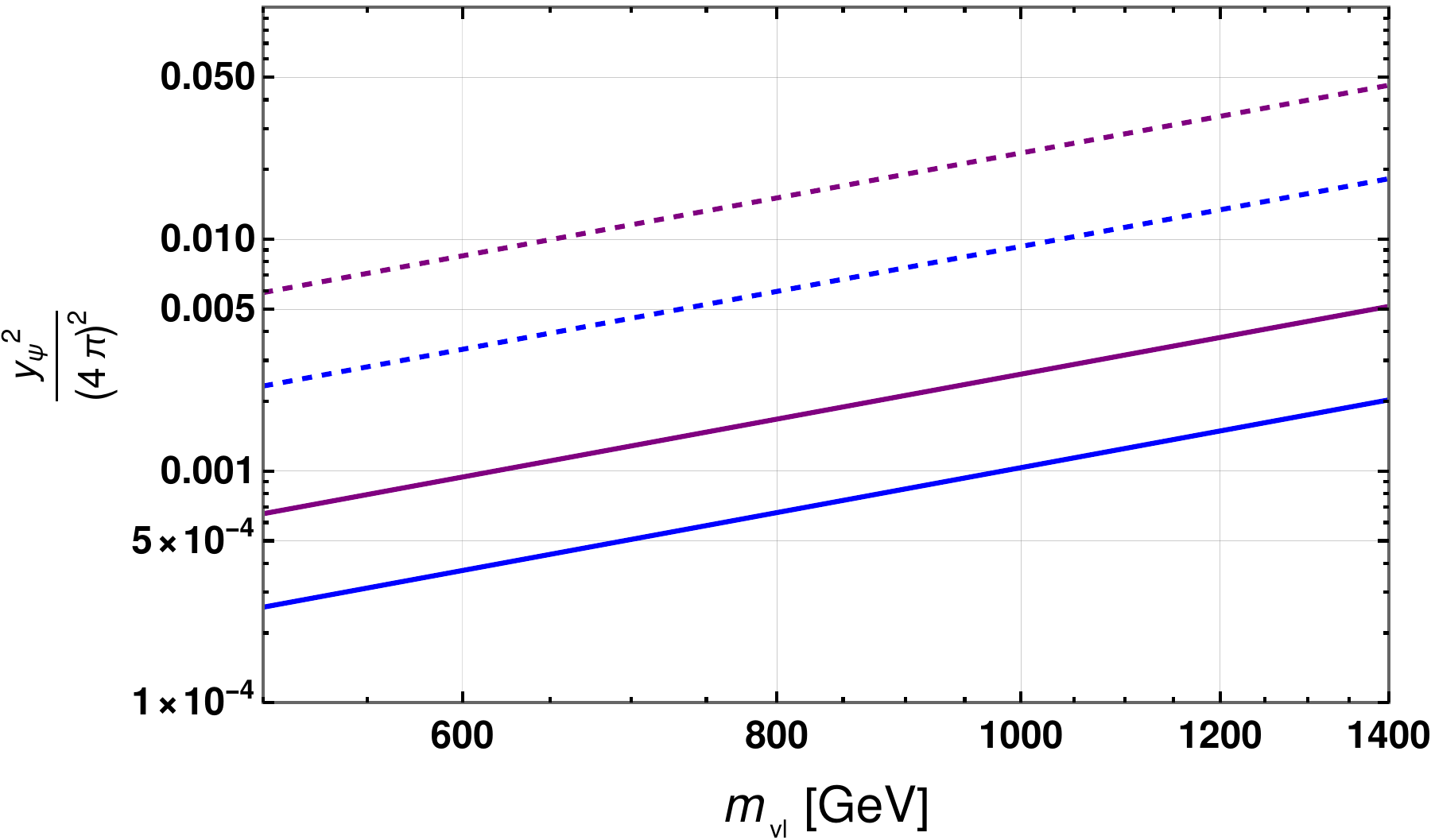}
    \caption{Variation of $\Delta a_\mu$ (left) and the Yukawa coupling (right) with $m_{vl}$. Solid (dashed) blue/purple curves denote $\sin\theta_R=0.1$ ($0.3$) for BP8/BP9. }
    \label{fig:g_2}
\end{figure}

 The contribution to $\Delta a_\mu$ from the exchange of the neutral scalars $(h_i,\eta_3)$ with the VLM, corresponding to the left diagram, is denoted by $\Delta a_{\mu}^{h_i,\eta_3}$ and is given by~\cite{Lee:2021gnw,Leveille:1977rc,Lindner:2016bgg} 
\begin{equation}
\Delta a_{\mu}=\frac{m_\mu^{2}}{8\pi^{2}}\sum_{i=h_i,\eta_3}\int_{0}^{1}dx\,\frac{|v_i^{\psi_2}|^2\left(x^2-x^3+\frac{m_{vl}}{m_\mu}x^2\right)+|d_i^{\psi_2}|^2\left(x^2-x^3-\frac{m_{vl}}{m_\mu}x^2\right)}{m_\mu^2x^2+\left(m_{vl}^2-m_\mu^2\right)x+m_{i}^2(1-x)}.
\end{equation}
Here, the scalar and pseudoscalar coupling factors, $v_i^{\psi_2}$ and $d_i^{\psi_2}$, can be obtained from Eqs.\,\ref{app:g-2}.
Fig.\,\ref{fig:g_2} shows the variation of $\Delta a_\mu$ (left panel) and the Yukawa coupling (right panel) with the VLM mass for two representative values of $\sin\theta_R$ at the benchmark points BP8 and BP9, as specified in the caption.
 We find that $\Delta a_\mu$ remains sufficiently small over the considered parameter space and hence does not alter the SM prediction for $a_\mu$. The other BPs considered in our analysis yield even more suppressed contributions to $\Delta a_\mu$ compared with BP8 and BP9, and are therefore also consistent with the SM prediction. Note that the increase in $\Delta a_\mu$ with $m_{vl}$ is driven by the enhancement of the Yukawa coupling, which dominates over the propagator suppression at larger masses.

\section{Dark Matter Phenomenology}
\label{dm-pheno}
In this section, we study the phenomenology of WIMP-like pNGB dark matter. We compute the thermal relic abundance using the \texttt{micrOMEGAs} package \cite{Belanger:2004yn, Alguero:2023zol}, with the model implemented in \texttt{FeynRules} \cite{Alloul:2013bka}. At tree level, the spin-independent DM–nucleon scattering amplitude vanishes because the DM candidate is a pNGB, as discussed below. We therefore consider the relic density constraint together with the loop-induced contributions to direct detection and delineate the resulting viable parameter space.

            As mentioned above, the CP-odd component $\eta_2$ of the scalar $S_2$ serves as the DM candidate, whose annihilation after spontaneous symmetry breaking proceeds via $s$-channel exchange of the three physical scalar states $h_1$, $h_2$, and $h_3$.
 The SM Higgs decays predominantly to $b\bar{b}$ and gauge bosons, while the other Higgs states $h_{2,3}$ can decay into pairs of lighter Higgs bosons, depending on their masses and mixing angles, as well as into SM final states through their mixing with the SM Higgs.  
In addition, $\eta_2$ annihilation can proceed through contact interactions into scalar pairs $h_i h_j$, as well as into $\eta_3\eta_3$, whenever kinematically allowed.
Therefore, the freeze-out relic abundance is governed predominantly by the masses and mixing angles in the CP-even scalar sector, as well as by the VEVs of the additional scalars, which determine the relevant interaction vertices and quartic couplings.

\begin{figure}
    \centering
    \includegraphics[width=0.48\linewidth]{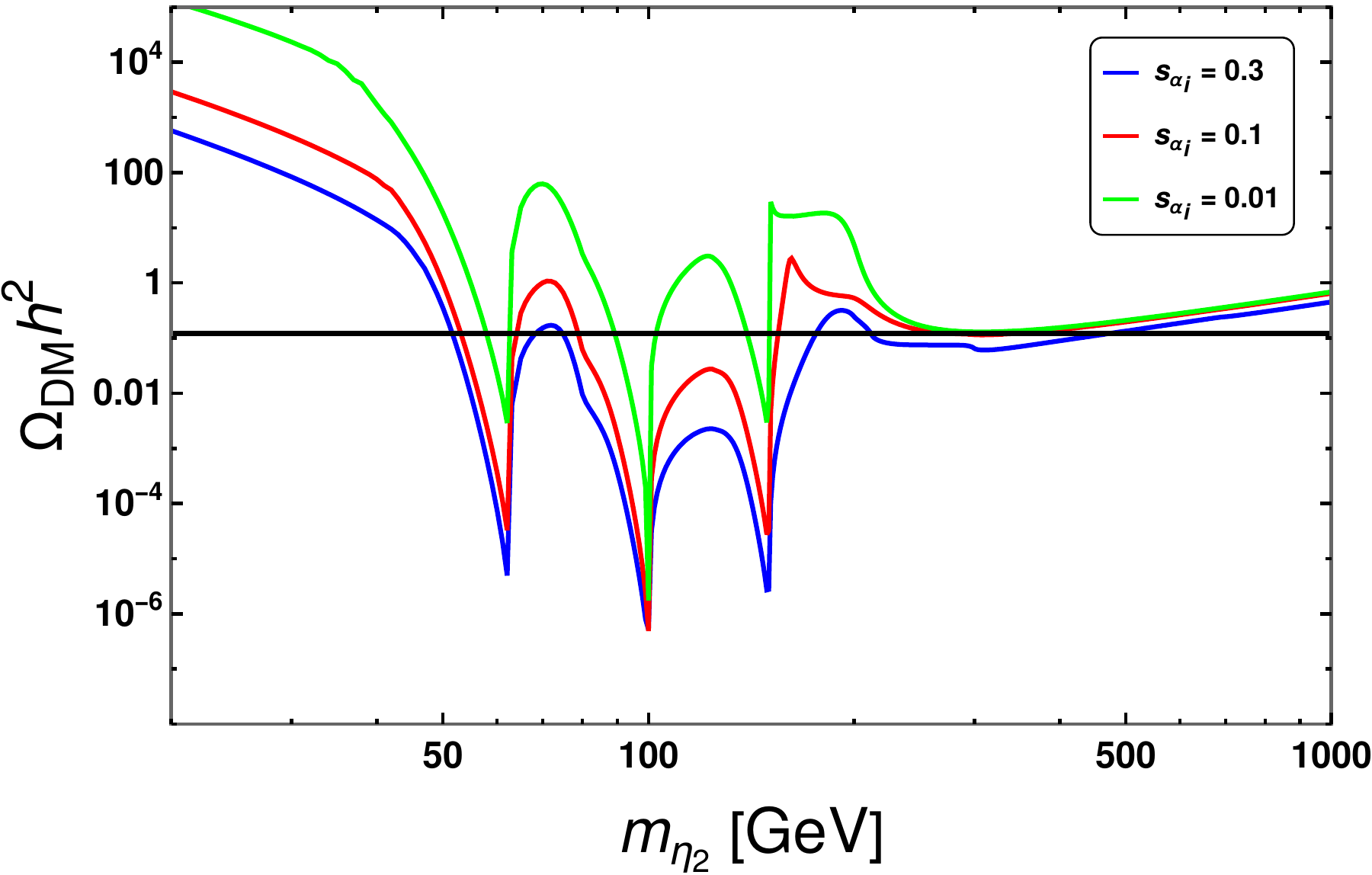}
    \includegraphics[width=0.48\linewidth]{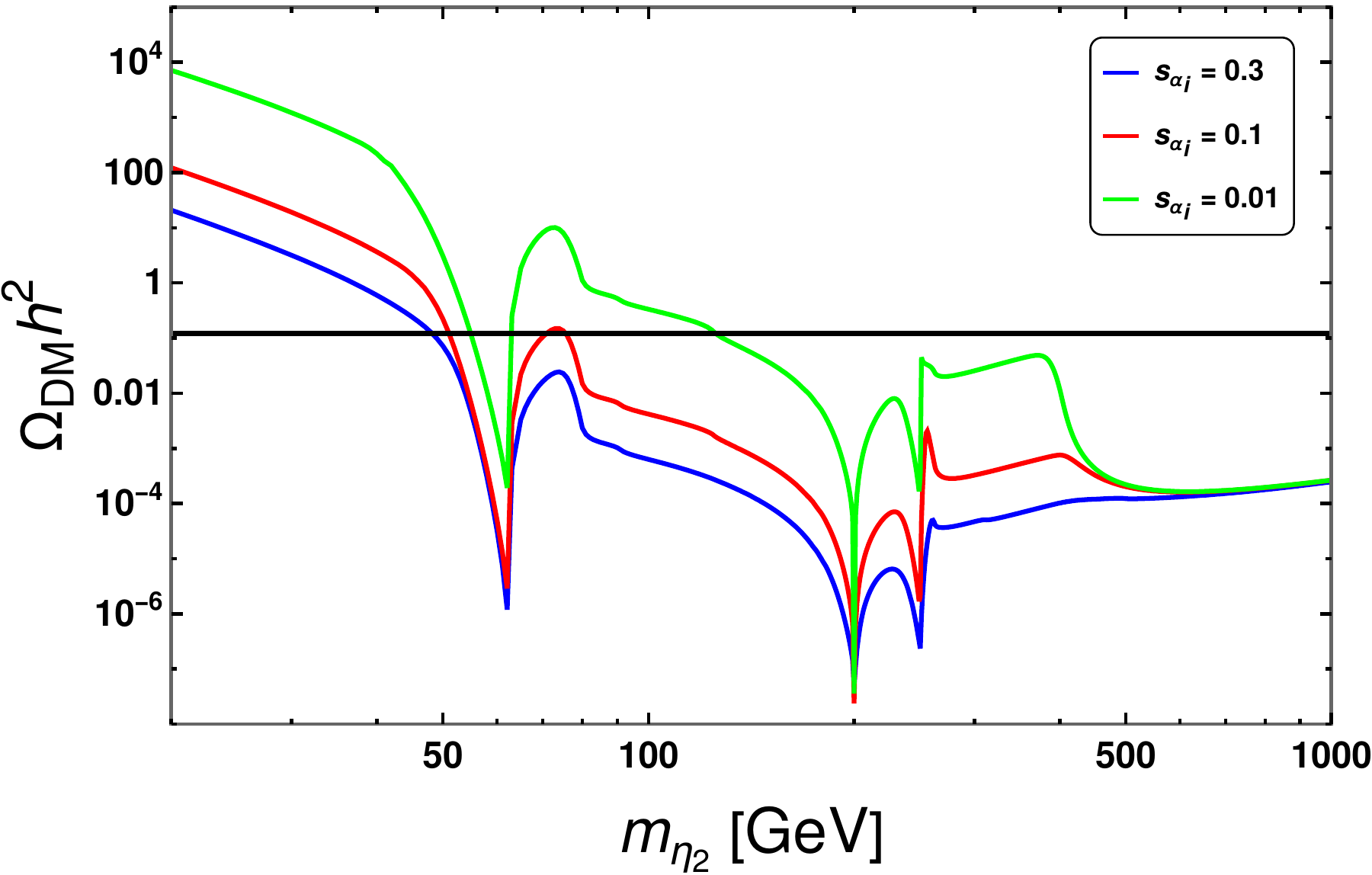}
    \caption{Variation of relic abundance with the DM mass $m_{\eta_2}$ for three representative mixing angles, as indicated in the legend. Left: $m_{h_2}=200,\text{GeV}$, $m_{h_3}=300,\text{GeV}$, $m_{\eta_3}=700,\text{GeV}$, $v_2=400,\text{GeV}$, $v_3=600,\text{GeV}$. Right: $m_{h_2}=400,\text{GeV}$, $m_{h_3}=500,\text{GeV}$, $m_{\eta_3}=700,\text{GeV}$, $v_2=100,\text{GeV}$, $v_3=400,\text{GeV}$. In both panels, $m_{vl}=1000,\text{GeV}$, $\sin\theta_R=0.1$. The black solid line denotes the observed relic abundance, $\Omega_{\rm DM}h^2=0.12\pm0.001$~\cite{Planck:2018vyg}. }
    \label{fig:relic}
\end{figure}

Fig.\,\ref{fig:relic} shows the dependence of the thermal relic abundance on the pNGB dark matter mass $m_{\eta_2}$ for two representative choices of the scalar sector parameters, shown in the left and right panels, and two sets of mixing angles indicated in the plot legends. Away from resonances, decreasing the scalar mixing suppresses the annihilation of $\eta_2$ into SM states and therefore increases the relic abundance. This behaviour is clearly visible from the hierarchy among the curves corresponding to different $\sin \alpha_i$. The characteristic dips in $\Omega_{\rm DM}h^2$ originate from resonantly enhanced $s$-channel annihilation through the three CP-even scalars, 
$$ \eta_2 \eta_2 \rightarrow h_i^{*} \rightarrow h_i h_j/W^+ W^-/ZZ,$$ 
when $2m_{\eta_2} \simeq m_{h_i}$. The resonance annihilation process can make the relic density of the remnant DM substantially smaller than the observed value even for small scalar mixing. Away from the resonance regions, the following $t$-channel process 
$$\eta_2 \eta_2 \rightarrow \eta_2^{*} \rightarrow h_i h_j $$
play important role in achieving required abundance of the relic DM. In particular, such additional annihilation channels involving even the heavier extended scalar sector become kinematically accessible and significantly affect the freeze-out abundance at larger DM masses. Consequently, the observed value $\Omega_{\rm DM}h^2\simeq0.12$ can be obtained in several disconnected mass regions, both around the scalar resonances and in the higher-mass regime.

\begin{figure}
    \centering
    \includegraphics[width=0.48\linewidth]{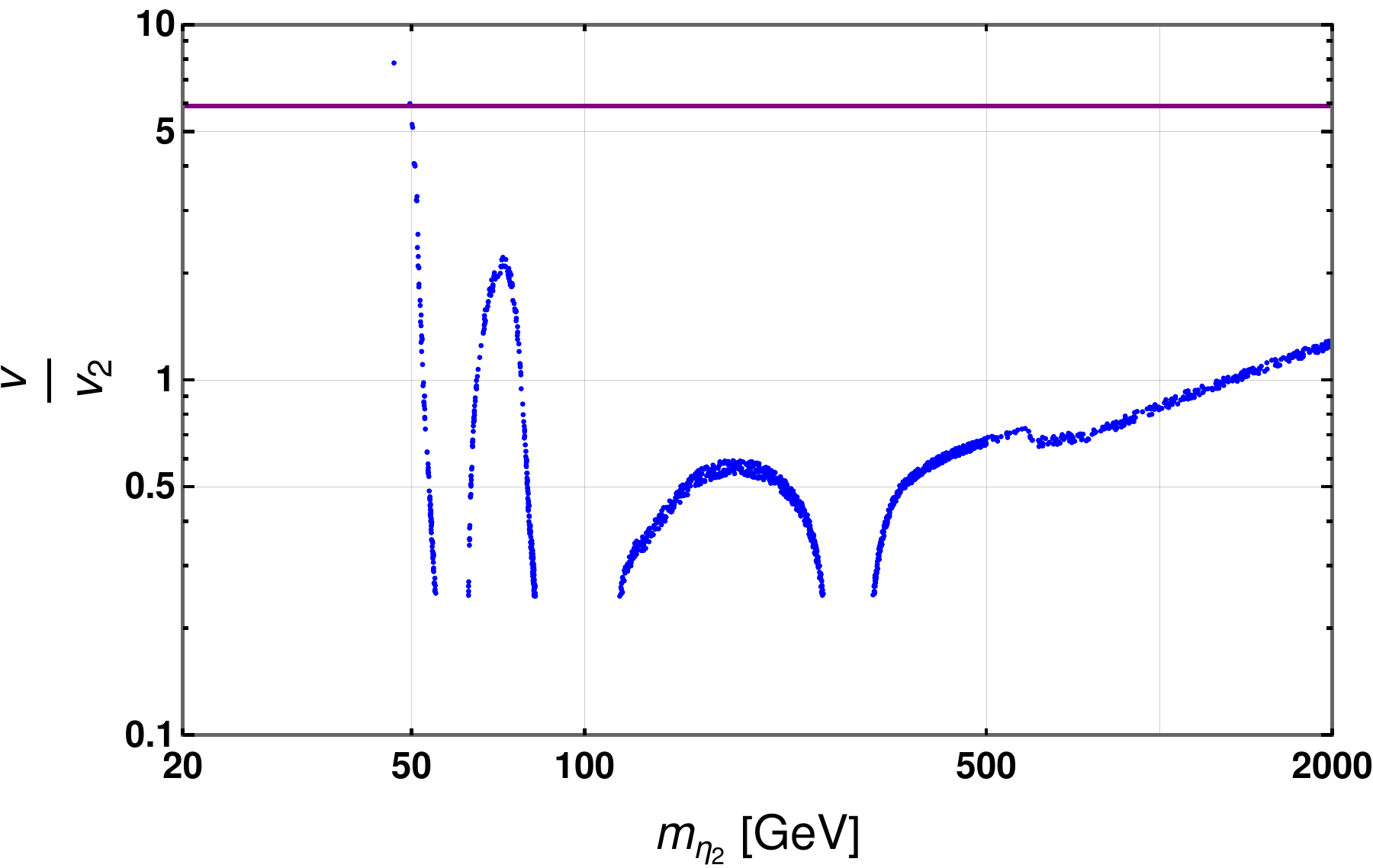}
    \includegraphics[width=0.48\linewidth]{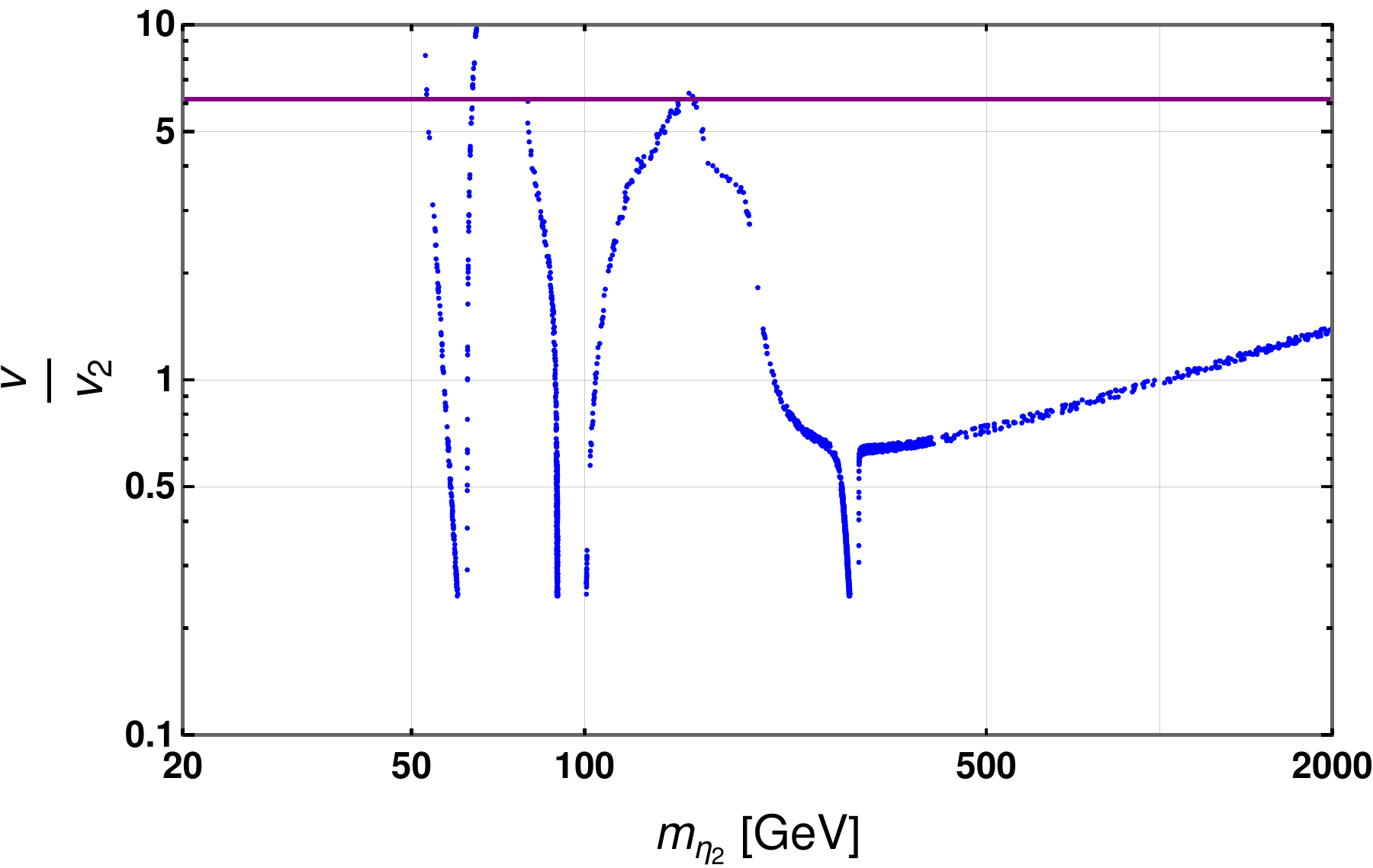}
    \caption{Variation of $v/v_2$ with the DM mass $m_{\eta_2}$ for $v_2\in[20,1000]$ GeV, with $m_{h_3}=600$ GeV, $m_{\eta_3}=700$ GeV, $v_3=500$ GeV, $m_{v_l}=1000$ GeV, and $\sin\theta_R=0.1$. The left and right panels correspond to $s_{\alpha_i}=0.1$ and $0.01$, respectively. All points satisfy the observed dark matter relic density, $\Omega_{\rm DM}h^2=0.12\pm0.001$~\cite{Planck:2018vyg}. The purple solid line denotes the perturbative unitarity bound $\lambda_{S_2} \leq 4\pi$, corresponding to an upper bound on $v/v_2$. }
    \label{fig:v2dm}
\end{figure}

We highlight the correlation between the $S_2$ symmetry breaking scale and the DM relic abundance in Fig.~\ref{fig:v2dm}.
 Since the interactions of the pNGB state with the CP-even scalar sector are controlled by the singlet scale $v_2$, decreasing $v_2$, or equivalently increasing $v/v_2$, generally enhances the DM annihilation rate. 
Away from resonant regions, the requirement of relic density provides a clear correlation between the DM mass $m_{\eta_2}$ and the ratio $v/v_2$, particularly through the annihilation channel $\eta_2\eta_2 \to h_2h_2$. As $m_{\eta_2}$ increases, a progressively stronger effective interaction is required throughout the contact term governed by the self-quartic coupling $\lambda_{S_2}$ to maintain an efficient thermal annihilation rate. Consequently, $v/v_2$ increases in the high-mass region, as $\lambda_{S_2}$ increases with decreasing $v_2$ when all other parameters are kept fixed. For very large DM masses, the value of $\lambda_{S_2}$ required to obtain the correct relic abundance eventually reaches the perturbative limit, which corresponds to a minimum allowed value of $v_2$ for fixed values of the other input parameters (see Eq.\,\eqref{eq:ls2}). In both panels, the perturbative unitarity limit is indicated by the horizontal purple line, with the slight difference between the two panels arising from their different choices of mixing angles. Note that, in the high-DM-mass region, the correct relic abundance can be achieved through the annihilation channels $\eta_2\eta_2 \to \eta_3\eta_3/h_3h_3$ via the corresponding contact interaction governed by the portal coupling $\lambda_{S_2S_3}$, which increases with decreasing $v_3$.
Near $m_{\eta_2}\simeq m_{h_i}/2$, resonant enhancement permits the observed relic density for considerably weaker effective couplings. Reducing the scalar mixing further suppresses annihilation into SM states, making the viable parameter space more localized around these resonant regions as shown in the right panel. A comparison of the two panels shows that, for a larger mixing angle, a smaller value of \(v/v_2\) is required to reproduce the observed relic density, as expected.

\begin{figure}
    \centering
    \includegraphics[width=0.49\linewidth]{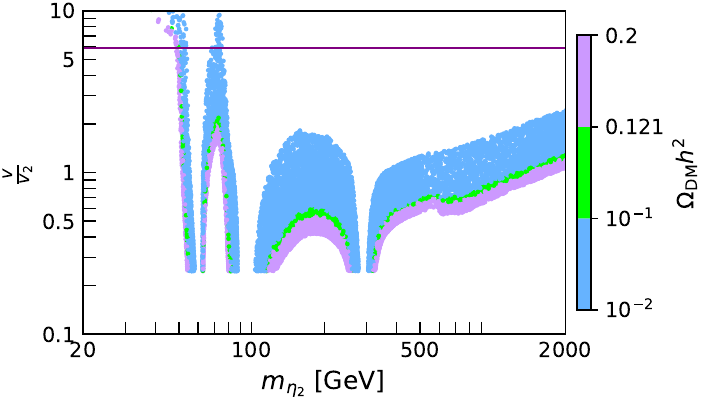}
    \includegraphics[width=0.49\linewidth]{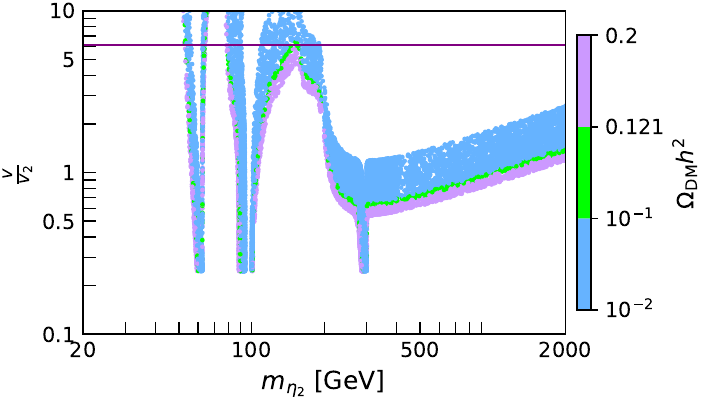}
    \caption{Same parameter space as in Fig.\,\ref{fig:v2dm}, with the color coding indicating the corresponding value of $\Omega_{\rm DM}h^2$ over the range $[0.01$--$0.2]$.}
    \label{fig:relic_band}
\end{figure}

Fig.\,\ref{fig:relic_band} displays the same parameter space features as Fig.\,\ref{fig:v2dm}; however, the color coding in this figure represents three distinct ranges of the relic abundance, as indicated in the bar legend. In particular, the green band corresponds to the range $0.1\leq\Omega_{\rm DM}h^2<0.121$, which encompasses the observed relic abundance. The figure demonstrates that, within the considered parameter space, the relic abundance tends to increase for smaller values of $v/v_2$. This behavior can be understood from the suppression of the relevant quartic coupling with increasing $v_2$, which leads to a reduction in the dark matter annihilation cross section. As a consequence, the reduced annihilation rate results in a larger relic abundance. Therefore, the region below the green band, corresponding to lower values of $v/v_2$, is characterized by an overabundance of dark matter, whereas the region above the green band corresponds to an underabundant relic density. 

\begin{figure}
    \centering
    \includegraphics[width=0.49\linewidth]{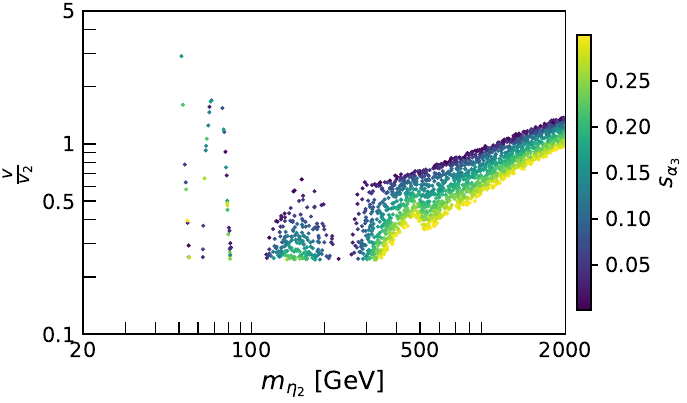}
     \includegraphics[width=0.49\linewidth]{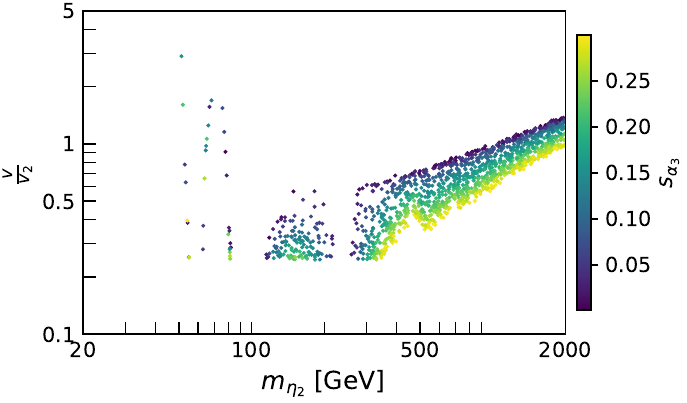}
    \caption{Parameter space for relic density for the range of $0.1\le \Omega_{\rm DM} h^2 \le 0.121$ in mass of DM and VEV of $S_2$ plane. Left panel for $\{s_{\alpha_1},s_{\alpha_2}\}= \{0.1,0.2\}$ and right for $\{s_{\alpha_1},s_{\alpha_2}\}= \{0.3,0.01\}$. For both plots, $\{m_{H_2},m_{H_3},m_{\eta_3},v_3\}=\{200,500,700,500\}$ GeV.}
    \label{fig:relic_alp3}
\end{figure}

Fig.\,\ref{fig:relic_alp3} shows the variation of $v/v_2$ with the dark matter mass $m_{\eta_2}$ in the parameter space consistent with the observed relic abundance, $0.1\leq\Omega_{\rm DM}h^2\leq0.121$. The color coding represents the value of the mixing angle $s_{\alpha_3}$, with $v_2$ scanned over the range $[20,1000]$ GeV, while the remaining parameters are fixed to the benchmark values specified in the caption. The left and right panels correspond to two different choices of $s_{\alpha_1}$ and $s_{\alpha_2}$. The figure demonstrates that, in order to obtain the observed relic abundance for smaller values of $v/v_2$, a relatively larger value of the mixing angle $s_{\alpha_3}$ is required, consistent with the behavior discussed previously. This correlation provides a complementary perspective on the results presented in Fig.\,\ref{fig:v2dm}, illustrating how the required mixing angle varies with the DM mass and $v_2$ for parameter points satisfying the observed relic density.

\begin{table}[ht]
    \centering
    \begin{tabular}{|c|c|c|c|c|c|c|c|c|c|c|c|}
    \hline
    \small Param & BP1 & BP2 & BP3 & BP4 & BP5 & BP6 & BP7 & BP8 & BP9 & BP10 & BP11 \\
    \hline
    $m_{h_2}$ & 400 & 195 & 200 & 200 & 350 & 200 & 400 & 377 & 200 & 200 & 200 \\
    \hline
    $m_{h_3}$ & 400 & 300 & 300 & 300 & 370 & 300 & 500 & 530 & 500 & 500 & 500 \\
    \hline
    $m_{\eta_2}$ & 500 & 500 & 370 & 370 & 550 & 370 & 700 & 124.6 & 150 & 150 & 150 \\
    \hline
    $m_{\eta_3}$ & 890 & 600 & 635 & 635 & 650 & 605 & 990 & 985 & 985 & 985 & 980 \\
    \hline
    $v_2$ & 650 & 350 & 388 & 430 & 520 & 410 & 600 & 150 & 500 & 320 & 280 \\
    \hline
    $v_3$ & 700 & 450 & 450 & 450 & 540 & 450 & 700 & 350 & 220 & 400 & 350 \\
    \hline
    $s_{\alpha_1}$ & 0.105 & 0.10 & 0.10 & 0.30 & 0.01 & 0.10 & 0.10 & 0.01 & 0.064 & 0.01 & 0.01 \\
    \hline
    $s_{\alpha_2}$ & 0.01 & 0.01 & 0.01 & 0.12 & 0.10 & 0.01 & 0.01 & 0.10 & 0.10 & 0.093 & 0.10 \\
    \hline
    $s_{\alpha_3}$ & 0.01 & 0.15 & 0.01 & 0.01 & 0.25 & 0.18 & 0.20 & 0.20 & 0.20 & 0.20 & 0.159 \\
    \hline
    \small $\Omega_{\rm DM}h^2$ & 0.12 & 0.12 & 0.12 & 0.12 & 0.12 & 0.12 & 0.12 & 0.12 & 0.12 & 0.12 & 0.12 \\
    \hline
    \end{tabular}
    \caption{Benchmark points satisfying the observed DM relic density. For all benchmark points, we fix $\sin\theta_R = 0.1$ and $m_{vl} = 1\,\mathrm{TeV}$. All masses and VEVs are given in GeV.}
    \label{tab:BpTab}
\end{table}

For a comprehensive analysis, we select 11 benchmark points (BPs) from the viable parameter space that satisfy the observed dark matter relic abundance, $\Omega_{\rm DM}h^2=0.12$, and exhibit a SFOPT, as discussed in the following section. Using the \texttt{micrOMEGAs} code, we calculate the dominant DM annihilation channels and their corresponding contributions to the relic abundance for the selected benchmark points, which are summarized in Tabs.\,\ref{tab:annhDM1} and \ref{tab:annhDM2}. For BP1–BP7, which correspond to comparatively higher DM masses away from the relevant resonance regions, the dominant DM annihilation channel is $h_2h_2$, mainly driven by the contact interaction governed by $\lambda_{S_2}$. The annihilation channels into SM states proceed through an $s$-channel scalar propagator, with comparatively larger $s_{\alpha_1}$ contributing to the corresponding DM–scalar couplings. In contrast, for BP8–BP11, the dominant annihilation channels of DM are to the final states of SM, particularly to $WW$, $ZZ$, and $h_1h_1$. This is due to the DM mass lying close to the SM-like Higgs mass, thereby enhancing the $s$-channel SM Higgs-mediated annihilation processes and making these channels dominant.

\begin{table}[httb]
\small
    \centering
    \begin{tabular}{|c|c|c|c|c|c|c|c|c|c|}
    \hline 
       BP1 & BP2 & BP3 & BP4 & BP5 & BP6 & BP7 \\
    \hline
     
     \shortstack{$h_2 h_2(36\%)$\\$W W(27\%)$\\$Z Z(13\%)$ \\$h_1 h_1(13\%)$}
    &  \shortstack{$h_2 h_2(86\%)$\\$h_3 h_3(4\%)$\\$h_2 h_3(4\%)$\\$h_1 h_2(3\%)$} & \shortstack{$h_2 h_2(90\%)$\\$h_1 h_2(3\%)$\\$W W(3\%)$\\[0.1cm]}  &  \shortstack{$h_2 h_2(41\%)$\\$h_1 h_2(24\%)$\\$WW(17\%)$\\$Z Z(9\%)$} & \shortstack{$h_2 h_2(88\%)$\\$h_2 h_3(12\%)$\\[0.3cm]} & \shortstack{$h_2 h_2(76\%)$\\$h_3 h_3(11\%)$\\$h_2 h_3(4\%)$\\$h_1 h_2(3\%)$} &  \shortstack{$h_2 h_2(64\%)$\\$W W(13\%)$\\$Z Z(6\%)$\\$h_1 h_1(6\%)$} \\
    \hline
    \end{tabular}
    \caption{Percentage contribution of the annihilation channels to the relic abundance for different BPs.}
    \label{tab:annhDM1}
\end{table}
\begin{table}[httb]
\small
    \centering
    \begin{tabular}{|c|c|c|c|c|c|c|c|c|c|}
    \hline 
       BP8 & BP9 & BP10 & BP11 \\
    \hline
     
     \shortstack{$W W(61\%)$\\$ZZ(26\%)$\\$h_1 h_1(13\%)$ }
    &  \shortstack{$W W(54\%)$\\$ZZ(24\%)$\\$h_1 h_1(22\%)$ } & \shortstack{$W W(51\%)$\\$h_1 h_1(26\%)$\\$ ZZ (23\%)$ } & \shortstack{$W W(53\%)$\\$h_1 h_1(24\%)$\\$ ZZ (23\%)$ } \\
    \hline
    \end{tabular}
    \caption{Percentage contribution of the annihilation channels to the relic abundance for different BPs. }
    \label{tab:annhDM2}
\end{table}

\subsection{Direct detection signature}
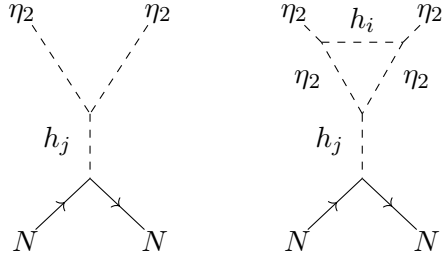
\begin{figure}[!htb]
\centering
\begin{tikzpicture}[scale=0.9]
\begin{scope}
\coordinate (C) at (0.00,0.25);
\draw[dashed] (-0.85,1.55)--(C);
\draw[dashed] (0.85,1.55)--(C);
\node at (-1.00,1.72){$\eta_2$};
\node at (1.00,1.72){$\eta_2$};
\draw[dashed] (C)--(0,-0.70);
\node[left] at (-0.12,-0.15){$h_j$};
\draw[antifermion] (0,-0.70)--(-0.80,-1.45);
\draw[fermion] (0,-0.70)--(0.80,-1.45);
\node at (-0.95,-1.62){$N$};
\node at (0.95,-1.62){$N$};
\end{scope}
\begin{scope}[xshift=4.0cm]
\coordinate (A) at (-0.60,1.30);
\coordinate (B) at (0.60,1.30);
\coordinate (C) at (0.00,0.25);
\draw[dashed] (A)--(B);
\node at (0.00,1.63){$h_i$};
\draw[dashed] (B)--(C);
\node[right] at (0.45,0.75){$\eta_2$};
\draw[dashed] (C)--(A);
\node[left] at (-0.45,0.75){$\eta_2$};
\draw[dashed] (-0.85,1.55)--(A);
\draw[dashed] (0.85,1.55)--(B);
\node at (-1.00,1.72){$\eta_2$};
\node at (1.00,1.72){$\eta_2$};
\draw[dashed] (C)--(0,-0.70);
\node[left] at (-0.12,-0.15){$h_j$};
\draw[antifermion] (0,-0.70)--(-0.80,-1.45);
\draw[fermion] (0,-0.70)--(0.80,-1.45);
\node at (-0.95,-1.62){$N$};
\node at (0.95,-1.62){$N$};
\end{scope}

\end{tikzpicture}
\caption{Tree-level (left) and representative one-loop topology of 9 distinct diagrams (right) containing the tree-level topology as a subdiagram, leading to a vanishing $\sigma_{\rm SI}$ through the same mechanism.}
\label{fig:tree}
\end{figure}

The SI scattering cross section of the DM off nucleons vanishes at tree level in the zero-momentum-transfer limit due to the pNGB nature of the DM candidate~\cite{Gross:2017dan}. To understand this property explicitly, it is convenient to work in the non-linear representation,
\begin{align}
S_2 &= \frac{1}{\sqrt{2}} \left(\frac{v_2}{\sqrt{2}} + s_2\right)
\exp\left[i \frac{\pi_a}{v_2}\right],
\end{align}
where $\pi_a$ corresponds to $\eta_2$ in the linear representation, and we have adopted the unitary gauge. In this representation, the vanishing of the cross section arises from the absence of the cubic $\pi_a\pi_a s_2$ coupling, rather than from a non-trivial cancellation among different diagrams (see left diagram of Fig.\,\ref{fig:tree}), as discussed in Ref.\,\cite{Cai:2021evx}.

Substituting the above form of $S_2$ into the Lagrangian in Eq.\,\eqref{eq:LagsoftU1}, we obtain
\begin{align}
 \mathcal{L} &\supset |\partial_\mu S_2|^2 - \frac{\mu_2}{2} \left(S_2^2 + \mathrm{h.c.}  \right)\\
& \supset 
\frac{1}{2 } \left(\partial_\mu \pi_a\right)^2 +  \mu_2 \pi_a^2 + \frac{s_2}{v_2}\left(\partial_\mu \pi_a\right)^2+  \frac{2 \mu_2}{v_2} s_2 \pi_a^2\, .
\end{align}
Using $m_{\eta_2}^2=-2\mu_2$, the cubic couplings are obtained up to total derivatives,
\begin{align}
\mathcal{L} &\supset
-\frac{1 }{v_2} \left[
s_2\pi_a \left(\partial_\mu\partial^\mu+m_{\eta_2}^2\right) \pi_a
+ \pi_a \left(\partial^\mu \pi_a\right) \left(\partial_\mu  s_2\right)
 \right]\,. \label{eq:lags_2} 
\end{align}
Here, the first term in Eq.\,\eqref{eq:lags_2} vanishes upon applying the on-shell condition,
\begin{align}
(\partial_\mu\partial^\mu+m_{\eta_2}^2)\pi_a=0.
\end{align}
Hence, the remaining cubic couplings are proportional to the momentum of the $s_2$, causing the tree-level DM--nucleon scattering cross section to vanish in the zero momentum transfer limit. This suppression is consistent with the current null results from direct detection experiments.

 We numerically calculate the \(t\)-channel contributions to DM--nucleon scattering, shown in the left panel of Fig.\,\ref{fig:tree}. At tree level, the scattering receives contributions from three $t$-channel diagrams mediated by the mass eigenstates $h_i$ ($i=1,2,3$), whose contributions undergo a non-trivial cancellation. Consequently, in the limit of vanishing momentum transfer, the scattering cross section is strongly suppressed to approximately $\mathcal{O}(10^{-80}\,\mathrm{cm}^2)$. This cancellation and suppression of the \(t\)-channel contributions is consistent with the result obtained above.

\begin{figure}[!htb]
\centering
\begin{tikzpicture}[scale=0.9]

\begin{scope}
\coordinate (A) at (-0.60,1.30);
\coordinate (B) at (0.60,1.30);
\coordinate (C) at (0.00,0.25);
\draw[dashed] (A)--(B);
\node at (0.00,1.63){$\eta_2$};
\draw[dashed] (B)--(C);
\node[right] at (0.45,0.75){$h_k$};
\draw[dashed] (C)--(A);
\node[left] at (-0.45,0.75){$h_j$};
\draw[dashed] (-0.85,1.55)--(A);
\draw[dashed] (0.85,1.55)--(B);
\node at (-1.00,1.72){$\eta_2$};
\node at (1.00,1.72){$\eta_2$};
\draw[dashed] (C)--(0,-0.70);
\node[left] at (-0.12,-0.15){$h_i$};
\draw[antifermion] (0,-0.70)--(-0.80,-1.45);
\draw[fermion] (0,-0.70)--(0.80,-1.45);
\node at (-0.95,-1.62){$N$};
\node at (0.95,-1.62){$N$};
\node at (0.,-1.92) {$(a)$};
\end{scope}
\begin{scope}[xshift=5.2cm]
\coordinate (V) at (0.,-0.70);
\coordinate (C) at (-0.6,0.85);
\draw[dashed] (-2.55,1.55) -- (-1.15,0.52);
\draw[dashed] (0.0,0.52) -- (0.85,1.55);
\draw[dashed] (C) circle (0.65);
\draw[dashed] (0.0,0.52) -- (V);
\draw[fermion] (-0.80,-1.45) -- (V);
\draw[fermion] (V) -- (0.80,-1.45);
\node at (-2.95,1.72) {$\eta_2$};
\node at (1.0,1.72) {$\eta_2$};
\node at (-1.2,1.65) {$\eta_2$};
\node at (-0.80,-0.1) {$h_j$};
\node at (0.35,-0.35) {$h_i$};
\node at (-0.95,-1.62) {$N$};
\node at (0.95,-1.62) {$N$};
\node at (0.,-1.92) {$(b)$};
\end{scope}
\begin{scope}[xshift=10.0cm]
\coordinate (V) at (0.,-0.70);
\coordinate (C) at (0.6,0.85);
\draw[dashed] (-1.2,1.55) -- (-0.0,0.52);
\draw[dashed] (1.17,0.52) -- (1.85,1.55);
\draw[dashed] (C) circle (0.65);
\draw[dashed] (0.0,0.52) -- (V);
\draw[fermion] (-0.80,-1.45) -- (V);
\draw[fermion] (V) -- (0.80,-1.45);
\node at (2.0,1.72) {$\eta_2$};
\node at (0.1,1.65) {$\eta_2$};
\node at (-1.5,1.72) {$\eta_2$};
\node at (0.80,-0.1) {$h_j$};
\node at (-0.35,-0.35) {$h_i$};
\node at (-0.95,-1.62) {$N$};
\node at (0.95,-1.62) {$N$};
\node at (0.,-1.92) {$(c)$};
\end{scope}
\end{tikzpicture}
\caption{One-loop topologies contributing to the spin-independent DM–nucleon DD cross section. The left and middle topologies generate 27 and 9 distinct diagrams, respectively, while the middle and right topologies yield identical matrix elements.}
\label{fig:one-loopDD}
\end{figure}
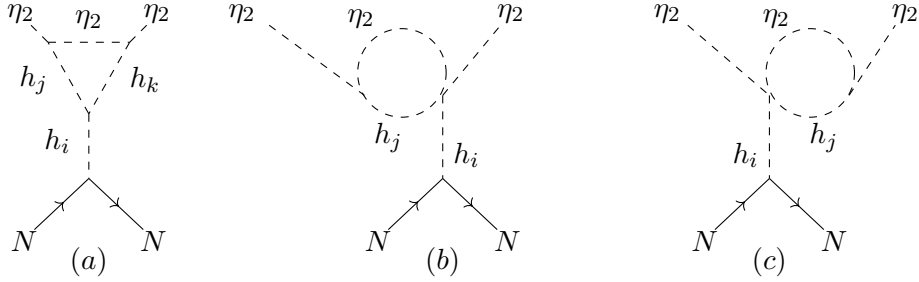

\subsubsection{Loop induced cross section}\label{sec:loop}
 At the one-loop level, the suppression of the spin-independent cross section is no longer guaranteed, since the explicit $U(1)$ breaking associated with the soft-breaking parameter $\mu_2$ makes the DM candidate a pNGB rather than an exact Nambu-Goldstone boson (NGB). While the tree-level amplitude is suppressed in the small momentum transfer limit as a consequence of the pNGB nature of the DM, loop corrections can generate additional effective interactions that are not subject to the same cancellation~\cite{Gross:2017dan,Azevedo:2018exj,Alanne:2020jwx}. Therefore, the one-loop contributions can lead to a non-vanishing spin-independent scattering cross section, even though it remains suppressed by the smallness of the explicit symmetry-breaking effects.

\begin{figure}
    \centering
    \includegraphics[width=0.48\linewidth]{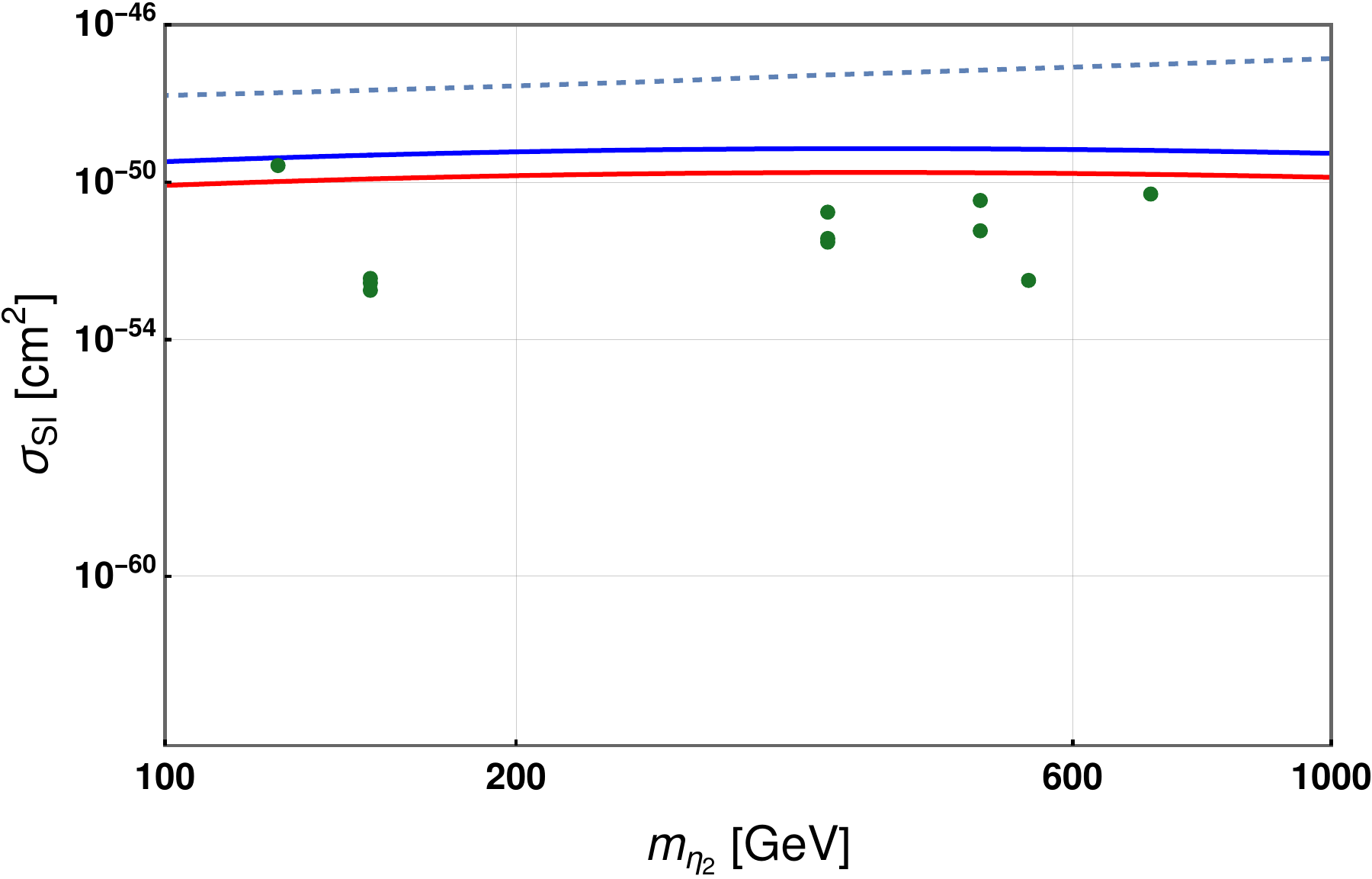}
    \includegraphics[width=0.48\linewidth]{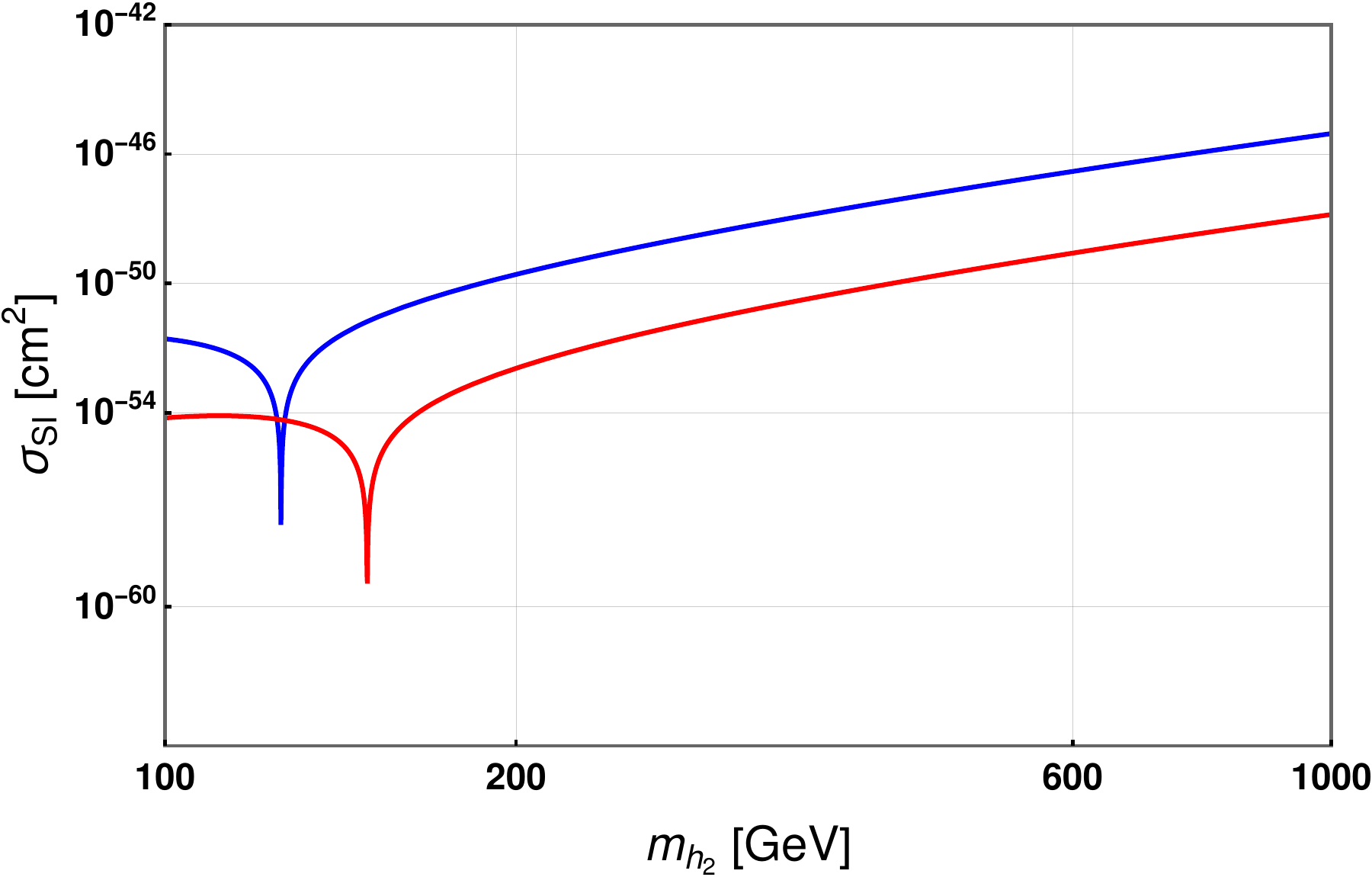}
   \caption{  
   Variation of the one-loop spin-independent cross section with the DM mass (left) for $(m_{h_2},m_{h_3},m_{\eta_3},v_2,v_3)=(200,300,800,250,400)$ GeV, and with $m_{h_2}$ (right) for $(m_{h_3},m_{\eta_2},m_{\eta_3},v_2,v_3)=(300,500,800,250,400)$ GeV. The blue and red curves correspond to $(s_{\alpha_1},s_{\alpha_2},s_{\alpha_3})=(0.2,0.1,0.1)$ and $(0.01,0.1,0.1)$, respectively. The dashed curve denotes the latest LZ limit~\cite{LZ:2024zvo}, while the dark-green dots mark the 11 benchmark points.}
    \label{fig:dd}
\end{figure}

In the limit $\mu_2 \to 0$, the DM candidate becomes a massless NGB, for which the scattering amplitude would be suppressed even at the loop level. Thus,
\begin{equation}
    \lim_{\mu_2\to 0}\sum_{\rm diagrams}\mathcal{M}(\mu_2)=0.
\end{equation}
This property allows us to simplify the loop calculation by writing~\cite{Abe:2022mlc}
\begin{equation}\label{eq:id}
    \sum_{\rm diagrams}\mathcal{M}(\mu_2)
    =
    \sum_{\rm diagrams}\left[\mathcal{M}(\mu_2)-\mathcal{M}(0)\right].
\end{equation}
Consequently, all $\mu_2$-independent contributions cancel, and only the $\mu_2$-dependent parts need to be evaluated. Note that the right topology in Fig.\,\ref{fig:tree} contains 9 distinct diagrams containing the tree-level diagram as a subdiagram. We numerically verify that their total contribution vanishes at the amplitude level, consistent with the tree-level cancellation. Therefore, the external leg corrections to the DM do not contribute to the scattering amplitude for zero momentum transfer.

The leading one-loop vertex correction contributions are shown in Fig.\,\ref{fig:one-loopDD}, yielding a nonzero, momentum-transfer independent SI cross section that depends on the DM mass. The explicit expressions for the individual loop contributions are given in App.\,\ref{app:dd}. Using Eq.\,\eqref{eq:id}, which ensures the cancellation of the ultraviolet (UV) divergences, the one-loop SI cross section can be written as
 \begin{equation}\label{eq:one_loop_X}
     \sigma_{\rm SI} \simeq \frac{m_N^4 f_N^2 }{4 \pi (m_{\eta_2} + m_N)^2 v^2} \left|\frac{1}{16 \pi^2} F(\mu_2)\right|^2,
 \end{equation}
where $m_N$ denotes the nucleon mass and $f_N$ is the effective Higgs--nucleon coupling. For numerical analysis, we adopt $m_N=0.946$ GeV~\cite{Alanne:2018zjm} and $f_N\simeq0.3$~\cite{Alarcon:2011zs,Alarcon:2012nr,Cline:2013gha}. Here, the loop function $F(\mu_2)$ of mass dimension $(-1)$ arising from the one-loop contributions shown in Fig.~\ref{fig:one-loopDD} is given in App.\,\ref{app:dd} in integral form. Note that in our model, the dependence $\mu_2$ enters only through $m_{\eta_2}$, such that the limit $\mu_2\to 0$ corresponds to $m_{\eta_2}\to 0$. 
%Also note that the ultraviolet (UV) divergences arising from the one-loop diagrams cancel upon employing the identity given in Eq.\,\eqref{eq:id}.  
The remaining contributions, including tadpole and self-energy diagrams as well as SM corrections, are suppressed or largely cancel among themselves~\cite{Abe:2024vxz,Azevedo:2018exj,Alanne:2020jwx}.

\begin{figure}
    \centering
    \includegraphics[width=0.48\linewidth]{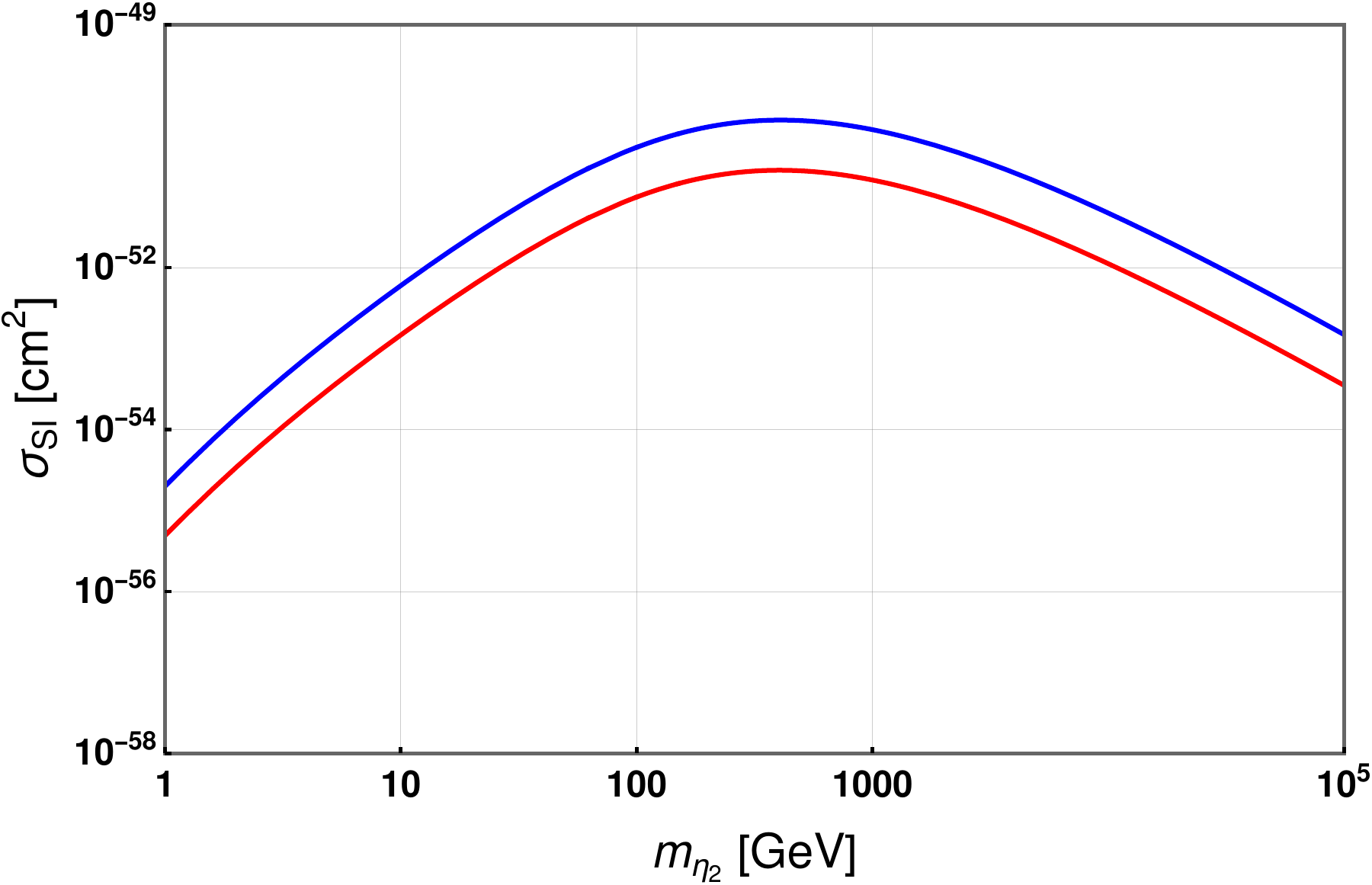}
    \includegraphics[width=0.48\linewidth]{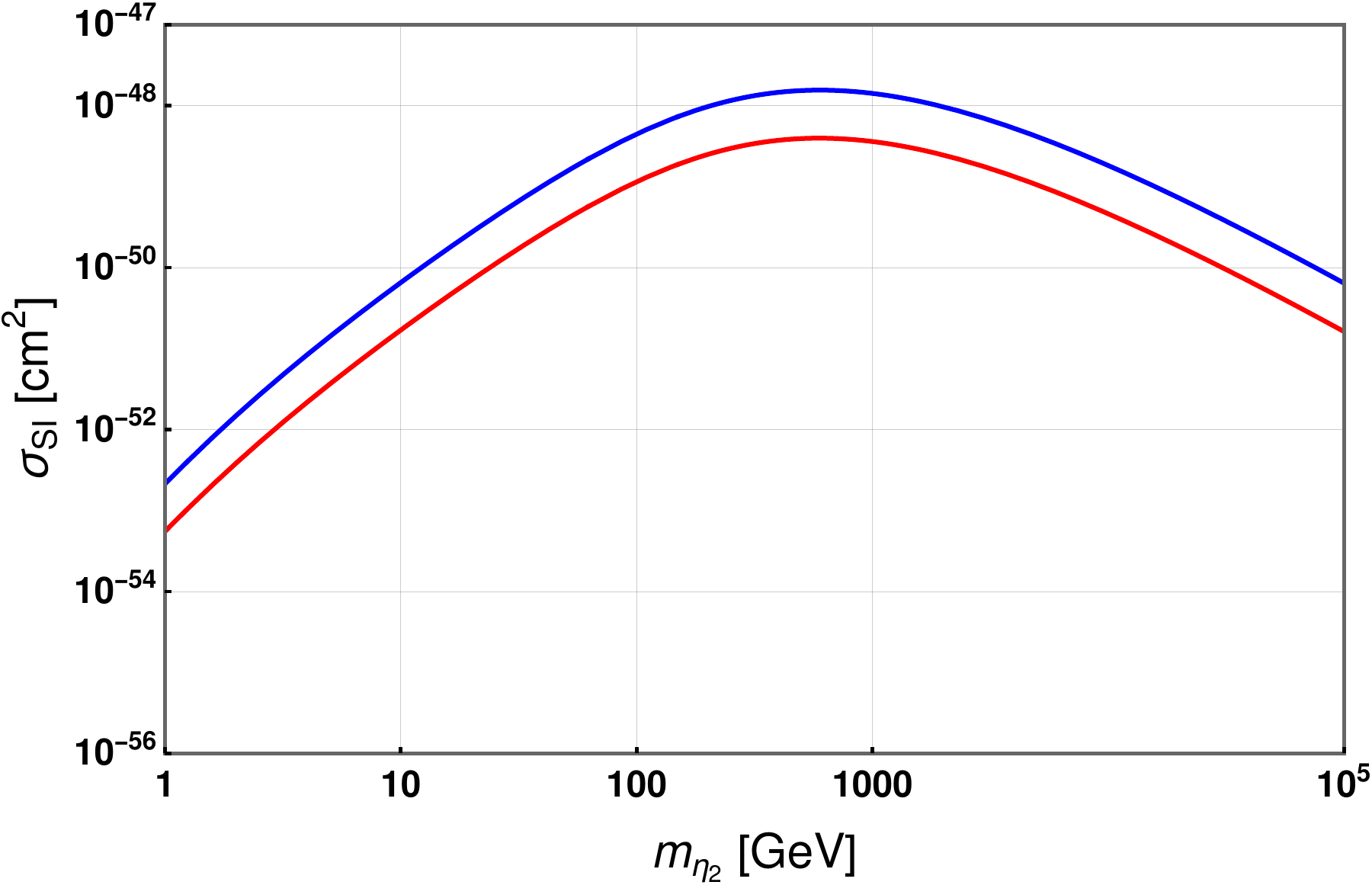}
   \caption{One-loop $\sigma_{\rm SI}$ as a function of the DM mass for $(m_{h_2},v_2)=(200,300)$ GeV (left panel) and $(m_{h_2},v_2)=(300,200)$ GeV (right panel).  For both panels, $(m_{h_2},m_{\eta_3}, v_3)=(300,800,400)$ GeV, and  the blue and red curves correspond to $(s_{\alpha_1},s_{\alpha_2},s_{\alpha_3})=(0.2,0.1,0.1)$ and $(0.01,0.1,0.1)$, respectively.}
    \label{fig:dd_scale}
\end{figure}
 The left panel of Fig.\,\ref{fig:dd} shows the loop-induced SI cross section as a function of the DM mass around the electroweak scale for two sets of scalar mixing angles, with the corresponding parameters specified in the legend. The cross section increases with the scalar mixing angles, as evident from the plot. In contrast to the highly suppressed tree-level contribution, the loop-induced SI cross section is significantly enhanced, although it remains well below the sensitivity of current direct detection experiments. The right panel shows the variation of the loop-induced $\sigma_{\rm SI}$ with $m_{h_2}$, with the remaining parameters fixed as specified in the legend. As $m_{h_2}$ increases, the portal couplings grow, leading to an enhancement of $\sigma_{\rm SI}$. For the two sets of scalar mixing angles, the cross section exhibits sharp dips due to accidental cancellations among the contributing amplitudes in specific regions of the parameter space.

The two panels of Fig.\,\ref{fig:dd_scale} show the variation of the loop-induced $\sigma_{\rm SI}$ with the DM mass over a broad range. As $m_{\eta_2}$ increases, $\sigma_{\rm SI}$ decreases due to the DM contribution in the loop propagator, while for decreasing $m_{\eta_2}$ it exhibits a progressively stronger suppression, as discussed~\cite{Azevedo:2018exj}. This behavior is expected as $m_{\eta_2}\to0$, where the soft global $U(1)$-breaking $\mu_2$ vanishes, and the pNGB DM becomes an exact NGB, causing the loop-induced $\sigma_{\rm SI}$ to approach zero, as discussed earlier. To illustrate the impact of the portal couplings on $\sigma_{\rm SI}$, the right panel corresponds to larger portal couplings, scaling as $m_{h_2}^2/v_2^2$ (see Eqs.\,\eqref{eq:ls2},\,\eqref{eq:lh2}) relative to the left panel. The values of $(m_{h_2},v_2)$ for the two panels are specified in the plot legend.
%%%%%%%%%%%%%%%%%%%%%%%%%%%%%%%%%%%%%%%%%%%%%%%%%%%%%%%%%%
%%%%%%%%%%%%%%%%%%%%%%%%%%%%%%%%%%%%%%%%%%%%%%%%%%%%%%
\section{Electroweak Phase Transition}
\label{ewpt}
Our model exhibits rich EWPT phenomenology due to three scalar fields and their non-trivial interactions. At sufficiently high temperatures, the electroweak symmetry is restored, while, as the Universe cools, the scalar fields can acquire temperature-dependent VEVs, driving the system toward the electroweak-broken phase. The additional scalar degrees of freedom can significantly modify the finite-temperature effective potential, leading to multiple local minima and potential barriers between them. Consequently, the model can exhibit a SFOEWPT as well as multi-step phase transitions involving intermediate vacuum configurations. The resulting thermal history is therefore considerably richer than in the Standard Model and depends sensitively on the scalar couplings and mass parameters of the extended potential.

\subsection{High temperature effective potential}
To investigate the thermal evolution of the scalar fields and the nature of the EWPT, we consider the one-loop finite-temperature effective potential. The temperature-dependent contributions from the scalar, gauge-boson, and fermion sectors\footnote{We neglect the contribution of the VLM $\psi$ to the one-loop effective potential, as its large field-dependent mass makes its thermal contribution exponentially suppressed over the temperature range of interest. Its effect has therefore negligible impact on the phase transition parameters.} are incorporated as~\cite{Dolan:1973qd,Weinberg:1974hy,Quiros:1999jp}
\begin{equation} \label{eq:finiteT}
\begin{split}
V^T_{\rm 1-loop} (h, s_2,s_3, T)
&= \frac{T^4}{2\pi^2} \left[ \sum_{B} n_B J_B \left(\frac{m_B^2(h,s_2,s_3)} {T^2}\right) - \sum_{F} n_F J_F \left(\frac{m_F^2(h)} {T^2}\right)  \right],
\end{split}
\end{equation}
The multiplicity factor is one for each scalar degree of freedom, six for the $W$ boson, three for the $Z$ boson, and twelve for the top quark\footnote{We neglect the contributions of the SM fermions other than the top quark to the one-loop finite-temperature correction, as their comparatively small Yukawa couplings result in much smaller field-dependent masses and hence numerically subdominant thermal contributions.}. The field-dependent masses entering these
thermal functions are given in App.\,\ref{tab:input_params}. The corresponding bosonic and fermionic thermal functions are given by
\begin{eqnarray}
J_{B/F}\left(\frac{m^2}{T^2}\right)
&=&\int_0^\infty dx\,x^2\log\left(1\mp e^{-\sqrt{x^2+m^2/T^2}}\right).
\end{eqnarray}
For $m^2 \gg T^2$, the thermal functions are exponentially suppressed by Boltzmann factors. Consequently, the contributions from particles much heavier than the relevant temperature are negligible and can be safely omitted from the finite-temperature effective potential. On the other hand, in the high-temperature regime, $T^2 \gg m^2$, the thermal functions admit the well-known high-temperature expansions \cite{Dolan:1973qd,Arnold:1992rz}
\begin{eqnarray}
J_B\left(\frac{m^2}{T^2}\right)
&=&-\frac{\pi^4}{45}+\frac{\pi^2}{12}\frac{m^2}{T^2}
-\frac{\pi}{6}\left(\frac{m^2}{T^2}\right)^{3/2}
-\frac{1}{32}\left(\frac{m^2}{T^2}\right)^2
\ln\left(\frac{m^2}{a_bT^2}\right)+\cdots,\\
J_F\left(\frac{m^2}{T^2}\right)
&=&\frac{7\pi^4}{360}-\frac{\pi^2}{24}\frac{m^2}{T^2}
-\frac{1}{32}\left(\frac{m^2}{T^2}\right)^2
\ln\left(\frac{m^2}{a_fT^2}\right)+\cdots.
\end{eqnarray}
Here, $a_f=\pi^2\exp(3/2-2\gamma_E)$ and
$a_b=16\pi^2\exp(3/2-2\gamma_E)$, with $\gamma_E$ denoting the
Euler--Mascheroni constant. In the present analysis, we neglect the zero-temperature Coleman--Weinberg correction and retain only the leading $\mathcal{O}(T^2)$ terms in the high-temperature expansion, which give the dominant thermal mass corrections. The resulting finite-temperature effective potential is therefore given by
\begin{eqnarray}\label{eq:effpot}
    V_{\rm eff}(h, s_2,s_3,T) &=& V_0(h,s_2,s_3)  + V^{\rm 1-loop}_{\rm high-T}(h,s_2,s_3,T).
\end{eqnarray}
The finite-temperature effective potential determines the thermal evolution of the vacuum structure of the theory. At a given temperature, the possible phases of the system are determined by the extrema of the finite-temperature effective potential. Thus, the extrema are obtained by solving
\begin{equation}
\begin{aligned}
\frac{\partial V_{\rm eff}(\phi_i,T)}{\partial \phi_i} &= 0,
&\qquad
V_{\rm eff}\left(\boldsymbol{\phi}_{\rm high}(T_c),T_c\right)
&=
V_{\rm eff}\left(\boldsymbol{\phi}_{\rm low}(T_c),T_c\right).
\end{aligned}
\label{eq:PText}
\end{equation}
where \(\phi_i\) collectively denotes the scalar fields. The solutions corresponding to local minima represent the possible phases of the system. In the present analysis, we distinguish these phases according to the magnitude of their VEVs, referring to them as the high-VEV and low-VEV phases, with the corresponding field configurations denoted by \(\boldsymbol{\phi}_{\rm high}(T)\) and \(\boldsymbol{\phi}_{\rm low}(T)\), respectively.

The first condition in Eq.\,\eqref{eq:PText} determines the extrema of the effective potential, while the second specifies the critical temperature \(T_c\), at which the two phases are degenerate in free energy. Away from the critical temperature, the phase with the lower value of the effective potential is thermodynamically favored, while the other phase is metastable as long as it remains a local minimum. For a first-order phase transition, the two phases are separated by a potential barrier, and the transition proceeds through the nucleation and subsequent expansion of bubbles of the thermodynamically favored phase.

 The order parameters characterizing the transition can be defined in terms of the changes in the VEVs between the high-VEV and low-VEV phases,
\begin{equation}
\xi_h=\frac{\Delta v_h(T_c)}{T_c},
\qquad
\xi_2=\frac{\Delta v_2(T_c)}{T_c},
\qquad
\xi_3=\frac{\Delta v_3(T_c)}{T_c},
\end{equation}
where
\begin{equation}
\Delta v_i(T)
=
\left|
v_i^{\rm high}(T)-v_i^{\rm low}(T)
\right|,
\end{equation}
with $v_h$, $v_2$, and $v_3$ denoting the VEVs of the Higgs field and the two additional scalar fields, respectively.

To study the EWPT, we consider 11 BPs consistent with the observed DM relic abundance. For each BP, we use the publicly available \texttt{CosmoTransitions} \cite{Wainwright:2011kj} package to trace the thermal evolution of the three-scalar potential and determine $T_c$ and the nucleation temperature $T_N$, discussed in the following subsection. At $T_c$ and $T_N$, we extract the VEVs of the three scalar fields in the high- and low-$T$ phases and construct the corresponding order parameters, which quantify the strength of the phase transition.
For BP1--BP7, the Higgs direction exhibits a crossover, while a SFOPT occurs along the $s_3$ direction with $\xi_3>1$. The tree-level barrier along the $s_3$ direction is generated by the cubic $\mu_3$ term, making the singlet sector responsible for the strong first-order behaviour rather than the electroweak direction. For BP8--BP9, SFOPT occurs along both the Higgs and $s_3$ directions, with $\xi_h>1$ and $\xi_3>1$, respectively, corresponding to an SFOEWPT with a simultaneous contribution from the $s_3$ direction. In this case, the $\mu_3$ term together with the relevant quartic couplings, particularly $\lambda_{HS_3}$ and $\lambda_{S_3}$, shapes the potential in the Higgs--$s_3$ field space.

For BP10--BP11, SFOPT occurs along both the $s_2$ and $s_3$ directions, with $\xi_2>1$ and $\xi_3>1$, respectively, while the Higgs direction remains a crossover. Here, the $\mu_3$ term, together with the singlet quartic couplings such as $\lambda_{S_2}$, $\lambda_{S_3}$, and $\lambda_{S_2S_3}$, determines the structure of the potential in the singlet-field directions. Thus, BP1--BP7, BP8--BP9, and BP10--BP11 exhibit distinct SFOPT patterns depending on the interplay of the cubic and quartic terms in the scalar potential.
%%%%%%%%%%%%%%%%%%%%%%%%%%%%%%%%%%%%%%%%%%%%%%%%%%%%%%%%%%%%%%%%%%%%%%%%%%%%%%%%%%%%%%%%%%%%%%%%%%%%%%%%%%%%%%%%%%%%%%%%%%%%%%%%%%%%%%%%%%%%%%%%
\subsection{Stochastic gravitational waves}

\begin{table}[ht]
\centering
\small
\setlength{\tabcolsep}{3.5pt}
\renewcommand{\arraystretch}{1.0}
\begin{tabular}{cccccccc}
\toprule
\textbf{PT param.} & \textbf{BP1} & \textbf{BP2} & \textbf{BP3} & \textbf{BP4} & \textbf{BP5} & \textbf{BP6}  \\
\midrule
$(h,s_2,s_3)|_{T_c}^{\rm high}$ & $(0, 606, 0)$ & $(0,251,0)$  & $(0,349,0)$ & $(0,372,0)$  & $(0,450,0)$  & $(0,340,0)$  \\
$(h,s_2,s_3)|_{T_c}^{\rm low}$  & $(0, 606,581)$& $(0,286,347)$ & $(0,352,359)$ & $(0,384,361)$& $(0,454,368)$ &  $(0,359,347)$ \\
$T_c$  & 381 & 287 & 270& 254 & 427 & 289  \\
$(\xi_h,\xi_{s_2},\xi_{s_3})$ & $(0,0,1.52)$ & $(0,0.12,1.21)$ & $(0,0.01,1.33)$& $(0,0.05,1.42)$& $(0,0.01,0.86)$ & $(0,0.07,1.2)$  \\
$(h,s_2,s_3)|_{T_n}^{\rm high}$ &$(0,616,0)$ & $(0,261,0)$ & $(0,358,0)$ & $(0,379,0)$ &  $(0,456,0)$& $(0,347,0)$  \\
$(h,s_2,s_3)|_{T_n}^{\rm low}$  & $(0,617,647)$ & $(0,300,377)$ & $(0,361,392)$ & $(0,393,397)$ & $(0,461,394)$ & $(0,538,576)$  \\
$T_n$ & 274 & 254& 228 & 202& 406 & 257  \\
$\alpha_n$ & 0.03 & 0.01 & 0.02 & 0.02 & 0.004 &0.01  \\
$\beta/H_n$ & 347& 1526 & 1023& 668 & 4174 & 1607 \\
\bottomrule
\end{tabular}
\caption{Phase transition parameters for BP1-BP6. Here, VEVs and temperature units of GeV.}
\label{tab:FOPT1}
\end{table}

\begin{table}[ht]
\centering
\small
\setlength{\tabcolsep}{3.5pt}
\renewcommand{\arraystretch}{1.0}
\begin{tabular}{cccccccc}
\toprule
\textbf{PT param.} & \textbf{BP7} & \textbf{BP8} & \textbf{BP9} & \textbf{BP10} & \textbf{BP11}  \\
\midrule
$(h,s_2,s_3)|_{T_c}^{\rm high}$ & $(0,499,0)$ & $(0,334,0)$  & $(0,400,0)$ & $(0,0,0)$  & $(0,0,0)$     \\
$(h,s_2,s_3)|_{T_c}^{\rm low}$  & $(0,525,534)$ & $(149,342,116)$ & $(133,475,185)$ & $(0,249,326)$& $(0,219,285)$ \\
$T_c$  & 460 & 87 & 100& 201 & 175   \\
$(\xi_h,\xi_{s_2},\xi_{s_3})$ & $(0,0.07,1.2)$ & $(1.71,0.09,1.33)$ & $(1.33,0.75,1.85)$& $(0,1.24,1.62)$& $(0,1.25,1.63)$  \\
$(h,s_2,s_3)|_{T_n}^{\rm high}$ &$(0,508,0)$ & $(0,335,0)$ & $(0,401,0)$ & $(0,0,0)$ &  $(0,0,0)$\\
$(h,s_2,s_3)|_{T_n}^{\rm low}$  & $(0,370,376)$ & $(169,344,123)$ & $(188,486,201)$ & $(0,287,360)$ & $(0,250,312)$   \\
$T_n$ & 416 & 80& 78 & 146& 131  \\
$\alpha_n$ & 0.01 & 0.35 & 0.51 & 0.13 & 0.15   \\
$\beta/H_n$ & 1897& 3208 & 538& 173 & 577 \\
\bottomrule
\end{tabular}
\caption{Phase transition parameters for BP7-BP11. Here, VEVs and temperature units are in GeV.}
\label{tab:FOPT2}
\end{table}

\begin{figure}
    \centering
    \includegraphics[width=0.48\linewidth]{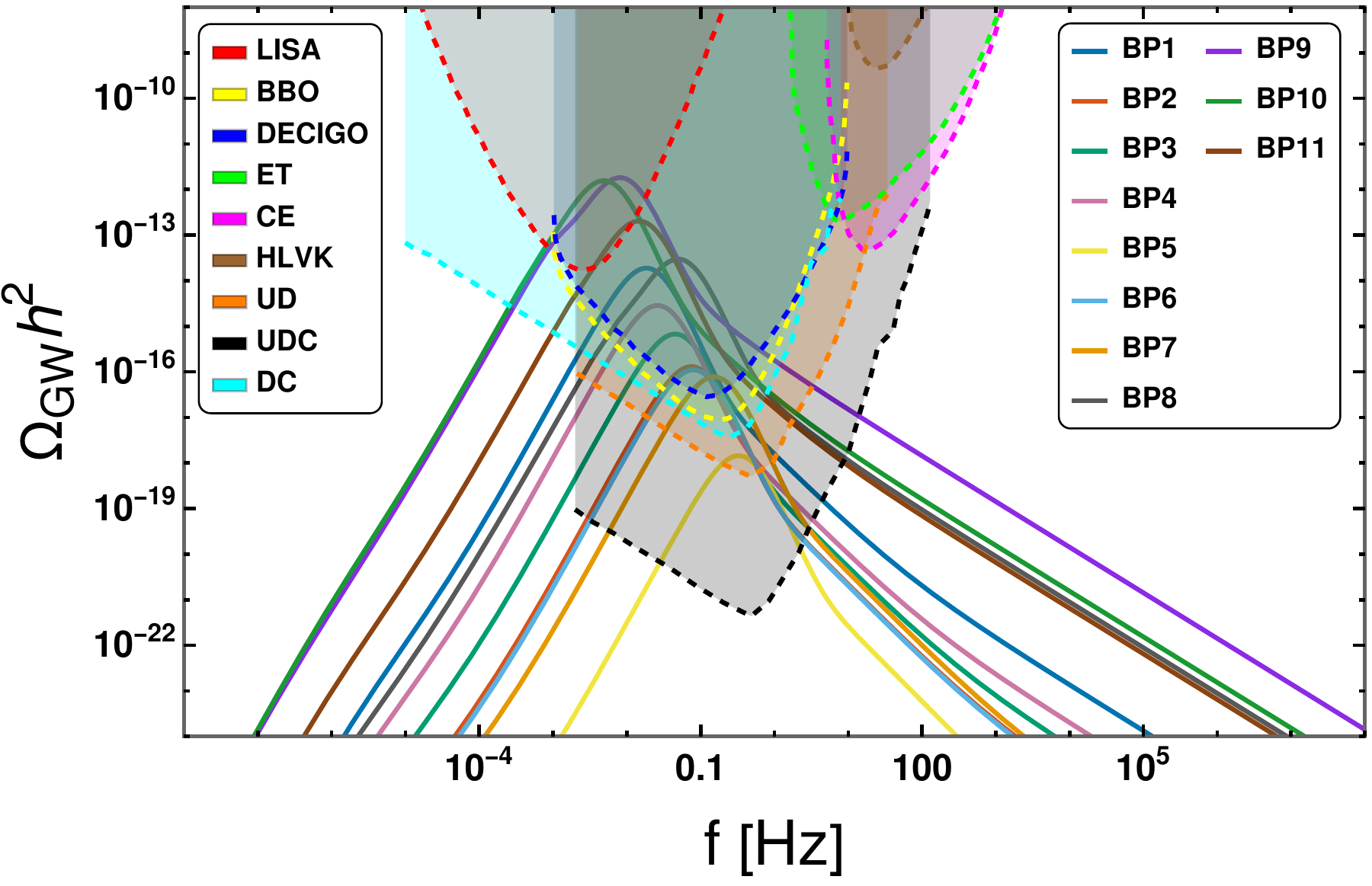}
    \includegraphics[width=0.48\linewidth]{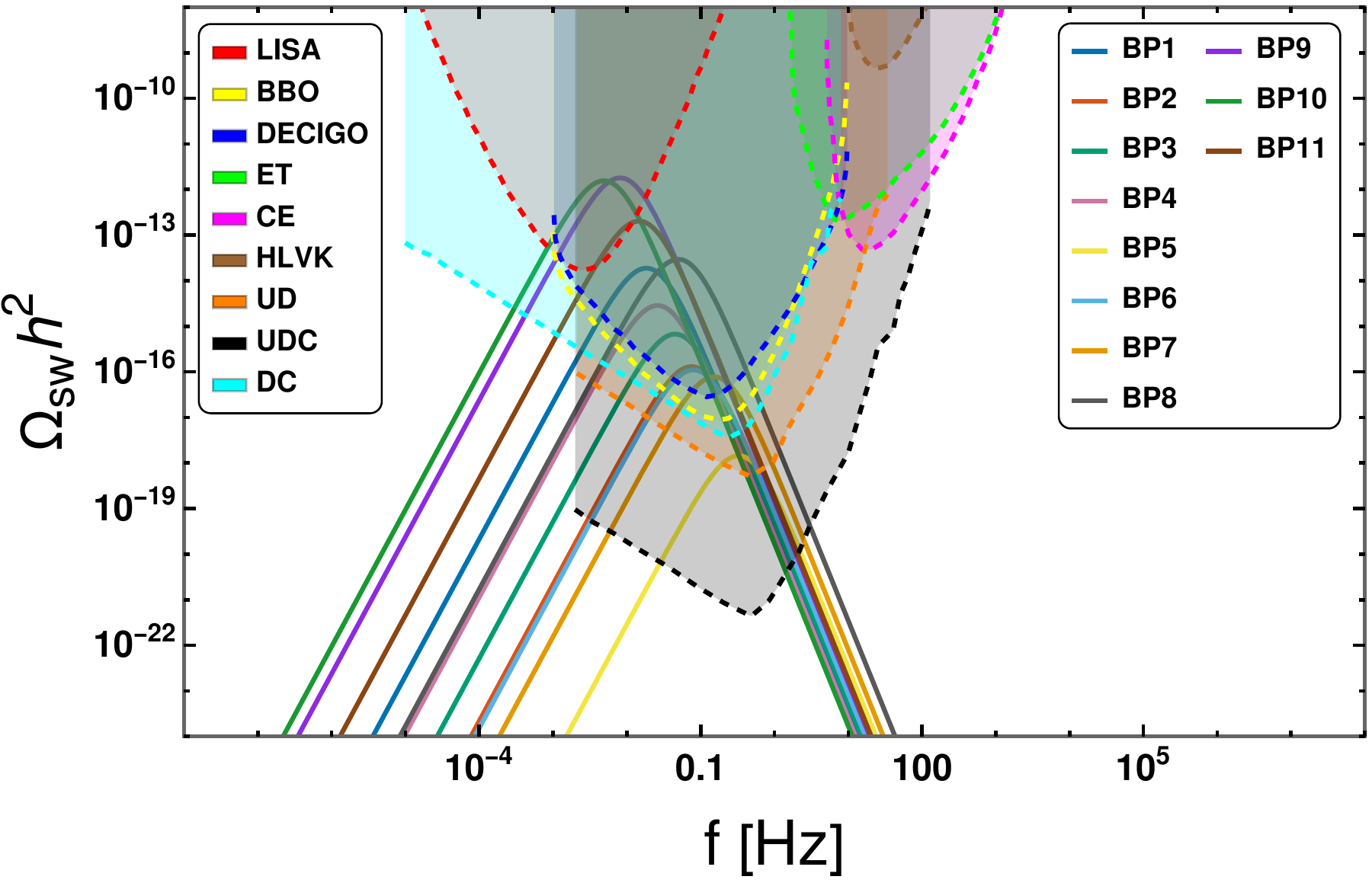}
    \caption{GW amplitude as a function of frequency. The left panel shows the total GW spectrum, while the right panel shows the sound wave contribution, which dominates among the three GW sources. The shaded regions represent the projected sensitivities of future GW detectors: LISA~\cite{LISA:2017pwj}, BBO~\cite{Yagi:2011wg}, the Einstein Telescope (ET)~\cite{Punturo:2010zz}, Cosmic Explorer (CE)~\cite{Reitze:2019iox}, the HLVK network (LIGO Hanford--Livingston, Virgo, and KAGRA)~\cite{LIGOScientific:2014pky,VIRGO:2014yos,KAGRA:2018plz}, DECIGO, Ultimate DECIGO (UD), DECIGO Correlation (DC), and Ultimate DECIGO Correlation (UDC)~\cite{Nakayama:2009ce}.}
\label{fig:gwplot}
\end{figure}

A first-order EWPT provides a possible source of a stochastic GW background. In contrast to a crossover transition, a first-order transition proceeds through the nucleation and subsequent expansion of bubbles of the broken electroweak phase within the metastable symmetric phase. Collisions between expanding bubbles, bulk fluid motion in the plasma, and associated magnetohydrodynamic turbulence can generate GWs. The resulting spectrum is determined primarily by the characteristic temperature and time scale of the transition, together with the amount of energy released during the transition. Consequently, the properties of the EWPT obtained from the finite-temperature effective potential can be directly connected to potentially observable GW signatures.

The nucleation temperature $T_n$ is defined as the temperature at which the probability for bubble nucleation becomes appreciable within a Hubble volume. The thermal tunnelling rate per unit volume is given by~\cite{Grojean:2006bp,Linde:1981zj}
\begin{equation}
    \Gamma(T)\simeq T^4
    \left(\frac{S_3(T)}{2\pi T}\right)^{3/2}
    e^{-S_3(T)/T},
\end{equation}
where $S_3(T)$ is the three-dimensional Euclidean bounce action~\cite{Linde:1981zj}. The nucleation temperature $T_n$ is determined by imposing the condition that approximately one bubble is nucleated per Hubble volume, leading to~\cite{Linde:1981zj,Mazumdar:2018dfl}
\begin{equation}
\frac{\Gamma(T_n)}{H_n^4}\sim 1,
\qquad\Longrightarrow\qquad
\frac{S_3(T_n)}{T_n}\simeq 140.
\end{equation}
Here, $H_n\equiv H(T_n)$ is the Hubble parameter at $T_n$.

Two important quantities entering the GW prediction are the strength parameter, $\alpha$, and the inverse duration parameter, $\beta/H_n$. The strength of the phase transition is characterized by~\cite{Kamionkowski:1993fg,Kehayias:2009tn}
\begin{equation}
    \alpha \equiv \frac{\epsilon(T_n)}{\rho_{\rm rad}(T_n)},\qquad
    \text{with}\qquad
    \rho_{\rm rad}(T_n)=\frac{\pi^2}{30}g_*(T_n)T_n^4,
\end{equation}
where $\rho_{\rm rad}(T_n)$ is the radiation energy density and $g_\ast(T_n)$ is the effective number of relativistic degrees of freedom at $T_n$. The latent heat is given by~\cite{Kamionkowski:1993fg,Kehayias:2009tn}
\begin{equation}
    \epsilon(T)  = \left[ \Delta V_{\rm eff} -T\frac{\partial \Delta V_{\rm eff}}{\partial T}\right]_{T=T_n}.
\end{equation}
Here, $\Delta V_{\rm eff}=V_{\rm eff}^{\rm false}-V_{\rm eff}^{\rm true}$ denotes the difference between the effective potentials of the false and true vacua. A larger value of $\alpha$ indicates a stronger phase transition, corresponding to a greater amount of energy released into the plasma and, consequently, a potentially larger GW signal.

The inverse duration parameter $\beta/H_n$ characterizes the time scale of the phase transition relative to the Hubble expansion rate and plays a crucial role in determining the characteristic scale of the GW spectrum. It is defined as~\cite{Nicolis:2003tg}
\begin{equation}
    \frac{\beta}{H_n}\equiv\left. T\frac{d(S_3/T)}{dT} \right|_{T=T_n}.
\end{equation}
A smaller $\beta/H_n$ corresponds to a slower transition with larger characteristic bubbles and a lower GW peak frequency, whereas a larger $\beta/H_n$ indicates a faster transition with smaller bubbles and a higher peak frequency.

The GW spectrum generated by the EWPT can schematically be written as the sum of contributions from bubble-wall dynamics, sound waves in the plasma, and magnetohydrodynamic turbulence~\cite{Ellis:2020awk,Caprini:2015zlo},
\begin{equation}
    \Omega_{\rm GW}h^2  \approx \Omega_{\rm col}h^2 + \Omega_{\rm sw}h^2 + \Omega_{\rm turb}h^2.
\end{equation}
Here, $h \equiv H_0/(100{\rm \,km\, s^{-1} Mpc^{-1}})$, with $H_0$ being the present-day Hubble constant~\cite{DES:2017txv}. The explicit expressions for the GW spectra associated with the three sources can be found in Ref.~\cite{Das:2026zuo}. For transitions in which the released vacuum energy is efficiently transferred to the surrounding plasma, the sound wave contribution typically dominates. These sound waves are generated in the plasma following bubble percolation~\cite{Hindmarsh:2013xza,Hindmarsh:2016lnk,Hindmarsh:2017gnf}, and give rise to the corresponding GW energy-density spectrum~\cite{Das:2026zuo}.
\begin{equation}\label{eq:GWswpart}
\Omega_{\rm sw} h^2 = 2.65 \times 10^{-6}\; \Upsilon(\tau_{\rm sw}) \left(  \frac{\beta}{H_n} \right)^{-1} v_w \left( \frac{\kappa_{\rm sw} \alpha_n}{1 + \alpha_n} \right)^2 \left( \frac{100}{g^{\ast}} \right)^{1/3} \left( \frac{f}{f_{\rm sw}} \right)^3 \left[ \frac{7}{4 + 3 \left( f/f_{\rm sw} \right)^2} \right]^{7/2},
\end{equation}
where $\kappa_{\rm sw}$ denotes the efficiency factor for the conversion of latent heat into bulk fluid motion~\cite{Kamionkowski:1993fg}, while the peak frequency
\begin{equation}\label{eq:PF2}
f_{\rm sw} = 1.9 \times 10^{-5} \left( \frac{1}{v_w} \right) \left( \frac{\beta}{H_n} \right) \left( \frac{T_n}{100 \, {\rm GeV}} \right) \left( \frac{g^{\ast}}{100} \right)^{1/6} \, {\rm Hz}.
\end{equation}
The factor $\Upsilon(\tau_{\rm sw})$ accounts for the finite sound wave lifetime~\cite{Hindmarsh:2017gnf}, while we take the relativistic wall velocity $v_w=1$~\cite{Kamionkowski:1993fg,Espinosa:2010hh}.

For the 11 BPs, we use the publicly available code \texttt{CosmoTransitions} to numerically evaluate the relevant phase transition parameters governing the GW signal, as summarized in Tab.\,\ref{tab:FOPT1}. Fig.\,\ref{fig:gwplot} shows the resulting GW spectra for all benchmark points, with the total GW contribution shown in the left panel and the dominant sound wave contribution shown in the right panel. For all BPs, the GW spectra peak in the frequency range from the mHz to the few-Hz regime, placing them within the sensitivity range of space- and ground-based GW observatories. Among these, BP7-BP11 yield the largest GW amplitudes, corresponding to relatively large $\alpha_n$ and smaller $\beta/H_n$, as expected from Eq.\,\eqref{eq:GWswpart}. This primarily arises from the strong FOPT along the two-field direction, with the corresponding order parameters exceeding unity, leading to a larger $\alpha_n$.
 Their predicted spectra fall within the sensitivity ranges of LISA, DECIGO, and BBO. Notably, BPs 7 and 8 exhibit the largest $\alpha_n$ among all BPs, associated with a SFOEWPT along the Higgs direction together with a transition along the $s_3$ direction. The remaining BPs yield comparatively smaller GW amplitudes due to their lower $\alpha_n$ and relatively larger $\beta/H_n$. For these BPs, the SFOEWPT is obtained primarily along the $s_3$ direction, while the resulting GW signals are accessible mainly to the more sensitive future detectors indicated in the plot legend, such as Ultimate DECIGO, Ultimate DECIGO correlation, DECIGO correlation, and other proposed next-generation observatories.

\section{Summary and Conclusions}
\label{summary}

In this work, we studied an extension of the SM with two gauge-singlet complex scalars, $S_2$ and $S_3$, charged under $\mathbb{Z}_2$ and $\mathbb{Z}_3$, respectively. The model also contains a muon-specific VLL, $\psi$, which carries the same $\mathbb{Z}_3$ charge as $S_3$, the same hypercharge as the right-handed muon, and a nonzero muon number. The pseudoscalar associated with the $\mathbb{Z}_2$ sector remains stable due to the CP symmetry of the potential and constitutes the pNGB DM candidate, with its mass generated by the soft breaking of the corresponding global $U(1)$ symmetry. Meanwhile, the pseudoscalar associated with $\mathbb{Z}_3$ becomes unstable through its Yukawa interaction with $\psi$ and the right-handed muon, resulting in an effectively single-component DM scenario.

The observed DM relic abundance is obtained through standard thermal freeze-out. 
%The pNGB DM remains in equilibrium with the thermal bath at early times and subsequently freezes out when its annihilation rate falls below the Hubble expansion rate. Viable regions of parameter space reproduce the observed relic density while satisfying the relevant theoretical and experimental constraints. 
The pNGB nature of the DM leads to a strong suppression of the tree-level SI DM--nucleon scattering amplitude. In the limit of small momentum transfer, relevant for direct-detection experiments, the contributions from the three scalar-exchange diagrams in the mass basis exhibit a cancellation, causing the leading tree-level amplitude to vanish at zero momentum transfer and leaving only momentum-suppressed terms. Consequently, the tree-level SI cross section is highly suppressed. At one loop, the resulting effective DM--nucleon interaction contains a nonzero contribution in the vanishing-momentum-transfer limit. Thus, unlike the tree-level amplitude, the leading loop-induced SI cross section is essentially independent of the small momentum transfer and is instead controlled by the DM mass and the masses, mixing angles, and couplings of the particles entering the loop. We find that this irreducible one-loop contribution remains compatible with current direct-detection observables, providing a viable parameter space for future experiments.

We have discussed perturbative unitarity and vacuum stability constraints on the model parameters. We have further imposed LHC monojet constraints, which restrict the missing-energy signatures associated with the DM sector while retaining phenomenologically viable regions. The VLM mixes with the muon through the VEV of $S_3$, leading to the relevant mass-basis interactions and modifying its collider phenomenology. 
The VLM sector also provides a distinctive collider probe of the framework. At a $\sqrt{s}=3~\mathrm{TeV}$ muon collider, VLM pair production is particularly promising, since the scalar-mediated $t$-channel contribution is not subject to the strong muon–VLM mixing suppression that limits associated production and can therefore substantially enhance the $vl^+vl^-$ production rate. For the benchmark considered in our detailed collider analysis, the subsequent decays $vl^\pm\to\mu^\pm h_3$, followed by the invisible decay $h_3\to\eta_2\eta_2$, lead to a characteristic $\mu^+\mu^-+\slashed{E}_T$ final state. Using an XGBoost-based multivariate analysis, we obtain a strong separation between the signal and the SM backgrounds. Optimizing the BDT response yields a purely statistical significance of $S/\sqrt{S+B}\simeq 12$ at $500~\mathrm{fb}^{-1}$ integrated luminosity. These results demonstrate that a multi-TeV muon collider can provide a sensitive and complementary probe of the muon-philic VLL sector.
%The VLM therefore provides an additional and distinctive probe at a future muon collider. At $\sqrt{s}=3$~TeV, scalar-mediated $t$-channel processes can significantly enhance the production of $vl^+ vl^-$ compared with the conventional $s$-channel contribution. Our multivariate analysis demonstrates the potential of a high-energy muon collider to probe this muon-philic VLL sector. Our multivariate analysis demonstrates the potential of a high-energy muon collider to probe the muon-philic sector.

Our model exhibits a rich scalar structure, which leads to a nontrivial thermal history of the Universe. We find viable BPs exhibiting a SFOEWPT with successful bubble nucleation, with the nature of the transition depending on the trajectory in the multi-field scalar field space. In some BPs, the EWPT proceeds along all three scalar directions, while in others the transition occurs predominantly along two directions or exclusively along the $\mathbb{Z}_3$-scalar direction. These different transition paths reflect the interplay among the scalar fields and can lead to distinct thermal histories and phase-transition dynamics. The first-order EWPT can generate a stochastic GW background, with long-lived sound waves in the plasma generally providing the dominant contribution, while bubble-wall collisions and subsequent magnetohydrodynamic turbulence contribute additional, typically subdominant components. For the viable BPs considered here, the predicted GW spectra can potentially be probed by future space-based interferometers such as LISA, DECIGO, and BBO, providing an additional window into the extended scalar sector and its thermal history.

In summary, this framework connects pNGB dark matter, a muon-philic VLL sector, and a strong first-order electroweak phase transition within a phenomenologically viable setup. The pNGB nature of the DM suppresses the tree-level spin-independent scattering amplitude, while the loop-induced contribution remains consistent with current direct-detection limits and the observed relic abundance. The dark sector is further testable through LHC monojet searches as well, whereas the muon-philic VLL provides a distinctive target for a future multi-TeV muon collider. At the same time, a SFOEWPT can generate an observable stochastic gravitational-wave signal. Taken together, direct detection, collider searches, and gravitational-wave observations provide complementary probes of the model and its underlying dark-sector and electroweak dynamics.

%In summary, this framework establishes a unified connection between pNGB DM, collider phenomenology of a muon-philic VLL, and the EWPT, while remaining consistent with the relevant theoretical and experimental constraints. The interplay among these sectors gives rise to complementary signatures across present and future experiments. In particular, the tree-level DM–nucleon scattering is highly suppressed, while the one-loop contribution provides the leading SI interaction and remains consistent with the latest DD bounds while maintaining the observed value of the relic abundance; the DM sector can be further probed through LHC monojet searches. The muon-philic nature of the VLL further provides promising prospects for enhanced production and distinctive signatures at a future multi-TeV muon collider. In addition, the SFOPT can generate a potentially observable stochastic GW background, offering an independent probe of the thermal history of the model. The combined sensitivity of DD experiments, relic density constraints, LHC monojet searches, future muon-collider experiments, and GW observatories therefore provides a powerful and complementary set of probes of the framework. With the continued improvement in experimental sensitivity, particularly at future high-energy colliders and GW observatories, a significant portion of the viable parameter space can be explored, providing the prospect of probing the underlying dynamics responsible for the DM, VLL, and EWPT phenomenology.

\section*{Acknowledgment}
The authors thank Prof. Kristjan Kannike for helpful communication and Subhajit Kala for useful discussions. JD acknowledges ANRF (formerly Science and Engineering Research Board (SERB)), Government of India, for the National Postdoctoral Fellowship (NPDF) grant No. PDF/2023/001540. SN acknowledges funding from Anusandhan National Research Foundation, Govt. of India, with grant number SUR/2022/001404 of SURE scheme.

\appendix
\section{One-loop DD cross section}\label{app:dd}
For the left and middle topologies of Fig.\,\ref{fig:one-loopDD}, the matrix elements
\begin{eqnarray}
 &&   i\mathcal{M}^{(a)}(\mu_2) = i^5 \frac{c_{h_j\eta_2\eta_2} c_{h_k\eta_2\eta_2}c_{h_ih_jh_k} c_{h_iqq}}{m_{h_i}^2}  \int \frac{d^4k}{i\pi^2} \frac{1}{k^2-m_{\eta_2}^2} \frac{1}{(p_1-k)^2-m_{h_j}^2} \frac{1}{(p_1-k)^2-m_{h_k}^2},\nonumber \\
  && i\mathcal{M}^{(b)}(\mu_2) = i\mathcal{M}^{(c)}(\mu_2)= i^4 \frac{c_{h_j\eta_2\eta_2} c_{h_i h_j \eta_2\eta_2} c_{h_iqq}}{m_{h_i}^2}  \int \frac{d^4k}{i\pi^2} \frac{1}{k^2-m_{\eta_2}^2} \frac{1}{(p_1-k)^2-m_{h_j}^2},
\end{eqnarray}
where $i,j,k=1,2,3$ run implicitly. Symbolically, the total amplitude can be written as
\begin{eqnarray}
    \mathcal{M}(\mu_2)= \sum_{\rm 27\,diag} \mathcal{M}^{(a)}(\mu_2) + 2 \sum_{\rm 9\,diag} \mathcal{M}^{(b)}(\mu_2).
\end{eqnarray}
Therefore, the loop function in Eq.\,\eqref{eq:one_loop_X} can be written as
\begin{eqnarray}
    F(\mu_2) = \mathcal{M}(\mu_2)- \mathcal{M}(0).
\end{eqnarray}
As mentioned earlier, the dependence on $\mu_2$ enters only through $m_{\eta_2}$. In particular, the limit $\mu_2 \to 0$ corresponds to $m_{\eta_2} \to 0$. Using \texttt{Package-X}~\cite{Patel:2015tea}, we evaluate these integrals to obtain the corresponding analytic expression.

Three- and four-point scalar vertices, with the corresponding factors relevant for the calculation of the diagrams in Fig.\,\ref{fig:one-loopDD}.
\begin{eqnarray}
    c_{h_1h_1h_1} &&=  -i\Big[
2\mu_3s_{\alpha_2}^3
+3\Big(
2\lambda_{S_3}s_{\alpha_2}^3v_3
+\lambda_{S_2S_3}s_{\alpha_1}s_{\alpha_2}^2v_2c_{\alpha_2}
+\lambda_{S_2S_3}s_{\alpha_1}^2s_{\alpha_2}v_3c_{\alpha_2}^2\nonumber\\
&&+2\lambda_{S_2}s_{\alpha_1}^3v_2c_{\alpha_2}^3
+2\lambda_Hvc_{\alpha_1}^3c_{\alpha_2}^3
+c_{\alpha_1}^2c_{\alpha_2}^2
\left(\lambda_{HS_3}s_{\alpha_2}v_3
+\lambda_{HS_2}s_{\alpha_1}v_2c_{\alpha_2}\right)\nonumber\\
&&+vc_{\alpha_1}c_{\alpha_2}
\left(\lambda_{HS_3}s_{\alpha_2}^2
+\lambda_{HS_2}s_{\alpha_1}^2c_{\alpha_2}^2\right)
\Big)
\Big], \nonumber\\
    c_{h_2h_2h_2} &&= -i\bigg[
2s_{\alpha_3}^3(\mu_3+3\lambda_{S_3}v_2)c_{\alpha_2}^3
-3s_{\alpha_3}^2c_{\alpha_2}^2
\Big(
\lambda_{S_2S_3}s_{\alpha_1}s_{\alpha_2}s_{\alpha_3}v_3
+\lambda_{HS_3}s_{\alpha_2}s_{\alpha_3}v c_{\alpha_1}\nonumber \\
&&+\lambda_{HS_3}s_{\alpha_1}v c_{\alpha_3}
-\lambda_{S_2S_3}v_3c_{\alpha_1}c_{\alpha_3}
\Big)
+3s_{\alpha_3}v_2c_{\alpha_2}
\Big[
2(-\lambda_{S_2S_3}+\lambda_{HS_3})
s_{\alpha_1}s_{\alpha_2}s_{\alpha_3}c_{\alpha_1}c_{\alpha_3}\nonumber \\
&&+c_{\alpha_1}^2
\Big(
\lambda_{HS_3}s_{\alpha_2}^2s_{\alpha_3}^2
+\lambda_{S_2S_3}c_{\alpha_3}^2
\Big)
+s_{\alpha_1}^2
\Big(
\lambda_{S_2S_3}s_{\alpha_2}^2s_{\alpha_3}^2
+\lambda_{HS_3}c_{\alpha_3}^2
\Big)
\Big]\nonumber\\
&&+3\Big[
c_{\alpha_1}^3
\Big(
-2\lambda_Hs_{\alpha_2}^3s_{\alpha_3}^3v
+\lambda_{HS_2}s_{\alpha_2}^2s_{\alpha_3}^2v_3c_{\alpha_3}
-\lambda_{HS_2}s_{\alpha_2}s_{\alpha_3}vc_{\alpha_3}^2
+2\lambda_{S_2}v_3c_{\alpha_3}^3
\Big)\nonumber\\
&&+s_{\alpha_1}^2c_{\alpha_1}
\Big(
-\lambda_{HS_2}s_{\alpha_2}^3s_{\alpha_3}^3v
+2(3\lambda_{S_2}-\lambda_{HS_2})
s_{\alpha_2}^2s_{\alpha_3}^2v_3c_{\alpha_3}
+2(-3\lambda_H+\lambda_{HS_2})
s_{\alpha_2}s_{\alpha_3}vc_{\alpha_3}^2\nonumber\\
&&+\lambda_{HS_2}v_3c_{\alpha_3}^3
\Big)
-s_{\alpha_1}^3
\Big(
2\lambda_{S_2}s_{\alpha_2}^3s_{\alpha_3}^3v_3
+\lambda_{HS_2}s_{\alpha_2}^2s_{\alpha_3}^2vc_{\alpha_3}
+\lambda_{HS_2}s_{\alpha_2}s_{\alpha_3}v_3c_{\alpha_3}^2
+2\lambda_Hvc_{\alpha_3}^3
\Big)\nonumber\\
&&-s_{\alpha_1}c_{\alpha_1}^2
\Big(
\lambda_{HS_2}s_{\alpha_2}^3s_{\alpha_3}^3v_3
+2(3\lambda_H-\lambda_{HS_2})
s_{\alpha_2}^2s_{\alpha_3}^2vc_{\alpha_3}\nonumber\\
&&+2(3\lambda_{S_2}-\lambda_{HS_2})
s_{\alpha_2}s_{\alpha_3}v_3c_{\alpha_3}^2
+\lambda_{HS_2}vc_{\alpha_3}^3
\Big)
\Big]
\bigg],\nonumber\\
    c_{h_3h_3h_3} &&= i\bigg[
-2(\mu_3+3\lambda_{S_3}v_3)c_{\alpha_2}^3c_{\alpha_3}^3
+3c_{\alpha_2}^2c_{\alpha_3}^2
\Big(
-\lambda_{HS_3}s_{\alpha_1}s_{\alpha_3}v
+\lambda_{S_2S_3}s_{\alpha_3}v_2c_{\alpha_1}\nonumber\\
&&+\lambda_{S_2S_3}s_{\alpha_1}s_{\alpha_2}v_2c_{\alpha_3}
+\lambda_{HS_3}s_{\alpha_2}v c_{\alpha_1}c_{\alpha_3}
\Big)
-3v_3c_{\alpha_2}c_{\alpha_3}
\bigg[
2(\lambda_{S_2S_3}-\lambda_{HS_3})
s_{\alpha_1}s_{\alpha_2}s_{\alpha_3}c_{\alpha_1}c_{\alpha_3}\nonumber\\
&&+s_{\alpha_1}^2
\Big(
\lambda_{HS_3}s_{\alpha_3}^2
+\lambda_{S_2S_3}s_{\alpha_2}^2c_{\alpha_3}^2
\Big)
+c_{\alpha_1}^2
\Big(
\lambda_{S_2S_3}s_{\alpha_3}^2
+\lambda_{HS_3}s_{\alpha_2}^2c_{\alpha_3}^2
\Big)
\bigg]\nonumber\\
&&+3\bigg[
s_{\alpha_1}^3
\Big(
-2\lambda_Hs_{\alpha_3}^3v
+\lambda_{HS_2}s_{\alpha_2}s_{\alpha_3}^2v_2c_{\alpha_3}
-\lambda_{HS_2}s_{\alpha_2}^2s_{\alpha_3}v c_{\alpha_3}^2
+2\lambda_{S_2}s_{\alpha_2}^3v_2c_{\alpha_3}^3
\Big)\nonumber\\
&&+s_{\alpha_1}c_{\alpha_1}^2
\Big(
-\lambda_{HS_2}s_{\alpha_3}^3v
+2(3\lambda_{S_2}-\lambda_{HS_2})
s_{\alpha_2}s_{\alpha_3}^2v_2c_{\alpha_3}
+2(-3\lambda_H+\lambda_{HS_2})
s_{\alpha_2}^2s_{\alpha_3}v c_{\alpha_3}^2\nonumber\\
&&+\lambda_{HS_2}s_{\alpha_2}^3v_2c_{\alpha_3}^3
\Big)
+c_{\alpha_1}^3
\Big(
2\lambda_{S_2}s_{\alpha_3}^3v_2
+\lambda_{HS_2}s_{\alpha_2}s_{\alpha_3}^2v c_{\alpha_3}
+\lambda_{HS_2}s_{\alpha_2}^2s_{\alpha_3}v_2c_{\alpha_3}^2
+2\lambda_Hs_{\alpha_2}^3v c_{\alpha_3}^3
\Big)\nonumber\\
&&+s_{\alpha_1}^2c_{\alpha_1}
\Big(
\lambda_{HS_2}s_{\alpha_3}^3v_2
+2(3\lambda_H-\lambda_{HS_2})
s_{\alpha_2}s_{\alpha_3}^2v c_{\alpha_3}
+2(3\lambda_{S_2}-\lambda_{HS_2})
s_{\alpha_2}^2s_{\alpha_3}v_2c_{\alpha_3}^2\nonumber\\
&&+\lambda_{HS_2}s_{\alpha_2}^3v c_{\alpha_3}^3
\Big)
\bigg]
\bigg],\nonumber\\
    c_{h_1h_1h_2} &&= -i\bigg[
s_{\alpha_3}v_3
\left(\lambda_{S_2S_3}s_{\alpha_1}^2
+\lambda_{HS_3}c_{\alpha_1}^2\right)c_{\alpha_2}^3
-s_{\alpha_2}^2
\Big(
\lambda_{S_2S_3}s_{\alpha_1}s_{\alpha_2}s_{\alpha_3}v_3
+\lambda_{HS_3}s_{\alpha_2}s_{\alpha_3}v c_{\alpha_1}\nonumber \\
&&+\lambda_{HS_3}s_{\alpha_1}v c_{\alpha_3}
-\lambda_{S_2S_3}v_3c_{\alpha_1}c_{\alpha_3}
\Big)
+c_{\alpha_2}^2
\bigg[
2\lambda_{S_2S_3}s_{\alpha_1}s_{\alpha_2}s_{\alpha_3}v_3
+(2\lambda_{HS_3}-3\lambda_{HS_2}s_{\alpha_1}^2)
s_{\alpha_2}s_{\alpha_3}v c_{\alpha_1}\nonumber\\
&&+2(3\lambda_{S_2}-\lambda_{HS_2})s_{\alpha_1}^2v_3c_{\alpha_1}c_{\alpha_3}
+c_{\alpha_1}^3
\Big(
-6\lambda_Hs_{\alpha_2}s_{\alpha_3}v
+\lambda_{HS_2}v_3c_{\alpha_3}
\Big)
-s_{\alpha_1}^3
\Big(
6\lambda_{S_2}s_{\alpha_2}s_{\alpha_3}v_3 \nonumber \\  
&&+\lambda_{HS_2}vc_{\alpha_3}
\Big)
+s_{\alpha_1}c_{\alpha_1}^2
\Big(
-3\lambda_{HS_2}s_{\alpha_2}s_{\alpha_3}v_3
-6\lambda_Hvc_{\alpha_3}
+2\lambda_{HS_2}vc_{\alpha_3}
\Big)
\bigg]\nonumber \\
&&+2s_{\alpha_2}c_{\alpha_2}
\bigg[
\mu_3s_{\alpha_2}s_{\alpha_3}
-v_3\Big(
(-3\lambda_{S_3}+\lambda_{S_2S_3}s_{\alpha_1}^2)
s_{\alpha_2}s_{\alpha_3}
+\lambda_{HS_3}s_{\alpha_2}s_{\alpha_3}c_{\alpha_1}^2 \nonumber \\
&&+(-\lambda_{S_2S_3}+\lambda_{HS_3})
s_{\alpha_1}c_{\alpha_1}c_{\alpha_3}
\Big)
\bigg]
\bigg], \nonumber\\
   c_{h_1h_1h_3}&&= i\bigg[
-v_3\left(\lambda_{S_2S_3}s_{\alpha_1}^2
+\lambda_{HS_3}c_{\alpha_1}^2\right)
c_{\alpha_2}^3c_{\alpha_3}
+s_{\alpha_2}^2
\Big(
-\lambda_{HS_3}s_{\alpha_1}s_{\alpha_3}v
+\lambda_{S_2S_3}s_{\alpha_3}v_2c_{\alpha_1}\nonumber\\
&&+\lambda_{S_2S_3}s_{\alpha_1}s_{\alpha_2}v_2c_{\alpha_3}
+\lambda_{HS_3}s_{\alpha_2}v c_{\alpha_1}c_{\alpha_3}
\Big)
-2s_{\alpha_2}c_{\alpha_2}
\bigg[
(\!-\lambda_{S_2S_3}+\lambda_{HS_3})
s_{\alpha_1}s_{\alpha_3}v_3c_{\alpha_1}\nonumber\\
&&+s_{\alpha_2}
\Big(
\mu_3+3\lambda_{S_3}v_3
-\lambda_{S_2S_3}s_{\alpha_1}^2v_3
-\lambda_{HS_3}v_3c_{\alpha_1}^2
\Big)c_{\alpha_3}
\bigg]
+c_{\alpha_2}^2
\bigg[
s_{\alpha_3}
\Big(
-\lambda_{HS_2}s_{\alpha_1}^3v \nonumber\\
&&+2(3\lambda_{S_2}-\lambda_{HS_2})s_{\alpha_1}^2v_2c_{\alpha_1}
+2(-3\lambda_H+\lambda_{HS_2})s_{\alpha_1}vc_{\alpha_1}^2
+\lambda_{HS_2}v_2c_{\alpha_1}^3
\Big)\nonumber\\
&&+s_{\alpha_2}
\Big(
-2\lambda_{S_2S_3}s_{\alpha_1}v_2
+6\lambda_{S_2}s_{\alpha_1}^3v_2
-2\lambda_{HS_3}vc_{\alpha_1}
+3\lambda_{HS_2}s_{\alpha_1}^2vc_{\alpha_1}
+3\lambda_{HS_2}s_{\alpha_1}v_2c_{\alpha_1}^2\nonumber\\
&&+6\lambda_Hvc_{\alpha_1}^3
\Big)c_{\alpha_3}
\bigg]
\bigg],\nonumber\\
    c_{h_1h_2h_2}&&= -i\bigg[
s_{\alpha_3}^2
\left(\lambda_{S_2S_3}s_{\alpha_1}v_2
+\lambda_{HS_3}vc_{\alpha_1}\right)c_{\alpha_2}^3
+2s_{\alpha_3}c_{\alpha_2}^2
\bigg[
\mu_3s_{\alpha_2}s_{\alpha_3}
-v_3\Big(
(-3\lambda_{S_3} \nonumber\\
&&+\lambda_{S_2S_3}s_{\alpha_1}^2)
s_{\alpha_2}s_{\alpha_3}
+\lambda_{HS_3}s_{\alpha_2}s_{\alpha_3}c_{\alpha_1}^2
+(-\lambda_{S_2S_3}+\lambda_{HS_3})
s_{\alpha_1}c_{\alpha_1}c_{\alpha_3}
\Big)
\bigg] \nonumber\\
&&+s_{\alpha_2}v_3
\bigg[
2(-\lambda_{S_2S_3}+\lambda_{HS_3})
s_{\alpha_1}s_{\alpha_2}s_{\alpha_3}c_{\alpha_1}c_{\alpha_3}
+c_{\alpha_1}^2
\Big(
\lambda_{HS_3}s_{\alpha_2}^2s_{\alpha_3}^2
+\lambda_{S_2S_3}c_{\alpha_3}^2
\Big)\nonumber\\
&&+s_{\alpha_1}^2
\Big(
\lambda_{S_2S_3}s_{\alpha_2}^2s_{\alpha_3}^2
+\lambda_{HS_3}c_{\alpha_3}^2
\Big)
\bigg]
+c_{\alpha_2}
\bigg[
s_{\alpha_1}
\Big(
-2(\lambda_{S_2S_3}-3\lambda_{S_2}s_{\alpha_1}^2)
s_{\alpha_2}^2s_{\alpha_3}^2v_2\nonumber\\
&&-2(\lambda_{HS_3}-\lambda_{HS_2}s_{\alpha_1}^2)
s_{\alpha_2}s_{\alpha_3}vc_{\alpha_3}
+\lambda_{HS_2}s_{\alpha_1}^2v_2c_{\alpha_3}^2
\Big)
+s_{\alpha_1}c_{\alpha_1}^2
\Big(
3\lambda_{HS_2}s_{\alpha_2}^2s_{\alpha_3}^2v_2\nonumber\\
&&+4(3\lambda_H-\lambda_{HS_2})
s_{\alpha_2}s_{\alpha_3}vc_{\alpha_3}
+(6\lambda_{S_2}-2\lambda_{HS_2})
v_2c_{\alpha_3}^2
\Big)
+c_{\alpha_1}^3
\Big(
6\lambda_Hs_{\alpha_2}^2s_{\alpha_3}^2v\nonumber\\
&&-2\lambda_{HS_2}s_{\alpha_2}s_{\alpha_3}v_2c_{\alpha_3}
+\lambda_{HS_2}vc_{\alpha_3}^2
\Big)
+c_{\alpha_1}
\Big(
(-2\lambda_{HS_3}+3\lambda_{HS_2}s_{\alpha_1}^2)
s_{\alpha_2}^2s_{\alpha_3}^2v
+2\big(
\lambda_{S_2S_3}\nonumber\\
&&+2(-3\lambda_{S_2}+\lambda_{HS_2})s_{\alpha_1}^2
\big)
s_{\alpha_2}s_{\alpha_3}v_2c_{\alpha_3}
+2(3\lambda_H-\lambda_{HS_2})
s_{\alpha_1}^2vc_{\alpha_3}^2
\Big)
\bigg]
\bigg],\nonumber\\
    c_{h_1h_3h_3} &&= -i\bigg[
\left(\lambda_{S_2S_3}s_{\alpha_1}v_2
+\lambda_{HS_3}vc_{\alpha_1}\right)
c_{\alpha_2}^3c_{\alpha_3}^2
+2c_{\alpha_2}^2c_{\alpha_3}
\bigg[
(-\lambda_{S_2S_3}+\lambda_{HS_3})
s_{\alpha_1}s_{\alpha_3}v_3c_{\alpha_1}\nonumber\\
&&+s_{\alpha_2}
\left(
\mu_3+3\lambda_{S_3}v_3
-\lambda_{S_2S_3}s_{\alpha_1}^2v_3
-\lambda_{HS_3}v_3c_{\alpha_1}^2
\right)c_{\alpha_3}
\bigg]
+c_{\alpha_2}
\bigg[
s_{\alpha_3}^2
\Big(
\lambda_{HS_2}s_{\alpha_1}^3v_2\nonumber\\
&&+2(3\lambda_H-\lambda_{HS_2})s_{\alpha_1}^2vc_{\alpha_1}
+2(3\lambda_{S_2}-\lambda_{HS_2})s_{\alpha_1}v_2c_{\alpha_1}^2
+\lambda_{HS_2}vc_{\alpha_1}^3
\Big)
+2s_{\alpha_2}s_{\alpha_3}\nonumber\\
&&\Big(
s_{\alpha_1}(\lambda_{HS_3}-\lambda_{HS_2}s_{\alpha_1}^2)v
-\left(\lambda_{S_2S_3}
+2(-3\lambda_{S_2}+\lambda_{HS_2})s_{\alpha_1}^2\right)
v_2c_{\alpha_1}\nonumber\\
&&+2(-3\lambda_H+\lambda_{HS_2})
s_{\alpha_1}vc_{\alpha_1}^2
+\lambda_{HS_2}v_2c_{\alpha_1}^3
\Big)c_{\alpha_3}
+s_{\alpha_2}^2
\Big(
-2\lambda_{S_2S_3}s_{\alpha_1}v_2
+6\lambda_{S_2}s_{\alpha_1}^3v_2\nonumber\\
&&-2\lambda_{HS_3}vc_{\alpha_1}
+3\lambda_{HS_2}s_{\alpha_1}^2vc_{\alpha_1}
+3\lambda_{HS_2}s_{\alpha_1}v_2c_{\alpha_1}^2
+6\lambda_Hvc_{\alpha_1}^3
\Big)c_{\alpha_3}^2
\bigg]\nonumber\\
&&+s_{\alpha_2}v_3
\bigg[
2(\lambda_{S_2S_3}-\lambda_{HS_3})
s_{\alpha_1}s_{\alpha_2}s_{\alpha_3}c_{\alpha_1}c_{\alpha_3}
+s_{\alpha_1}^2
\Big(
\lambda_{HS_3}s_{\alpha_3}^2
+\lambda_{S_2S_3}s_{\alpha_2}^2c_{\alpha_3}^2
\Big)\nonumber\\
&&+c_{\alpha_1}^2
\Big(
\lambda_{S_2S_3}s_{\alpha_3}^2
+\lambda_{HS_3}s_{\alpha_2}^2c_{\alpha_3}^2
\Big)
\bigg]
\bigg],\nonumber\\
    c_{h_1h_2h_3} &&= i\bigg[
-s_{\alpha_3}
\left(\lambda_{S_2S_3}s_{\alpha_1}v_2
+\lambda_{HS_3}vc_{\alpha_1}\right)
c_{\alpha_2}^3c_{\alpha_3}
+c_{\alpha_2}^2
\bigg[
(\lambda_{S_2S_3}-\lambda_{HS_3})
s_{\alpha_1}s_{\alpha_3}^2v_3c_{\alpha_1}\nonumber\\
&&+2s_{\alpha_2}s_{\alpha_3}
\Big(-\mu_3+v_3(-3\lambda_{S_3}
+\lambda_{S_2S_3}s_{\alpha_1}^2
+\lambda_{HS_3}c_{\alpha_1}^2)\Big)c_{\alpha_3}
+(-\lambda_{S_2S_3}+\lambda_{HS_3})
s_{\alpha_1}v_3c_{\alpha_1}c_{\alpha_3}^2
\bigg]\nonumber\\
&&+s_{\alpha_2}v_3
\bigg[
s_{\alpha_1}^2
(\lambda_{HS_3}-\lambda_{S_2S_3}s_{\alpha_2}^2)
s_{\alpha_3}c_{\alpha_3}
+(\lambda_{S_2S_3}-\lambda_{HS_3}s_{\alpha_2}^2)
s_{\alpha_3}c_{\alpha_1}^2c_{\alpha_3}\nonumber\\
&&+(\lambda_{S_2S_3}-\lambda_{HS_3})
s_{\alpha_1}s_{\alpha_2}c_{\alpha_1}
(-s_{\alpha_3}^2+c_{\alpha_3}^2)
\bigg]
+c_{\alpha_2}
\bigg[
c_{\alpha_1}^3
\Big(
-\lambda_{HS_2}s_{\alpha_2}s_{\alpha_3}^2v_2\nonumber\\
&&+(\lambda_{HS_2}-6\lambda_Hs_{\alpha_2}^2)
s_{\alpha_3}vc_{\alpha_3}
+\lambda_{HS_2}s_{\alpha_2}v_2c_{\alpha_3}^2
\Big)
+c_{\alpha_1}
\Big(
\big(\lambda_{S_2S_3}+2(-3\lambda_{S_2}+\lambda_{HS_2})s_{\alpha_1}^2\big)
s_{\alpha_2}s_{\alpha_3}^2v_2\nonumber\\
&&+\big(
6\lambda_Hs_{\alpha_1}^2
+2\lambda_{HS_3}s_{\alpha_2}^2
-\lambda_{HS_2}s_{\alpha_1}^2(2+3s_{\alpha_2}^2)
\big)
s_{\alpha_3}vc_{\alpha_3}
-\big(\lambda_{S_2S_3}+2(-3\lambda_{S_2}\nonumber\\
&&+\lambda_{HS_2})s_{\alpha_1}^2\big)
s_{\alpha_2}v_2c_{\alpha_3}^2
\Big)
+s_{\alpha_1}c_{\alpha_1}^2
\Big(
2(3\lambda_H-\lambda_{HS_2})
s_{\alpha_2}s_{\alpha_3}^2v
+(6\lambda_{S_2}-\lambda_{HS_2}(2+3s_{\alpha_2}^2))
s_{\alpha_3}v_2c_{\alpha_3}\nonumber\\
&&+2(-3\lambda_H+\lambda_{HS_2})
s_{\alpha_2}vc_{\alpha_3}^2
\Big)
+s_{\alpha_1}
\Big(
(-\lambda_{HS_3}+\lambda_{HS_2}s_{\alpha_1}^2)
s_{\alpha_2}s_{\alpha_3}^2v
+\big(
\lambda_{HS_2}s_{\alpha_1}^2\nonumber\\
&&+2(\lambda_{S_2S_3}-3\lambda_{S_2}s_{\alpha_1}^2)s_{\alpha_2}^2
\big)
s_{\alpha_3}v_2c_{\alpha_3}
+(\lambda_{HS_3}-\lambda_{HS_2}s_{\alpha_1}^2)
s_{\alpha_2}vc_{\alpha_3}^2
\Big)
\bigg]
\bigg], \nonumber\\
    c_{h_2h_2h_3} &&= i\bigg[
-2s_{\alpha_3}^2
\left(\mu_3+3\lambda_{S_3}v_3\right)
c_{\alpha_2}^3c_{\alpha_3}
+s_{\alpha_1}c_{\alpha_1}^2
\bigg[
2(-3\lambda_H+\lambda_{HS_2})
s_{\alpha_2}^2s_{\alpha_3}^3v \nonumber\\
&&+s_{\alpha_2}
(-12\lambda_{S_2}+4\lambda_{HS_2}
+3\lambda_{HS_2}s_{\alpha_2}^2)
s_{\alpha_3}^2v_2c_{\alpha_3}
+(-3\lambda_{HS_2}
+12\lambda_Hs_{\alpha_2}^2 \nonumber\\
&&-4\lambda_{HS_2}s_{\alpha_2}^2)
s_{\alpha_3}vc_{\alpha_3}^2
+2(3\lambda_{S_2}-\lambda_{HS_2})
s_{\alpha_2}v_2c_{\alpha_3}^3
\bigg]
+s_{\alpha_1}^3
\bigg[
-\lambda_{HS_2}s_{\alpha_2}^2s_{\alpha_3}^3v \nonumber\\
&&+2s_{\alpha_2}(-\lambda_{HS_2}
+3\lambda_{S_2}s_{\alpha_2}^2)
s_{\alpha_3}^2v_2c_{\alpha_3}
+2(-3\lambda_H
+\lambda_{HS_2}s_{\alpha_2}^2)
s_{\alpha_3}vc_{\alpha_3}^2
+\lambda_{HS_2}s_{\alpha_2}v_2c_{\alpha_3}^3
\bigg] \nonumber\\
&&+s_{\alpha_1}^2c_{\alpha_1}
\bigg[
2(3\lambda_{S_2}-\lambda_{HS_2})
s_{\alpha_2}^2s_{\alpha_3}^3v_2
+s_{\alpha_2}
(-12\lambda_H+4\lambda_{HS_2}
+3\lambda_{HS_2}s_{\alpha_2}^2)
s_{\alpha_3}^2vc_{\alpha_3}\nonumber\\
&&+(3\lambda_{HS_2}
-12\lambda_{S_2}s_{\alpha_2}^2
+4\lambda_{HS_2}s_{\alpha_2}^2)
s_{\alpha_3}v_2c_{\alpha_3}^2
+2(3\lambda_H-\lambda_{HS_2})
s_{\alpha_2}vc_{\alpha_3}^3
\bigg]\nonumber\\
&&+c_{\alpha_1}^3
\bigg[
\lambda_{HS_2}s_{\alpha_2}^2s_{\alpha_3}^3v_2
+2s_{\alpha_2}(-\lambda_{HS_2}
+3\lambda_Hs_{\alpha_2}^2)
s_{\alpha_3}^2vc_{\alpha_3}
+2(3\lambda_{S_2}
-\lambda_{HS_2}s_{\alpha_2}^2)
s_{\alpha_3}v_2c_{\alpha_3}^2\nonumber\\
&&+\lambda_{HS_2}s_{\alpha_2}vc_{\alpha_3}^3
\bigg]
-v_3c_{\alpha_2}
\bigg[
2(\lambda_{S_2S_3}-\lambda_{HS_3})
s_{\alpha_1}s_{\alpha_2}s_{\alpha_3}c_{\alpha_1}
(s_{\alpha_3}^2-2c_{\alpha_3}^2)\nonumber\\
&&+c_{\alpha_1}^2c_{\alpha_3}
\Big(
(-2\lambda_{S_2S_3}
+3\lambda_{HS_3}s_{\alpha_2}^2)s_{\alpha_3}^2
+\lambda_{S_2S_3}c_{\alpha_3}^2
\Big)
+s_{\alpha_1}^2c_{\alpha_3}
\Big(
(-2\lambda_{HS_3}
+3\lambda_{S_2S_3}s_{\alpha_2}^2)s_{\alpha_3}^2\nonumber\\
&&+\lambda_{HS_3}c_{\alpha_3}^2
\Big)
\bigg]
+s_{\alpha_3}c_{\alpha_2}^2
\bigg[
c_{\alpha_1}
\Big(
\lambda_{S_2S_3}s_{\alpha_3}^2v_2
+3\lambda_{HS_3}s_{\alpha_2}s_{\alpha_3}vc_{\alpha_3}
-2\lambda_{S_2S_3}v_2c_{\alpha_3}^2
\Big)\nonumber\\
&&+s_{\alpha_1}
\Big(
-\lambda_{HS_3}s_{\alpha_3}^2v
+3\lambda_{S_2S_3}s_{\alpha_2}s_{\alpha_3}v_2c_{\alpha_3}
+2\lambda_{HS_3}vc_{\alpha_3}^2
\Big)
\bigg]
\bigg],\nonumber\\
    c_{h_2h_3h_3} &&= -i\bigg[
2s_{\alpha_3}(\mu_3+3\lambda_{S_3}v_3)c_{\alpha_2}^3c_{\alpha_3}^2
+s_{\alpha_1}^2c_{\alpha_1}
\bigg(
2(-3\lambda_H+\lambda_{HS_2})s_{\alpha_2}s_{\alpha_3}^3v
+(3\lambda_{HS_2}-12\lambda_{S_2}s_{\alpha_2}^2\nonumber\\
&&+4\lambda_{HS_2}s_{\alpha_2}^2)s_{\alpha_3}^2v_2c_{\alpha_3}
+s_{\alpha_2}(12\lambda_H-4\lambda_{HS_2}-3\lambda_{HS_2}s_{\alpha_2}^2)s_{\alpha_3}vc_{\alpha_3}^2
+2(3\lambda_{S_2}-\lambda_{HS_2})s_{\alpha_2}^2v_2c_{\alpha_3}^3
\bigg)\nonumber \\
&&+c_{\alpha_1}^3
\bigg(
-\lambda_{HS_2}s_{\alpha_2}s_{\alpha_3}^3v
+2(3\lambda_{S_2}-\lambda_{HS_2}s_{\alpha_2}^2)s_{\alpha_3}^2v_2c_{\alpha_3}
+2s_{\alpha_2}(\lambda_{HS_2}-3\lambda_Hs_{\alpha_2}^2)s_{\alpha_3}vc_{\alpha_3}^2 \nonumber \\
&&+\lambda_{HS_2}s_{\alpha_2}^2v_2c_{\alpha_3}^3
\bigg)
-s_{\alpha_1}^3
\bigg(
\lambda_{HS_2}s_{\alpha_2}s_{\alpha_3}^3v_2
+2(3\lambda_H-\lambda_{HS_2}s_{\alpha_2}^2)s_{\alpha_3}^2vc_{\alpha_3} \nonumber\\
&&-2s_{\alpha_2}(\lambda_{HS_2}-3\lambda_{S_2}s_{\alpha_2}^2)s_{\alpha_3}v_2c_{\alpha_3}^2
+\lambda_{HS_2}s_{\alpha_2}^2vc_{\alpha_3}^3
\bigg)
+s_{\alpha_1}c_{\alpha_1}^2
\bigg(
2(-3\lambda_{S_2}+\lambda_{HS_2})s_{\alpha_2}s_{\alpha_3}^3v_2 \nonumber \\
&&+(-3\lambda_{HS_2}+12\lambda_Hs_{\alpha_2}^2-4\lambda_{HS_2}s_{\alpha_2}^2)s_{\alpha_3}^2vc_{\alpha_3}
+s_{\alpha_2}(12\lambda_{S_2}-4\lambda_{HS_2}-3\lambda_{HS_2}s_{\alpha_2}^2)s_{\alpha_3}v_2c_{\alpha_3}^2 \nonumber\\
&&+2(-3\lambda_H+\lambda_{HS_2})s_{\alpha_2}^2vc_{\alpha_3}^3
\bigg)
+v_3c_{\alpha_2}
\bigg[
-2(\lambda_{S_2S_3}-\lambda_{HS_3})
s_{\alpha_1}s_{\alpha_2}c_{\alpha_1}c_{\alpha_3}
(-2s_{\alpha_3}^2+c_{\alpha_3}^2)\nonumber\\
&&+s_{\alpha_1}^2s_{\alpha_3}
\Big(
\lambda_{HS_3}s_{\alpha_3}^2
+(-2\lambda_{HS_3}+3\lambda_{S_2S_3}s_{\alpha_2}^2)c_{\alpha_3}^2
\Big)
+c_{\alpha_1}^2
\Big(
\lambda_{S_2S_3}s_{\alpha_3}^3
+(-2\lambda_{S_2S_3}s_{\alpha_3}\nonumber\\
&&+3\lambda_{HS_3}s_{\alpha_2}^2s_{\alpha_3})c_{\alpha_3}^2
\Big)
\bigg]
+c_{\alpha_2}^2c_{\alpha_3}
\bigg[
c_{\alpha_1}
\Big(
-2\lambda_{S_2S_3}s_{\alpha_3}^2v_2
-3\lambda_{HS_3}s_{\alpha_2}s_{\alpha_3}vc_{\alpha_3}
+\lambda_{S_2S_3}v_2c_{\alpha_3}^2
\Big)\nonumber\\
&&-s_{\alpha_1}
\Big(
-2\lambda_{HS_3}s_{\alpha_3}^2v
+3\lambda_{S_2S_3}s_{\alpha_2}s_{\alpha_3}v_2c_{\alpha_3}
+\lambda_{HS_3}vc_{\alpha_3}^2
\Big)
\bigg]
\bigg],\nonumber\\
    c_{h_1\eta_2 \eta_2} && = -i\Big(
2c_{\alpha_2}\lambda_{S_3}s_{\alpha_1}v_2
+\lambda_{S_2S_3}s_{\alpha_2}v_3
+c_{\alpha_1}c_{\alpha_2}\lambda_{HS_3}v
\Big),\nonumber\\
   c_{h_2 \eta_2 \eta_2}&&= -i\Big(
2c_{\alpha_1}c_{\alpha_3}\lambda_{S_3}v_2
-2\lambda_{S_3}s_{\alpha_1}s_{\alpha_2}s_{\alpha_3}v_2
+c_{\alpha_2}\lambda_{S_2S_3}s_{\alpha_3}v_3
-c_{\alpha_3}\lambda_{HS_3}s_{\alpha_1}v\nonumber \\
&&-c_{\alpha_1}\lambda_{HS_3}s_{\alpha_2}s_{\alpha_3}v
\Big),\nonumber\\
    c_{h_3 \eta_2 \eta_2} &&= i\Big(
2c_{\alpha_3}\lambda_{S_3}s_{\alpha_1}s_{\alpha_2}v_2
+2c_{\alpha_1}\lambda_{S_3}s_{\alpha_3}v_2
-c_{\alpha_2}c_{\alpha_3}\lambda_{S_2S_3}v_3 \nonumber\\
&&+c_{\alpha_1}c_{\alpha_3}\lambda_{HS_3}s_{\alpha_2}v
-\lambda_{HS_3}s_{\alpha_1}s_{\alpha_3}v
\Big),\nonumber\\
    c_{h_1 h_1 \eta_2 \eta_2} && = -i\left(\lambda_{S_2S_3}s_{\alpha_2}^2
+2\lambda_{S_2}s_{\alpha_1}^2c_{\alpha_2}^2
+\lambda_{HS_2}c_{\alpha_1}^2c_{\alpha_2}^2\right),\nonumber\\
   c_{h_2 h_2 \eta_2 \eta_2} &&=-i\Big[
s_{\alpha_3}^2
\Big(2\lambda_{S_2}s_{\alpha_1}^2s_{\alpha_2}^2
+\lambda_{S_2S_3}c_{\alpha_2}^2\Big)
+2(-2\lambda_{S_2}+\lambda_{HS_2})
s_{\alpha_1}s_{\alpha_2}s_{\alpha_3}
c_{\alpha_1}c_{\alpha_3}\nonumber \\
&&+\lambda_{HS_2}s_{\alpha_1}^2c_{\alpha_3}^2
+c_{\alpha_1}^2
\Big(\lambda_{HS_2}s_{\alpha_2}^2s_{\alpha_3}^2
+2\lambda_{S_2}c_{\alpha_3}^2\Big)
\Big],\nonumber\\
c_{h_3 h_3 \eta_2 \eta_2} &&= -i\Big[
2(2\lambda_{S_2}-\lambda_{HS_2})
s_{\alpha_1}s_{\alpha_2}s_{\alpha_3}
c_{\alpha_1}c_{\alpha_3}
+\lambda_{S_2S_3}c_{\alpha_2}^2c_{\alpha_3}^2\nonumber \\
&&+s_{\alpha_1}^2
\Big(\lambda_{HS_2}s_{\alpha_3}^2
+2\lambda_{S_2}s_{\alpha_2}^2c_{\alpha_3}^2\Big)
+c_{\alpha_1}^2
\Big(2\lambda_{S_2}s_{\alpha_3}^2
+\lambda_{HS_2}s_{\alpha_2}^2c_{\alpha_3}^2\Big)
\Big],\nonumber\\
    c_{h_1 h_2 \eta_2 \eta_2} &&= i\,c_{\alpha_2}\Big[
-\left(\lambda_{S_2S_3}-2\lambda_{S_2}s_{\alpha_1}^2\right)s_{\alpha_2}s_{\alpha_3}
+\lambda_{HS_2}s_{\alpha_2}s_{\alpha_3}c_{\alpha_1}^2\nonumber\\
&&+\left(-2\lambda_{S_2}+\lambda_{HS_2}\right)
s_{\alpha_1}c_{\alpha_1}c_{\alpha_3}
\Big],\nonumber\\
    c_{h_1 h_3 \eta_2 \eta_2} &&= i\,c_{\alpha_2}\left[
(2\lambda_{S_2}-\lambda_{HS_2})s_{\alpha_1}s_{\alpha_3}c_{\alpha_1}
+s_{\alpha_2}\left(-\lambda_{S_2S_3}
+2\lambda_{S_2}s_{\alpha_1}^2
+\lambda_{HS_2}c_{\alpha_1}^2\right)c_{\alpha_3}
\right],\nonumber\\
    c_{h_2 h_3 \eta_2 \eta_2} &&= i\Big[
(2\lambda_{S_2}-\lambda_{HS_2}s_{\alpha_2}^2)
s_{\alpha_3}c_{\alpha_1}^2c_{\alpha_3}
+s_{\alpha_3}
\Big(s_{\alpha_1}^2
(\lambda_{HS_2}-2\lambda_{S_2}s_{\alpha_2}^2)
-\lambda_{S_2S_3}c_{\alpha_2}^2\Big)c_{\alpha_3} \nonumber \\
&&+(2\lambda_{S_2}-\lambda_{HS_2})
s_{\alpha_1}s_{\alpha_2}c_{\alpha_1}
(-s_{\alpha_3}^2+c_{\alpha_3}^2)
\Big].
\end{eqnarray}
The multiplicative factors of the couplings of the CP-even scalar with the quarks are
\begin{eqnarray}
  &&  c_{h_1 q q}=-i\,c_{\alpha_1}c_{\alpha_2}, \nonumber\\
   &&  c_{h_2 q q}=i\left(c_{\alpha_3}s_{\alpha_1}+c_{\alpha_1}s_{\alpha_2}s_{\alpha_3}\right),\nonumber\\
    &&  c_{h_3 q q}= i\left(c_{\alpha_1}c_{\alpha_3}s_{\alpha_2}-s_{\alpha_1}s_{\alpha_3}\right).
\end{eqnarray}

\subsection*{Relevant couplings for muon collider processes and muon $g-2$ }
The coupling factors for the Yukawa interactions involved in muon collider processes and muon $g-2$ are given by
\begin{align}\label{app:g-2}
& \eta_3\,\bar{\mu}\,\mu :\,  -\frac{\cos\theta_R\,\sin\theta_L}{\sqrt{2}}\,y_\psi \gamma^5, \nonumber\\
& \eta_3\,\bar{vl}\,vl :\,  \frac{\cos\theta_L\,\sin\theta_R}{\sqrt{2}}\,y_\psi \gamma^5, \nonumber\\
& \eta_3\,\bar{\mu}\,vl :\,  -\frac{y_\psi}{\sqrt{2}}\left(\cos\theta_L\cos\theta_R\,P_L +\sin\theta_L\sin\theta_R\,P_R
\right), \nonumber\\
& h_1\,\bar{\mu}\,\mu :\, \frac{i}{\sqrt{2}}\,\cos\theta_R \Big(y_\psi s_{\alpha_2}\sin\theta_L - c_{\alpha_1}c_{\alpha_2}\cos\theta_L \,y_{\mu^\prime}\Big),\nonumber\\
&    h_2\,\bar{\mu}\,\mu :\, \frac{i}{\sqrt{2}}\,\cos\theta_R\left[y_\psi c_{\alpha_2}s_{\alpha_3}\sin\theta_L+y_{\mu^\prime}\cos\theta_L(c_{\alpha_3}s_{\alpha_1}+c_{\alpha_1}s_{\alpha_2}s_{\alpha_3})\right],\nonumber\\
 &    h_3\,\bar{\mu}\,\mu :\, \frac{i}{\sqrt{2}}\cos\theta_R\left[y_\psi c_{\alpha_2}c_{\alpha_3}\sin\theta_L+y_{\mu^\prime}\cos\theta_L(c_{\alpha_1}c_{\alpha_3}s_{\alpha_2}-s_{\alpha_1}s_{\alpha_3})\right],\nonumber\\
 &    h_1\,\bar{vl}\,vl :\, -\frac{i}{\sqrt{2}}\sin\theta_R(y_\psi s_{\alpha_2}\cos\theta_L + y_{\mu^\prime} c_{\alpha_1}c_{\alpha_2}\sin\theta_L),\nonumber\\
  &   h_2\,\bar{vl}\,vl :\, -\frac{i}{\sqrt{2}}\sin\theta_R\left[y_\psi c_{\alpha_2}s_{\alpha_3}\cos\theta_L-y_{\mu^\prime}\sin\theta_L(c_{\alpha_3}s_{\alpha_1}+c_{\alpha_1}s_{\alpha_2}s_{\alpha_3})\right],\nonumber\\
   & h_3\,\bar{vl}\,vl :\, -\frac{i}{\sqrt{2}}\sin\theta_R\left[y_\psi c_{\alpha_2}c_{\alpha_3}\cos\theta_L+y_{\mu^\prime}\sin\theta_L(s_{\alpha_1}s_{\alpha_3}-c_{\alpha_1}c_{\alpha_3}s_{\alpha_2})\right],\nonumber\\
   &h_1\,\bar{\mu}\,vl :\, -\frac{i}{\sqrt{2}}\Big[\cos\theta_R\Big(y_\psi\cos\theta_Ls_{\alpha_2}+y_{\mu^\prime} c_{\alpha_1}c_{\alpha_2}\sin\theta_L\Big) P_L \nonumber \\
   &\qquad\qquad + \sin\theta_R\Big(y_{\mu^\prime} c_{\alpha_1}c_{\alpha_2}\cos\theta_L - y_\psi s_{\alpha_2}\sin\theta_L\Big) P_R\Big],\nonumber\\
   &h_2\,\bar{\mu}\,vl :\, -\frac{i}{\sqrt{2}}\left[\cos\theta_R\Big(y_\psi c_{\alpha_2}s_{\alpha_3}\cos\theta_L-y_{\mu^\prime}\sin\theta_L(c_{\alpha_3}s_{\alpha_1}+c_{\alpha_1}s_{\alpha_2}s_{\alpha_3})\Big)P_L\right.\nonumber\\&
   \qquad \qquad -\left.\sin\theta_R\Big(y_\psi \sin\theta_L c_{\alpha_2}s_{\alpha_3}+y_{\mu^\prime} \cos\theta_L(c_{\alpha_3}s_{\alpha_1}+c_{\alpha_1}s_{\alpha_2}s_{\alpha_3})\Big)P_R \right],\nonumber\\
   &h_3\,\bar{\mu}\,vl :\, -\frac{i}{\sqrt{2}}\left[\cos\theta_R\Big(y_\psi \cos\theta_L c_{\alpha_2}c_{\alpha_3}+y_{\mu^\prime}\sin\theta_L(s_{\alpha_1}s_{\alpha_3}-c_{\alpha_1}c_{\alpha_3}s_{\alpha_2})\Big)P_L\right.\nonumber\\&
   \qquad \qquad -\left. y_{\mu^\prime}\sin\theta_R \Big(y_\psi\sin\theta_Lc_{\alpha_2}c_{\alpha_3}+y_{\mu^\prime}\cos\theta_L(c_{\alpha_1}c_{\alpha_3}s_{\alpha_2}-s_{\alpha_1}s_{\alpha_3})\Big)P_R\right].
\end{align}
The coupling factors of the VLM and muon to the $Z$ boson and photon are given by:
 \begin{eqnarray}
   & & Z\, \bar{\mu}\, \mu : \,  -\frac{i e\gamma^\mu}{2 c_w s_w}\Big[\Big(-2\sin^2\theta_L\,s_w^2+\cos^2\theta_L\left(c_w^2-s_w^2\right)\Big)P_L -2 s_w^2P_R\Big], \nonumber\\
   %%%%
   & & Z\, \bar{vl}\, vl : \, -\frac{i e\gamma^\mu}{2 c_w s_w}\Big[\Big(c_w^2\sin^2\theta_L-\left(2\cos^2\theta_L+\sin^2\theta_L\right)s_w^2\Big)P_L-2 s_w^2 P_R
\Big],\nonumber\\
   & & Z\, \bar{\mu}\, vl : \,  -\frac{i\,e\,\cos\theta_L\sin\theta_L}{2c_w s_w}\,\gamma^{\mu}P_L, \nonumber\\
   & & A\, \bar{\mu}\, \mu : \, -ie\gamma^\mu, \nonumber\\
   & & A\, \bar{vl}\, vl : \, -ie\gamma^\mu ,
\end{eqnarray}
where $s_w\equiv\sin\theta_w$, $c_w\equiv\cos\theta_w$, with $\theta_w$ denoting the weak mixing angle.

\section{Field dependent masses}
At high temperatures, mixing occurs between the three CP-even scalars. The field-dependent symmetric mass-squared matrix is 
\begin{equation}\label{eq:massMatrix}
   M^2({h,s_2,s_3})= \begin{pmatrix}
        m_{hh}^2 & m^2_{hs_2} & m^2_{hs_3}\\
        m_{hs_2}^2 & m_{s_2s_2}^2 & m_{s_2s_3}^2\\
        m_{hs_3 }^2 & m_{s_2s_3}^2 & m_{s_3 s_3}^2
    \end{pmatrix}.
\end{equation}
In terms of background fields, the matrix elements are
\begin{align}
&m_{hh}^2  = \mu_H^2+ 3 \lambda_H\, h^2 +\frac{\lambda _{HS_2}}{2} s_2 ^2 +\frac{\lambda_{HS_3}}{2}s_3^2, \n\\
&m_{s_2s_2}^2= \mu_{S_2}^2+ 3  \lambda_{S_2}\, s_2^2  +\frac{\lambda_{HS_2}}{2}h^2 +\frac{\lambda _{S_2 S_3}}{2} s_3 ^2+\mu_2, \n\\
&m_{s_3s_3}^2 = \mu_{S_3}^2 + 3 \lambda_{S_3}\, s_3 ^2 + \frac{\lambda_{HS_3}}{2} h^2 + \frac{\lambda_{S_2 S_3}}{2}s_2^2 + 2\mu_3\, s_3, \n\\
&m_{hs_2}^2 = \lambda_{HS_2} \, h\, s_2 , \quad
m_{hs_3}^2 = \lambda_{HS_3}\, h\, s_3 ,\quad
m_{s_2s_3}^2= \lambda_{S_2 S_3}\, s_2 \,s_3.
\end{align}
 The eigenvalues of the scalar mass matrix contribute to the finite temperature corrections. 
The field-dependent masses of the CP-odd scalars $\chi_i$, $\eta_2$, and $\eta_3$ are
\begin{eqnarray}
    && m_{\chi_i}^2({h,s_2,s_3}) = \mu_H^2 +\lambda_H h^2 + \frac{\lambda_{HS_2} }{2}s_2^2+  \frac{\lambda_{HS_3} }{2} s_3^2,  \\
    && m_{\eta_2}^2({h,s_2,s_3}) = \mu_{S_2}^2+\lambda_{S_2} s_2^2+\frac{\lambda_{HS_2}}{2} h^2+\frac{\lambda_{S_2 S_3}}{2}s_3 ^2 - \mu_2,\\
    && m_{\eta_3}^2({h,s_2,s_3}) = \mu_{S_3}^2+\lambda_{S_3} s_3^2+\frac{\lambda_{HS_3}}{2} h^2+\frac{\lambda_{S_2 S_3}}{2}s_2 ^2 - 2\mu_3\, s_3,
\end{eqnarray}
On the other hand, the field-dependent masses of the top quark and electroweak gauge bosons are
\begin{eqnarray}
   m_t^2(h) = \frac{1}{2} y_t^2 h^2,\quad m_W^2(h) = \frac{1}{4} g^2 h^2,\quad m_Z^2(h) = \frac{1}{4} (g^2 + g^{\prime 2}) h^2,
\end{eqnarray}
where the $g$ and $g^\prime$ are gauge couplings associated with $SU(2)_L$ and $U(1)_Y$ group, and $y_t$ is the top Yukawa coupling.

\label{Bibliography}
\bibliographystyle{JHEP}
\bibliography{Refs}

\end{document}